\documentclass[11pt,letterpaper]{article}
\usepackage{jheppub}

\usepackage[percent]{overpic}
\usepackage{graphicx}
\usepackage{bbm}
\usepackage{subfigure}
\usepackage{amsmath}
\usepackage{amssymb}
\usepackage{mathtools}
\usepackage{amsthm}
\usepackage{mathrsfs}
\usepackage{marvosym}
\usepackage{dsfont}
\usepackage[labelformat=simple]{subcaption}
\usepackage{xcolor}
\usepackage{braket}
\usepackage{cleveref}
\usepackage{comment}
\usepackage{float}
\usepackage{fancybox}
\usepackage[skins,theorems]{tcolorbox}
\tcbset{highlight math style={enhanced,
		colframe=black,colback=white,arc=0pt,boxrule=1pt}}
\usepackage{overpic}

\usepackage{subcaption} 

\allowdisplaybreaks

\definecolor{dark-gray}{gray}{0.20}
\definecolor{gray}{gray}{0.30}
\definecolor{light-gray}{gray}{0.80}
\definecolor{dark-red}{rgb}{0.7,0,0}
\definecolor{dark-green}{rgb}{0.1,0.4,0}
\definecolor{dark-blue}{rgb}{0.3,0.3,0.7}
\definecolor{light-blue}{rgb}{0.8,0.8,1}
\definecolor{swamp}{RGB}{240, 199, 197}

\newcommand{\be}{\begin{equation}}
	\newcommand{\ee}{\end{equation}}

\def\be{\begin{equation}}
	\def\ee{\end{equation}}
\def\bea{\begin{eqnarray}}
	\def\eea{\end{eqnarray}}

\newcommand{\beq}{\begin{equation}}  \newcommand{\eeq}{\end{equation}}
\newcommand{\bal}{\begin{aligned}}   \newcommand{\eal}{\end{aligned}}
\def\beqa{\begin{eqnarray}}
	\def\eeqa{\end{eqnarray}}

\DeclareMathOperator{\csch}{csch}

\def\Mpf{M_{\text{Pl,}\, 4}}

\def\LQG{\Lambda_{\text{QG}}}

\numberwithin{equation}{section}

\def\simleq{\; \raise0.3ex\hbox{$<$\kern-0.75em
		\raise-1.1ex\hbox{$\sim$}}\; }
\def\simgeq{\; \raise0.3ex\hbox{$>$\kern-0.75em
		\raise-1.1ex\hbox{$\sim$}}\; }

\numberwithin{equation}{section}

\hypersetup{
	colorlinks=true,
	linkcolor=dark-blue,
	citecolor=dark-red,
	urlcolor=dark-green,
	linktoc=page
}

\theoremstyle{remark}

\newtheoremstyle{named}{}{}{\itshape}{}{\bfseries}{.}{.5em}{#3}
\theoremstyle{named}

\title{\centering Black Hole Entropy, Quantum Corrections\\ and EFT Transitions}

\author{Alberto Castellano$^{1,2}$,} 
\author{Matteo Zatti$^{3}$}
\affiliation{$^1$Enrico Fermi Institute \& Kadanoff Center for Theoretical Physics,\\
University of Chicago, Chicago, IL 60637, USA}
\affiliation{$^2$Kavli Institute for Cosmological Physics,\\
University of Chicago, Chicago, IL
60637, USA}
\affiliation{$^3$Max-Planck-Institut f\"ur Physik (Werner-Heisenberg-Institut),\\
Boltzmannstrasse 8, 85748 Garching bei M\"unchen, Germany}

\emailAdd{acastellano@uchicago.edu, zatti@mpp.mpg.de}

\abstract{We revisit and study quantum corrections to the supersymmetric entropy of BPS black holes in 4d $\mathcal{N}=2$ effective field theories (EFTs), which can be obtained from Type IIA string theory compactified on a Calabi–Yau threefold. Macroscopically, these corrections arise from an infinite series of higher-derivative F-terms that encode certain modifications to the two-derivative supergravity effective action. Within the large volume regime, we analyze in detail the moduli dependence of these semi-classical contributions and explore their implications for the black hole entropy. As a byproduct, we show that the latter captures, in a rather intricate way, the transition between four- and five-dimensional dual EFT descriptions. In fact, the expansion parameter $\alpha$ controlling the relevant asymptotic series can be related to the ratio of the black hole horizon and the Kaluza-Klein length-scale, given here by the inverse D0-brane mass. Furthermore, we are able to resum the series into a well-behaved convergent expression for all values of $\alpha$. This demonstrates, in turn, that (stable) black holes can, indeed, probe scales besides the quantum gravity cutoff. More precisely, by examining two representative BPS systems ---the D0-D2-D4 and D2-D6 black hole solutions--- we explicitly illustrate how highly non-local yet perturbative quantum effects resolve the divergences, ultimately leading to a well-defined entropy function. Additionally, in special cases, we show that one can take a suitable decompactification limit to 5d and verify that the corrected entropy function reproduces the exact microstate counting of the underlying five-dimensional black string. Our results also clarify the role of certain non-perturbative quantum corrections, which, remarkably, do not modify any of our prior conclusions.}

\begin{document}
	\makeatletter
	\let\old@fpheader\@fpheader
	\renewcommand{\@fpheader}{\vspace*{-0.1cm} \hfill EFI-25-2\\ \vspace* {-0.1cm} \hfill MPP-2025-13}
	\makeatother
	
	\maketitle
	\setcounter{page}{1}
	\pagenumbering{roman}

	\hypersetup{
		pdftitle={Black Hole Entropy, Quantum Corrections \& EFT Transitions},
		pdfauthor={Alberto Castellano, Matteo Zatti},
		pdfsubject={}
	}

\newcommand{\remove}[1]{\textcolor{red}{\sout{#1}}}

\newpage
\pagenumbering{arabic} 

\section{Introduction and Summary}
\label{s:intro}

Black holes serve as central objects both in classical General Relativity and quantum gravity, offering key insights into the fundamental nature of spacetime and high energy physical phenomena. One of the most profound results uncovered in our quest to understand black hole physics is the Bekenstein-Hawking formula \cite{Bekenstein:1972tm,Hawking:1975vcx}, which relates the entropy of a black hole to one-quarter of its event horizon area (in units where $G_N=1$). In the context of string theory and supergravity theories, supersymmetric black holes oftentimes provide a controlled environment which is ideal for studying quantum corrections to this semi-classical relation. The latter turn out to be of significant interest, since they may guide us towards a deeper understanding of the fundamental (i.e., microscopic) degrees of freedom of quantum gravity.

\medskip

As is widely expected ---given the non-renormalizability of Einstein's gravity at the quantum level, the structure of any gravitational Effective Field Theory (EFT) should be such that the suppression of generic higher-derivative and higher-curvature operators relative to the Einstein-Hilbert term is dictated by some specific energy scale \cite{vandeHeisteeg:2022btw,Cribiori:2022nke,vandeHeisteeg:2023ubh,vandeHeisteeg:2023dlw,Castellano:2023aum}, namely the quantum gravity or species cut-off \cite{Dvali:2007hz,Dvali:2009ks,Dvali:2010vm, Dvali:2012uq} (see also \cite{Castellano:2024bna} for a comprehensive treatment of this subject). However, it is also known that the same kind of higher-dimensional operators in EFTs coupled to gravity ---e.g., those arising from string or M-theory--- quite often exhibit an explicit suppression by (even parametrically) lower scales, such as the Kaluza-Klein (KK) mass \cite{Castellano:2023aum,Aoufia:2024awo, Calderon-Infante:2025ldq}. In fact, a simple realization of this scenario is given by four-dimensional $\mathcal{N}=2$ supergravities obtained from compactifying Type IIA string theory on a Calabi--Yau threefold. There, the lightest KK scale can sometimes correspond to the D0-brane mass, whose associated species cutoff is given by certain 5d Planck scale \cite{Castellano:2022bvr}. What happens then, when we reach the KK scale, is that the 4d description breaks down, signaling that we should switch to the dual five-dimensional EFT arising from considering M-theory on the \emph{same} Calabi--Yau space. An interesting question that one can ask, given this state of affairs, is whether and how such EFT transitions could be characterized using the thermodynamics of existing black hole solutions in the theory. Indeed, one may expect that, in general, the solutions themselves must be subject to some kind of phase transition. For instance, these could correspond to transitions of the Gregory-Laflamme \cite{Gregory:1993vy,Gregory:1994bj} or Horowitz-Polchinski \cite{Horowitz:1996nw,Horowitz:1997jc} type.\footnote{See \cite{Brustein:2021ifl,Chen:2021emg,Chen:2021dsw,Urbach:2022xzw,Balthazar:2022szl,Balthazar:2022hno,Ceplak:2023afb,Herraez:2024kux,Albertini:2024hwi,Chu:2024ggi,Emparan:2024mbp,Ceplak:2024dxm,Bedroya:2024igb,Chu:2025fko} for recent developments regarding this kind of transitions in quantum gravity and string theory.} Nevertheless, by sticking to stable BPS black holes it might be possible that some of these solutions do not actually suffer from any such instability, and that the associated thermodynamic quantities smoothly interpolate between possibly very different EFT regimes. This last possibility is particularly compelling because it would mean that for certain non-perturbative gravitational objects in the theory, we would be able to describe in detail their behavior within the transition regime. Also, this means that one could define a family of solutions which explicitly interpolates and glues between two complementary EFT descriptions living in different number of spacetime dimensions. The goal of the present work is to explore this latter scenario.

For this purpose, we investigate the behavior of a restricted set of quantum corrections to the black hole entropy in 4d $\mathcal{N}=2$ supersymmetric effective field theories, focusing on the convergence properties of their associated perturbative expansions. More concretely, the black hole solutions we consider are BPS configurations \cite{Bogomolny:1975de,Prasad:1975kr}, and to them we can associate some indexed entropy, $\mathcal{S}_{\rm BH}$, that can be determined solely as a function of its gauge charges \cite{LopesCardoso:1998tkj,LopesCardoso:1999cv,LopesCardoso:1999fsj,LopesCardoso:1999xn}, and which is moreover protected by supersymmetry. The main reason for choosing this particular set-up is that, according to the literature (see, e.g., \cite{Mohaupt:2000mj, Ooguri:2004zv} and references therein), it is strongly believed that all relevant corrections to the aforementioned quantity are already captured by the low energy (supergravity) EFT in the form of an infinite number of higher-dimensional and higher-curvature local BPS operators involving the (anti-self-dual parts of the) graviton and graviphoton field strengths. Such operators contribute non-trivially to the entropy of supersymmetric black holes \cite{Wald:1993nt,Iyer:1994ys}, and in fact can be seen to exhibit an interesting behavior for certain values of the black hole charges \cite{Cribiori:2022nke, Cribiori:2023ffn, Calderon-Infante:2023uhz,Basile:2023blg, Basile:2024dqq, Bedroya:2024ubj,Herraez:2024kux, Calderon-Infante:2025pls}. In this work, we show that the quantum corrections to the entropy organize themselves into an asymptotic series whose complex expansion parameter $\alpha$ is related to the ratio between the M-theory circle radius ---computed as the inverse D0-brane mass, and the size of the black hole horizon. We therefore identify a transition regime corresponding to $|\alpha| = \mathcal{O}(1)$, which is equivalent to considering black hole solutions whose radius becomes comparable to that of the M-theory circle. Moreover, for all values of $\alpha$, we are able to perform a resummation of the underlying asymptotic series, ultimately providing explicit and convergent expressions. 

To illustrate this point, we analyze in detail two sub-classes of BPS solutions of the attractor mechanism close to the large radius point, namely D0-D2-D4 brane configurations, and systems exhibiting only D2- and D6-brane charges. In both cases, the highly non-local perturbative effects induced by the infinite tower of light Kaluza-Klein states are crucial to cancel the UV divergences exhibited by the four-dimensional EFT, and they in turn allow us to resum the asymptotic series in an exact manner ---following the same strategy as in \cite{Gopakumar:1998ii,Gopakumar:1998jq}. Interestingly, we can also evaluate the most dominant non-perturbative effects, but they do not seem to play a major role neither in the construction of the solutions solving the attractor equations, nor in the cancellation of the divergences of the entropy. In fact, the two examples analyzed in this work turn out being somewhat complementary. Indeed, from a computational perspective, the resummation procedure that we use for the D0-D2-D4 works as long as $\alpha$ is not purely imaginary. The study of the D2-D6 configuration is therefore crucial to explain how to extend our contour prescription to a more general choice of charges.

In the case of the D0-D2-D4 configuration we end up with a quantum-corrected entropy that is well-defined for all values of the now real parameter $\alpha$. In particular, we show that for $\alpha = \mathcal{O}(1)$ the asymptotic series stops being valid (for any of its finite-order truncations), thereby reflecting that the four-dimensional EFT breaks down. Therefore, upon crossing this transition regime, the solution itself should be most naturally regarded as a five-dimensional black string wrapped on the extra circle. Moreover, in the limit of large $\alpha$, we recover an infinitely extended black string living in five non-compact dimensions. Crucially, we show how the resummed version of the higher-derivative corrections to the entropy get diluted ---except for one particular term--- so as to precisely reproduce the exact microstate counting of the five-dimensional black string, which is also in agreement with other macroscopic computations in 5d $\mathcal{N}=1$ supergravity. This provides, in turn, a highly non-trivial check of our result. On the other hand, for the D2-D6 configuration, even if we are able to describe analytically the transition regime, the five-dimensional uplift of the 4d black hole crucially carries Kaluza--Klein monopole charge, and it exists only as long as there is a compact $\mathbf{S}^1$ direction in the theory. In this second case, we are still capable to glue two different EFT descriptions but we cannot explore the purely five-dimensional regime (i.e., the large $|\alpha|$ limit).

\medskip

The outline of the paper is as follows. In Section \ref{s:entropyBPSBHs} we introduce the main ingredients of 4d $\mathcal{N}=2$ supergravity field theories coupled to gravity, which is the set-up where our discussion will be placed. We also review the precise mathematical description of a class of supersymmetric black hole solutions, whose physical properties are entirely determined by the so-called attractor mechanism \cite{Ferrara:1995ih,Strominger:1996kf,Ferrara:1996dd,Ferrara:1996um}. To make things more concrete, we specialize the formulae to black hole solutions belonging to the large volume regime. This discussion includes a detailed account of the relevant higher-derivative gravitational operators that control the deviations of the black hole entropy from the semi-classical area law. In Sections \ref{s:BHs&EFTtransitions} and \ref{s:other4dBHs} we analyze, respectively, the D0-D2-D4 and the D2-D6 configurations. This constitutes the main body of our work. In both cases, we first introduce and review their classical two-derivative description, and subsequently discuss the leading-order perturbative corrections close to the large radius point. We also describe their behavior in the transition regime, where the asymptotic series of quantum corrections becomes naively divergent. For each family of solutions, we further explain how to incorporate the most relevant (perturbative) non-local and non-perturbative corrections. Finally, in Section \ref{s:other4dBHs} we also comment on possible obstructions which can arise with other type of solutions. We conclude in Section \ref{s:conclusions} with some final remarks and future directions.

\section{Review: BPS Black Holes in Four Dimensions}
\label{s:entropyBPSBHs}

\subsection{4d $\mathcal{N}=2$ supergravity and higher-derivative corrections}
\label{ss:4dsugrahigherderivatives}

We consider hereafter 4d $\mathcal{N}=2$ set-ups arising from Type IIA string theory compactified on a Calabi--Yau threefold $X_3$. The corresponding bosonic part of the two-derivative action reads as follows \cite{Bodner:1990zm}
\begin{equation}\label{eq:IIAaction4d}
	\begin{aligned}
		\ S\, =\, & \frac{1}{2\kappa^2_4} \int \mathcal{R} \star 1 + \frac{1}{2}\, \text{Re}\, \mathcal{N}_{AB} F^A \wedge F^B + \frac{1}{2}\, \text{Im}\, \mathcal{N}_{AB} F^A \wedge \star F^B \\
		& - \frac{1}{\kappa^2_4} \int G_{a\bar b}\, d z^a\wedge \star d\bar z^b + h_{pq}\, d \xi^p \wedge \star d \xi^q\, ,
	\end{aligned}
\end{equation}
with $A, B=0,1, \ldots, h^{1,1}(X_3)$. We denote by $z^a= b^a + i t^a$, $a= 1, \ldots, h^{1,1}(X_3)$, the scalar fields describing the complexified K\"ahler (or vector multiplet) moduli space of the theory, whereas $\xi^p$, $p= 1, \ldots, h^{2,1}(X_3) +1$, belong to the hypermultiplets instead. The field strengths $F^B=dA^B$ correspond to $U(1)$ gauge bosons normalized so that they have integrally-quantized charges. In this work, we will restrict ourselves to the vector multiplet sector, since the black hole solutions we are most interested in here only depend on the latter. 

The vector moduli space is mathematically described as a projective special Kähler manifold \cite{deWit:1980lyi,deWit:1984wbb,deWit:1984rvr,Cremmer:1984hj}, whose metric tensor $G_{a\bar b}=\partial_a \partial_{\bar{b}} K$ can be derived from the following  K\"ahler potential
\begin{equation} \label{eq:kahlerpotential}
K = - \log i\left( \bar{X}^A \mathcal{F}_{A}- X^A \bar{\mathcal{F}}_{A}  \right)\, ,
\end{equation}
where a certain set of local projective coordinates $X^A$ have been introduced \cite{Strominger:1990pd,Ceresole:1995ca,Craps:1997gp}. In terms of these, the Kähler moduli are most easily expressed as the quotients
\begin{equation}\label{eq:specialcoords}
    z^a = \frac{X^a}{X^0}\, ,
\end{equation}
given a local patch where $X^0$ is nowhere vanishing. This also implies that the entire geometry of the vector multiplet moduli space can be encoded into a holomorphic function $\mathcal{F}(X^A)$, usually referred to as the prepotential \cite{Craps:1997gp,Craps:1997nv}. This function is moreover homogeneous of degree two, meaning that it satisfies $\mathcal{F}= \frac{1}{2} X^A \mathcal{F}_A$, where $\mathcal{F}_A = \partial_{X^A} \mathcal{F}$.

In addition, due to the constraints of $\mathcal{N}=2$ supersymmetry, the complexified gauge kinetic function $\mathcal{N}_{AB}$ appearing in \eqref{eq:IIAaction4d} is determined by the Kähler structure moduli through the expression
\begin{equation}
    \mathcal{N}_{AB} = \overline{\mathcal{F}}_{AB} + 2i \frac{(\text{Im}\, \mathcal{F})_{AC} X^C (\text{Im}\, \mathcal{F})_{BD} X^D}{X^C (\text{Im}\, \mathcal{F})_{CD} X^D}\, ,
\end{equation}
where \(\mathcal{F}_{KL} = \partial_{X^K} \partial_{X^L} \mathcal{F}\).

\medskip

On the other hand, these theories are known to present ---beyond the two-derivative Lagrangian \eqref{eq:IIAaction4d}--- interesting higher-dimensional and higher-curvature corrections. Some of these terms are furthermore $\frac12$-BPS which means, in practice, that they are protected by supersymmetry from receiving certain quantum corrections. This implies, in turn, that their dependence with respect to the moduli fields $z^a$ can sometimes be determined exactly. Using standard $\mathcal{N}=2$ superspace notation, they can be written as follows \cite{Bershadsky:1993ta, Bershadsky:1993cx,Antoniadis:1993ze,Antoniadis:1995zn}\footnote{Note that the $g=0$ contribution gives precisely the prepotential term in $\mathcal{N}=2$ supergravity, upon identifying $\mathcal{F}_0(\mathcal{X}^A) \equiv \mathcal{F}(\mathcal{X}^A)$ as functions of the chiral superfields \eqref{eq:scalarsuperfields}.}
\begin{equation}\label{eq:superspacelagrangian}
	\mathcal{L}_{\rm h.d.}\, \supset\, -\frac{i}{2}\sum_{g\geq 1}\int \text{d}^4\theta\, \mathcal{F}_g (\mathcal{X}^A)\, \left(\mathcal{W}^{ij} \mathcal{W}_{ij}\right)^{g}\ +\ \text{h.c.}\, ,
\end{equation}
where $\mathcal{F}_g (\mathcal{X}^A)$ is a chiral superfield that is related to the $g$-loop topological free energy of the closed superstring, $\theta_{\alpha}$ denote the fermionic superspace coordinates (of negative chirality) 
and 
\begin{equation}\label{eq:Weylsuperfield}
	\mathcal{W}_{\mu \nu}^{ij} = W^{ij, -}_{\mu \nu} - \mathcal{R}^-_{\mu \nu \rho \sigma} \theta^i \sigma^{\rho \sigma} \theta^j + \ldots\, ,
\end{equation}
is the Weyl superfield \cite{deWit:1979dzm,Bergshoeff:1980is}. The latter transforms under the $SO(2)$ antisymmetric representation in the $i,j = 1, 2$, indices and moreover depends on the anti-self-dual components of the graviphoton field-strength \cite{Ceresole:1995ca}
\begin{equation}\label{eq:graviphoton}
	W^-_{\mu \nu} = 2 i e^{K/2} \text{Im}\, \mathcal{N}_{AB} X^AF^{B, -}_{\mu \nu}\, ,\qquad W^{ij, -}_{\mu \nu}=\frac{\epsilon^{ij}}{2} W^{-}_{\mu \nu}\, ,
\end{equation}
as well as that of the Riemann tensor.
Performing the integration over the fermionic variables, one obtains several terms entering in the bosonic action. For instance, upon combining the lowest components (i.e., $\theta$-independent) in the superfield expansion of $\mathcal{F}_g(\mathcal{X}_A)$ and $\mathcal{W}^{2g-2}$ with the $\theta^2$-term in \eqref{eq:Weylsuperfield} squared, one obtains operators within \eqref{eq:superspacelagrangian} of the form \cite{Bergshoeff:1980is,LopesCardoso:1998tkj}
\begin{equation}\label{eq:GVterms}
	\mathcal{L}_{\rm h.d.}\, \supset\, -\frac{i}{2}\sum_{g\geq 1}\mathcal{F}_g(X^A)\, \mathcal{R}_-^2\, W_-^{2g-2}\ +\ \text{h.c.}\, ,
\end{equation}
with $X^A$ denoting the bottom (i.e., scalar) constituents of the reduced chiral superfields \cite{deRoo:1980mm}
\begin{equation}
\label{eq:scalarsuperfields}
	\mathcal{X}^A = X^A + \frac12 \epsilon_{ij} \theta^i \sigma^{\mu \nu} \theta^j \left( F^{A, -}_{\mu \nu} - i e^{K/2}\bar{X}^A W_{\mu \nu}^-\right) + \ldots\, ,
\end{equation}
and where the precise index contractions appearing in \eqref{eq:GVterms} can be deduced from eqs. \eqref{eq:superspacelagrangian} and \eqref{eq:Weylsuperfield}. Let us remark that not all the purely bosonic terms that can be extracted from the superspace Lagrangian \eqref{eq:superspacelagrangian} are quadratic in the Riemann tensor. In fact, if instead of using the $\theta^2$-component of $\mathcal{W}^{ij}_{\mu \nu}$ we rather insert the maximal $\theta^4$-term ---which contains a piece proportional to the anti-self dual combination of the antisymmetric tensor $\nabla_{[\mu} \nabla^{\sigma} W^{kl, -}_{\sigma |\nu]}$, we obtain a local operator in the action $W^{ij, -}_{\mu \rho} \nabla^{\rho} \nabla^{\sigma} W^{kl, -}_{\sigma \nu} \epsilon_{ik} \epsilon_{jl}$ that is quadratic in the graviphoton field strength and moreover contains two covariant derivatives \cite{Bergshoeff:1980is,LopesCardoso:1998tkj}. Such a term would then be linear in the Riemann tensor, and in fact turns out being the only one contributing to the entropy at the four-derivative level \cite{LopesCardoso:1998tkj} (see discussion in Section \ref{ss:entropyBPSBHs} below).

\medskip

Interestingly, as originally noticed in \cite{Gopakumar:1998ii,Gopakumar:1998jq}, one can compute all perturbative and non-perturbative stringy $\alpha'$-corrections in $\mathcal{F}_g(X^A)$ for $g\geq 0$ using the duality between Type IIA string theory on $X_3$ and M-theory compactified on $X_3 \times \mathbf{S}^1$. This exploits the fact that the string coupling belongs to a hypermultiplet, which is decoupled from the vector multiplets at the two-derivative level \cite{Candelas:1990pi}, such that it can be freely tuned at will. Hence, for a single hypermultiplet of mass $m=|Z|$ in 4d Planck units, with $Z = e^{K/2} \left(p^A \mathcal{F}_{A}-q_A X^A\right)$
being its central charge, one indeed obtains a generating function via a Schwinger-like one-loop computation as follows (see, e.g., \cite{Bastianelli:2008cu, Dunne:2004nc} and references therein) 
\begin{equation}\label{eq:generatingseries}
\begin{aligned}
	\sum_{g\geq 0} \delta \mathcal{F}^{(\rm hyp)}_g (X^A)\, W_-^{2g-2} &= -\frac{1}{4} \int_{i 0^+}^{i \infty}\frac{\text{d}\tau}{\tau} \frac{1}{\sin^2{\frac{\tau W_- \bar Z}{2}}} e^{-\tau |Z|^2}\\
    &= \frac{1}{4} \int_{0^+}^{ \infty}\frac{\text{d} \tau}{\tau} \sum_{g\geq0} \frac{(-1)^{g} 2^{2g} (2g-1) B_{2g}}{(2g)!} \left( \frac{\tau W_-}{2}\right)^{2g-2} e^{-\tau Z}\, +\, \mathcal{O}\left(e^{-\frac{Z}{W_-}}\right)\, , 
\end{aligned}
\end{equation}
where the integration along the positive imaginary axis follows from causality \cite{Chadha:1977my}. To reach the second equality we have first rescaled the proper time $\tau$,\footnote{\label{fnote:subtletychangevar}The change of variables $\tau \to \tau /\bar Z$ actually introduces some subtleties due to the infinitely many poles in the complex $\tau$-plane exhibited by the one-loop determinant \eqref{eq:generatingseries}. We refer to Section \ref{ss:nonlocalnonpertD0brane} as well as to \cite{Hattab:2024ewk,Hattab:2024ssg} for independent and complementary discussions on this important issue.} 
subsequently performed a perturbative expansion using the Laurent series for $\csc^2(x)$ around $x=0$, given by\footnote{The quantities $B_{2g}$ denote the Bernouilli numbers, which read as
\begin{equation}\label{eq:bernouilli}
	B_{2g}= \frac{(-1)^{g+1} 2 (2g)!}{(2\pi)^{2g}} \zeta(2g)\, .
\end{equation}
}
\begin{equation}\label{eq:expansionsin^-2(x)}
    \frac{1}{\sin^{2}(x)} = \sum_{n=0}^{\infty} \frac{2^{2n}(2n-1)}{(2n)!} (-1)^{n-1} B_{2n} x^{2n-2}\,, 
\end{equation} 
and finally we deformed the contour towards the real axis. We have moreover added some exponential correction in eq. \eqref{eq:generatingseries} to remind us that the one-loop calculation captures non-perturbative effects as well, such as Schwinger pair production \cite{Schwinger:1951nm}. Notice that the coupling of the BPS particle to the graviphoton field involves the anti-holomorphic piece of the central charge, the reason being that the supersymmetric background where the one-loop calculation is carried out requires a (constant) complex-valued anti-self-dual field strength \cite{Dedushenko:2014nya}. 

Let us stress that the notation $\delta \mathcal{F}_g$ used in \eqref{eq:generatingseries} is meant to indicate that the Schwinger integral does not capture a priori the full exact form of the higher-derivative Wilson coefficients \eqref{eq:superspacelagrangian}, but rather the quantum (loop) corrections due to the BPS spectrum in the theory.

\subsection{An exact entropy formula for BPS black holes}\label{ss:entropyBPSBHs}

An interesting class of geometrical objects that one can construct within these theories are supersymmetric black holes. An explicit analysis of this type of solutions can be found in e.g., \cite{Mohaupt:2000mj,Moore:2004fg}. They moreover exhibit certain universal features, such as the stabilization of the moduli fields ---which couple to the electromagnetic background turned on by the black hole charges $\left(q_A, p^B \right)$--- at the horizon locus, according to the so-called \emph{attractor mechanism} \cite{Ferrara:1995ih,Strominger:1996kf,Ferrara:1996dd,Ferrara:1996um}. Importantly for us, this analysis can be extended beyond the two-derivative level \cite{Behrndt:1996jn,LopesCardoso:1998tkj,Behrndt:1998eq,LopesCardoso:1999cv,LopesCardoso:1999fsj,LopesCardoso:2000qm}, also including the higher-curvature corrections discussed in the previous section. This is what we review next.

\medskip

For convenience, we introduce some rescaled quantities as follows \cite{Behrndt:1996jn,LopesCardoso:1998tkj} (we henceforth suppress the anti-self-dual subindex in the graviton and graviphoton field strengths)
\begin{equation}
\label{eq:rescaledvars}
	Y^A = e^{\mathscr{K}/2} \Bar{\mathscr{Z}} X^A\, , \qquad \Upsilon = e^{\mathscr{K}} \bar{\mathscr{Z}}^2 W^2\, ,
\end{equation}
where $\mathscr{Z}$ defines a generalized black hole central  charge (cf. eq. \eqref{eq:nearHorizonmetric})
\begin{equation}\label{eq:generalizedcentralchargeBH}
\begin{aligned}
    \mathscr{Z} &= \bar{\mathscr{Z}}^{-1} \left( p^A F_A (Y, \Upsilon) -q_A Y^A\right) \,, \\
    |\mathscr{Z}|^2 &= p^A F_A(Y, \Upsilon)-q_A Y^A = e^{\mathscr{K}} \left| p^A F_{A}(X, W^2) -q_A X^A \right|^2\, ,
\end{aligned}
\end{equation}
and $\mathscr{K}$ determines the following symplectic invariant combination 
\begin{equation}\label{eq:generalizedkahlerpot}
	e^{-\mathscr{K}}= i \bar{X}^A F_A(X, W^2) - i X^A \bar{F}_A (\bar{X}, \bar{W}^2)\, ,
\end{equation}
which has a functional form clearly reminiscent of the K\"ahler potential, cf. eq. \eqref{eq:kahlerpotential}. 
%
%
Here, $F(Y, \Upsilon)$ denotes a generalization of the holomorphic prepotential associated to the underlying 4d $\mathcal{N}= 2$ theory (see discussion after \eqref{eq:specialcoords}) that includes the effects of higher-derivative terms, namely
\begin{align}
\label{eq:generalizedprepotential}
    F(X, W^2) = \sum_{g=0}^{\infty} F_g(X^A) W^{2g}\, .
\end{align}
The coefficients $F_g(X^A)$ can be directly related to the topological closed string amplitudes \cite{Witten:1988xj,Witten:1991zz,Labastida:1991qq,Labastida:1994ss}, and we defined
\begin{equation}
    F_A(X, W^2) = \frac{\partial F(X, W^2)}{\partial X^A}\, ,
\end{equation}
in eqs. \eqref{eq:generalizedcentralchargeBH}-\eqref{eq:generalizedkahlerpot} above. In the following, we will find convenient to rescale the generalized prepotential \eqref{eq:generalizedprepotential} by the quantity $C^2= e^{\mathscr{K}} \Bar{\mathscr{Z}}^2$, such that \cite{Ooguri:2004zv} 
\begin{equation}\label{eq:normalizedgeneralizedprepotential}
    F(Y, \Upsilon) := C^2 F(X, W^2) = \sum_{g=0}^{\infty} F_g(Y^A) \Upsilon^{g}\, , \qquad \text{with}\ \ F_g(Y^A) = (-1)^g\, 2^{-6g} \mathcal{F}_{g} (Y^A)\, ,
\end{equation}
where the last equality follows from the homogeneity properties of $F(X, W^2)$. 

Physically, the quantity $\mathscr{Z}$ controls the warp factor of the metric in the BPS black hole background \cite{Ferrara:1997yr,LopesCardoso:1998tkj}, whose near-horizon line element reads (using isotropic coordinates)
\begin{equation} \label{eq:nearHorizonmetric}
    ds^2= -e^{2 U(r)}dt^2+ e^{-2U(r)} \left( dr^2 + r^2 d\Omega_2^2\right)\, , \qquad \text{with}\ \ e^{-2U(r)} =\frac{|\mathscr{Z}|^2 \kappa_4^2}{8\pi r^2}\, ,
\end{equation}
thus also incorporating the effect of the higher-derivative chiral terms captured by eq. \eqref{eq:superspacelagrangian}. The attractor equations then determine the values for the moduli fields $Y^A$ when evaluated at the horizon to be fixed by \cite{Behrndt:1998eq,LopesCardoso:2000qm}
\begin{equation}\label{eq:attractoreqs}
\begin{aligned}
	i p^A &= Y^A - \bar{Y}^A\, ,\\
    i q_A &= F_A(Y, \Upsilon)- \bar{F}_A (\bar{Y}, \bar{\Upsilon})\, ,
\end{aligned}
\end{equation}
whereas $\Upsilon$ is set to $-64$. 

Finally, let us state the \emph{quantum entropy formula} for BPS black holes with the near-horizon geometry given by (\ref{eq:nearHorizonmetric}), which may be expressed as follows \cite{LopesCardoso:1998tkj} 
\begin{equation}\label{eq:entropy}
    \mathcal{S}_{\rm BH} = \pi \left[ |\mathscr{Z}|^2 + 4 \text{Im}\, \left( \Upsilon \partial_{\Upsilon} F(Y, \Upsilon) \right) \right]\, ,
\end{equation}
and is therefore entirely determined by the black hole charges via \eqref{eq:attractoreqs}. The first term in \eqref{eq:entropy} coincides with the value of the horizon area divided by $4 G_4$, hence providing for the Bekenstein-Hawking contribution to the entropy, whilst the second piece captures further quantum corrections. Notice that both terms are sensitive to the higher-derivative operators shown in \eqref{eq:superspacelagrangian}.

\medskip

We conclude this section by giving some details on the quantum entropy formula presented above. The non-interested reader can safely skip this discussion. First of all, let us note that (\ref{eq:entropy}) has been computed using Wald's prescription \cite{Wald:1993nt,Iyer:1994ys} within the restricted framework of conformal off-shell $\mathcal{N} = 2$ supergravity coupled to $n_V+1$ vector multiplets \cite{Ferrara:1977ij,deWit:1979dzm,deWit:1980lyi,deWit:1984rvr,deWit:1984wbb}, which reduces to the more familiar 4d $\mathcal{N}= 2$ (Poincaré) supergravity only after partial gauge fixing.\footnote{The relation between conformal and Poincaré (extended) supegravity requires the introduction of an additional vector multiplet that can be used to gauge-fix dilatation invariance \cite{deWit:1980lyi}. See also \cite{VanProeyen:1983wk,deWit:1984hw} for early reviews on the topic.} This formalism can be used, in turn, to derive the attractor equations (\ref{eq:attractoreqs}) as well as the near-horizon metric (\ref{eq:nearHorizonmetric}). 
This means, consequently, that (\ref{eq:entropy}) 
provides the macroscopic entropy associated to BPS black hole solutions in Calabi--Yau compactifications of Type IIA string theory, when restricting ourselves to the gravity and vector multiplet sectors.\footnote{As is well known, the two-derivative theory can be truncated consistently. The quantum corrections to the hypermultiplet sector, on the other hand, are not fully known \cite{Antoniadis:1993ze, Robles-Llana:2006vct} and thus we cannot verify whether higher-derivative terms involving those will obstruct the truncation.} Notice, however, that upon doing so we might be missing some contributions due to non-chiral higher-derivative operators (i.e., those intrinsically defined as integrals over full superspace) in the vector-multiplet sector, as well as analogous hypermultiplet-dependent terms in the 4d effective action. A large class of the former type of couplings were already shown to give a vanishing contribution to the black hole entropy \cite{deWit:2010za,Murthy:2013xpa}, hence suggesting that this could always be the case. As for the latter, in \cite{LopesCardoso:2000qm} it was explicitly checked that adding neutral hypermultiplets in the form of gauge-fixed, superconformal multiplets \cite{deWit:1999fp} does not affect neither the attractor mechanism, nor the BPS near-horizon geometry. The analysis therein was carried out by considering perturbative $\mathcal{R}^2$-corrections. However, there is no guarantee that this will still work at all orders in perturbation theory. In fact, the authors of \cite{Ooguri:2004zv} argue ---also providing some amount of evidence--- that the exact black hole entropy should depend on the background hypermultiplet vevs. Crucially, though, the generalized prepotential \eqref{eq:normalizedgeneralizedprepotential} controlling the quantum entropy formula above is sensitive to the number of hypermultiplets but not to their vevs (see discussion after eq. (\ref{eq:D0mass}) in the next section). Thus, from the macroscopic perspective it is not clear whether we could be missing some additional operators contributing to the black hole entropy, namely if \eqref{eq:entropy} would be the end result of applying Wald's procedure in the \emph{full} Type IIA string theory. In \cite{Ooguri:2004zv} a detailed analysis of the origin of (\ref{eq:entropy}) was performed and they suggested that it is computing instead a protected supersymmetric index. This idea has been supported by explicitly matching the black hole free energy\footnote{This corresponds to the Legendre dual of the entropy and it gives the leading-order contribution to the gravitational path integral.} with a supersymmetric index defined within the CFT living on the branes sourcing the BPS black hole background. In particular, the alternating signs of the terms which add up to give the index should account for the cancellation of the dependence on hypermultiplet vevs of the BPS states degeneracy. In what follows, we will not be concerned about whether (\ref{eq:entropy}) is truly computing an entropy or a protected supersymmetric index in Type IIA, and we will just focus on its properties along certain decompactification limits. With this subtlety in mind, we will refer to (\ref{eq:entropy}) simply as the \emph{BPS black hole entropy}.

It is also worth mentioning that in \cite{Ooguri:2004zv} they revise the relation between macroscopic entropy and microstate counting performed in \cite{Strominger:1996sh}. We present the subtlety following the modern review of \cite{Zaffaroni:2019dhb} (see also references therein). What one should truly compute in order to compare the macroscopic and microscopic dual descriptions of the system is the partition function $\mathcal{Z}$. Such quantity can be formally defined via some path integral or as a microscopic generating function, respectively. In the latter case, it is not defined with a micro-canonical ensemble (i.e., with both electric $q$ and magnetic $p$ charges fixed), but rather with a mixed ensemble (fixed magnetic charges $p$ and electric potentials $\phi$). Thus, for a supersymmetric system, the partition function would have the form\footnote{In the general case, there would also be some dependence on the angular momentum, which we omit here.} 
\begin{equation}
    \mathcal{Z} = \text{Tr}\bigg[ e^{i q \phi}\bigg]_{\text{susy}} = \sum_q \Omega (p,q) \, e^{i q \phi}  \,,
\end{equation}
where $\Omega (p,q)$ is an integer counting the number of supersymmetric microstates with fixed $p$ and $q$, and the trace is taken over  states which are annihilated by the supercharges. On the other hand, the microscopic entropy is usually defined as the logarithm of the number of microstates (with fixed charges) and reads 
\begin{equation} \label{eq:microentropy}   \mathcal{S}_{\text{micro}}\, = \, \log \Omega (p,q) \,,
\end{equation}
whereas the macroscopic entropy instead computes the Legendre dual of the partition function
\begin{equation}\label{eq:macroentropy}
\mathcal{S}_{\text{BH}}\, = \, \log \mathcal{Z} - i q \phi  \,.
\end{equation}
Switching to the gravitational representation of $\mathcal{Z}$, the entropy $\mathcal{S}_{\text{BH}}$ can be readily identified with the Legendre dual of the quantum-corrected free energy. It is therefore nothing but the BH entropy as computed by Wald's prescription applied to the full quantum theory. As noticed in \cite{Ooguri:2004zv}, these definitions match only to leading order in the large electric charge expansion, which is the regime considered in \cite{Strominger:1996sh}.\footnote{This is motivated by the fact that, in general, a Laplace transform is not the inverse of a Legendre transform, and viceversa.} This will also be the regime considered throughout this work. Therefore, it is not surprising that the macroscopic computation eventually reproduces an exact result obtained via some microstate counting (see discussion in Section \ref{sss:gluing5d}).
However, in practice, $\mathcal{Z}$ cannot be easily computed, and one rather replaces it with a supersymmetric index $\mathcal{Z}_{\text{index}}$. A simple way to construct such an object (if a microscopic model is accessible) is via the insertion of a $(-1)^F$ factor
\begin{equation}
  \mathcal{Z}_{\text{index}} = \text{Tr}\bigg[ (-1)^F e^{i q \phi}\bigg]_{\text{susy}}\,,
\end{equation}
with $F$ being some $\mathbb{Z}_2$-graded (i.e., fermionic) operator. The advantage of using the index is that it can also be evaluated from the macroscopic perspective. It is indeed an euclidean path integral with proper boundary conditions (which can be explicitly determined in concrete examples). In general, a supersymmetric index does not coincide with the partition function. However, in particular setups one can actually prove that the index provides a good approximation to the partition function. In essence, what one has to ensure is that there are no large cancellations between the different supersymmetric states over which we are tracing.\footnote{In some examples this is automatically realized thanks to the symmetries exhibited by the configuration. In other case, the cancellation is avoided if the chemical potentials involved have complex phases. In general, though, there is not a unique, unambiguous prescription to find an appropriate $(-1)^F$ operator.} What \cite{Ooguri:2004zv} suggests then is that, despite \eqref{eq:entropy} not being constructed as an index, it truly computes the Legendre dual of $\mathcal{Z}_{\text{index}}$ in the context of Type IIA string theory, and is therefore protected. Moreover, in the large charge expansion, we would also have $\mathcal{Z}_{\text{index}}\sim \mathcal{Z} $, so that it really captures the same quantity as  $\mathcal{S}_{\text{micro}}$ and $\mathcal{S}_{\text{BH}}$, cf. eqs. \eqref{eq:microentropy} and \eqref{eq:macroentropy}. 

\subsection{The large volume approximation}\label{ss:LargevolAprox}

Up to now, our discussion has been somewhat general and thus model-independent. This is due to the fact that in all previous relations we have expressed every physical quantity in terms of an undetermined prepotential (or generalization thereof), which, as already stressed, must be a holomorphic and homogeneous function of the fields $X^A$, but is otherwise arbitrary. In the present section we will exploit our knowledge about string theory and particularize the description to the Type IIA large volume/radius regime, which is defined by having $z^a \to \infty$ for all $a= 1, \ldots, h^{1,1}(X_3)$. The reason being that there one can use very explicit formulae which are valid regardless of the specific Calabi--Yau threefold under consideration. In addition, this provides us with a useful scheme in which we can organize the different contributions appearing both in the prepotential and the relevant black hole observables, separating them between classical and purely stringy corrections.

\subsubsection{Leading-order corrections to the generalized prepotential}\label{sss:largevolprepotential}

Let us first discuss how the genus-$g$ terms within the generalized prepotential \eqref{eq:generalizedprepotential} get simplified when evaluated at large volume. For the genus-0 contribution, one obtains (using string units)
\begin{equation}\label{eq:prepotentialIIA}
\begin{aligned}
\mathcal{F} (X^A) = &-\frac{1}{6} \mathcal{K}_{abc} \frac{X^a X^b X^c}{X^0} + K_{ab}^{(1)} X^a X^b + K_{a}^{(2)} X^0 X^a+ K^{(3)} (X^0)^2\\
&- \frac{(X^0)^2}{(2 \pi i)^3}\sum_{\boldsymbol{k} > \boldsymbol{0}} n_{\boldsymbol{k}}^{(0)} \sum_{m \geq 1} \frac{1}{m^3} e^{2\pi i m k_a z^a}\, , 
\end{aligned}
\end{equation} 
with the different quantities appearing above being topological, such that they can be expressed in terms of an integral basis of harmonic 2-forms $ \{\omega_a\} \in H^{1,1}(X_3, \mathbb{Z})$ as follows
\begin{equation}
 \mathcal{K}_{abc}= \omega_a \cdot \omega_b \cdot \omega_c \, ,\qquad K_{a}^{(2)}=\frac{1}{24} c_2(TX_3) \cdot \omega_a\, ,\qquad K^{(3)}= \frac{i \zeta(3)}{2(2\pi)^3} \chi_E(X_3)\, ,
\end{equation}
whereas $K_{ab}^{(1)}$ can be fixed instead by requiring good symplectic transformation properties of the underlying period vector \cite{deWit:1992wf,Harvey:1995fq}. Similarly, the quantity $\chi_E(X_3) = 2 \left( h^{1,1}(X_3)-h^{2,1} (X_3)\right)$ denotes the Euler characteristic of the threefold. Lastly, the coefficients $n_{\boldsymbol{k}}^{(0)}$ are known as genus-zero Gopakumar--Vafa invariants and count, for each positive homology representative $\boldsymbol{k}=k_a\gamma^a \in H_2^+(X_3,\mathbb{Z})$, the indexed degeneracy of BPS D2-brane states wrapped on 2-cycles within the corresponding supersymmetric class \cite{Gopakumar:1998ii, Gopakumar:1998jq}.

On the other hand, the (holomorphic part of the) genus-1 topological string amplitude can be expanded around the large radius point as \cite{Bershadsky:1993ta, Bershadsky:1993cx,Katz:1999xq} 
\begin{equation}
\mathcal{F}_1 (X^A) = \frac{1}{24} \int_{X_3} J_{\mathbb{C}}\wedge c_2(TX_3)\, +\, \mathcal{O} \left(e^{2 \pi i z^a} \right) = \frac{1}{24} c_{2,a}\, z^a\, +\, \mathcal{O} \left(e^{2 \pi i z^a} \right)\, ,
\label{eq:F1LCS}
\end{equation}
where $J_{\mathbb{C}} = z^a\,  \omega_a= (b^a + it^a)\,  \omega_a$ is the complexified K\"ahler 2-form of the Calabi--Yau threefold, whilst $c_2(TX_3)$ denotes the second Chern class of its tangent bundle. This contribution can be easily understood as coming from the dimensional reduction of an analogous four-derivative, curvature squared operator in 5d $\mathcal{N}=1$ supergravity \cite{Grimm:2017okk}.

For higher-genus terms, the leading contribution corresponds to constant maps from the worldsheet to the Calabi--Yau threefold. These can be equivalently determined from the dual M-theory perspective as a Schwinger-loop calculation associated to the tower of D0 bound states, whose masses in string units are given by
\begin{equation}
	\label{eq:D0mass}
	m_n = 2\pi |n|\, \frac{m_s}{g_s} = |n|\, m_{\rm D0}\, ,
\end{equation}
where $n\in \mathbb{Z}$ is the 0-brane charge. Therefore, upon substituting this into \eqref{eq:generatingseries} and performing the integral ---taking account that each D0-brane yields $-\frac{\chi(X_3)}{2}$ times the contribution of a single hypermultiplet \cite{Gopakumar:1998jq}--- as well as the infinite sum, one finds (in units of $m_{\rm D0}/2\pi$)
\begin{equation}\label{eq:Fg>1LCS}
	\begin{aligned}
		\mathcal{F}_{g>1} (X^A)\, & \supset\, \frac{\chi(X_3)}{2} (-1)^{g-1}2 (2g-1) \frac{\zeta(2g) \zeta(3-2g)}{(2\pi)^{2g}}\, (X^0)^{2-2g}\\
		&=\, \chi(X_3)\frac{2(2g-1) \zeta(2g) \zeta(2g-2) \Gamma(2g-2)}{(2\pi)^{4g-2}}\, (X^0)^{2-2g}\, ,
	\end{aligned}
\end{equation}
which gives precisely the dominant result along this limit \cite{Gopakumar:1998ii, Marino:1998pg}. Note that in order to reach the second equality we have used the identity $\zeta(s)= 2^s \pi^{s-1} \sin \left( \frac{\pi s}{2}\right) \Gamma(1-s) \zeta(1-s)$.

\medskip

Putting everything together, we thus conclude that the generalized prepotential \eqref{eq:normalizedgeneralizedprepotential}, when expanded around the large volume point, can be well-approximated by the function
\begin{equation}
\label{eq:holomorphicprepotential@largevol}
	F(Y, \Upsilon) = \frac{D_{a b c} Y^a Y^b Y^c}{Y^0} + d_a\, \frac{Y^a}{Y^0}\, \Upsilon + G(Y^0, \Upsilon)\, +\, \mathcal{O} \left(e^{2 \pi i z^a} \right)\, ,
\end{equation}
where $D_{abc}$ and $d_a$ are related to topological data of the underlying Calabi--Yau threefold 
\begin{equation}
	\label{eq:topologicalquantites}
	D_{a b c} = -\frac16 \mathcal{K}_{abc}\, , \qquad d_a = -\frac{1}{24} \frac{1}{64} c_{2,a}\, .
\end{equation}
Notice that the first two terms in \eqref{eq:holomorphicprepotential@largevol} capture the leading-order contribution to $F(Y, \Upsilon)$ at $g=0, 1$, respectively, whereas the function $G(Y^0, \Upsilon)$ rather corresponds to the one-loop determinant \eqref{eq:generatingseries} of the D0-branes. The latter reads as follows
\begin{equation}\label{eq:Gfn}
	G(Y^0, \Upsilon) = -\frac{i}{2 (2\pi)^3}\, \chi_E(X_3)\, (Y^0)^2 \sum_{g=0, 2, 3, \ldots} c^3_{g-1}\, \alpha^{2g} + \ldots\, ,
\end{equation}
where we defined
\begin{equation}\label{eq:Gfactors}
	c^3_{g-1} = (-1)^{g-1} 2 (2g-1) \frac{\zeta(2g) \zeta(3-2g)}{(2\pi)^{2g}}\, , \qquad \alpha^2 = -\frac{1}{64}\, \frac{\Upsilon}{(Y^0)^2}\, .
\end{equation}
The notation is chosen so as to reflect the fact that $c^3_{g-1}$ actually corresponds to the (integrated) third power of the Chern class associated to the Hodge bundle over the moduli space of Riemann surfaces of genus $g$ \cite{Bershadsky:1993cx,Faber:1998gsw}. The ellipsis in \eqref{eq:Gfn}, on the other hand, are meant to indicate that there would be a priori further non-analytic terms around $\alpha=0$. These should moreover capture the highly non-local and non-perturbative properties of the Schwinger one-loop determinant, see discussion after eq. \eqref{eq:expansionsin^-2(x)}.

\subsubsection{Black hole solutions in the large volume patch}
\label{sss:BHsLargeVol}

Before closing this chapter, let us apply our general considerations for the thermodynamics associated to the BPS black hole solutions described in Section \ref{ss:entropyBPSBHs} within the present, more restricted context. In particular, we want to show explicitly how the stabilization equations and the entropy formula get simplified when focusing on black hole solutions pertaining to the large radius regime. We build on the results and use the notation of \cite{LopesCardoso:1999fsj}. 

\medskip

First, notice that given the form \eqref{eq:holomorphicprepotential@largevol} of the generalized holomorphic prepotential at large volume, the derivatives with respect to the (rescaled) chiral coordinates $Y^a$ take the following simple form
\begin{align}
	F_a(Y, \Upsilon) = \frac{1}{Y^0} \left( 3 D_{abc} Y^b Y^c + d_a \Upsilon\right)\, .
\end{align}
This implies that the attractor equations \eqref{eq:attractoreqs} for the electric charges $q_a$ do not depend on the details of the function $G(Y^0, \Upsilon)$, i.e., 
\begin{align}\label{eq:qaequation}
	q_a = -\frac{i}{|Y^0|^2} \left( d_a \left( \bar{Y}^0 \Upsilon - Y^0 \bar{\Upsilon}\right) + 3  D_{abc} \left(Y^b Y^c \bar{Y}^0 - \bar{Y}^b \bar{Y}^c Y^0\right)\right)\, ,
\end{align}
whilst that of $q_0$ reads as 
\begin{align}\label{eq:q0equation}
	q_0 = \frac{i}{(Y^0)^2} \left( D_{abc} Y^a Y^b Y^c + Y^a d_a \Upsilon\right) - i\, \frac{\partial G(Y^0, \Upsilon)}{\partial Y^0}\ +\ \text{h.c.}\ .
\end{align}
Substituting these into \eqref{eq:generalizedcentralchargeBH}, one finds \cite{LopesCardoso:1999fsj}
\begin{equation}\label{eq:Zequation}
\begin{aligned}
	|\mathscr{Z}|^2 &= i D_{abc} \left( \frac{3 Y^a Y^b \bar{Y}^c}{Y^0} - \frac{3 \bar{Y}^a \bar{Y}^b Y^c}{\bar{Y}^0} - \frac{ Y^a Y^b Y^c\bar{Y}^0}{(Y^0)^2} + \frac{ \bar{Y}^a \bar{Y}^b \bar{Y}^c Y^0}{(\bar{Y}^0)^2}\right)\\
	& + id_a \left( \frac{\bar{Y}^a \Upsilon}{Y^0} - \frac{Y^a \bar{\Upsilon}}{\bar{Y}^0} - \frac{Y^a \bar{Y}^0 \Upsilon}{(Y^0)^2} + \frac{\bar{Y}^a Y^0 \bar{\Upsilon}}{(\bar{Y}^0)^2}\right)\\
	& + \frac{i}{2} \left( Y^0 + \bar{Y}^0\right) \left( \frac{\partial G(Y^0, \Upsilon)}{\partial Y^0} - \frac{\partial \bar{G}(\bar{Y}^0, \bar{\Upsilon})}{\partial \bar{Y}^0}\right) + \frac{p^0}{2} \left( \frac{\partial G(Y^0, \Upsilon)}{\partial Y^0} + \frac{\partial \bar{G}(\bar{Y}^0, \bar{\Upsilon})}{\partial \bar{Y}^0}\right)\, ,
\end{aligned}
\end{equation}
for the generalized central charge of the supersymmetric black holes, where one should understand that the value for the moduli are fixed by the attractor equations and $\Upsilon = -64$ at the horizon. For the entropy, one obtains instead \cite{LopesCardoso:1999fsj} 
\begin{equation}\label{eq:BHentropylargevol}
	\mathcal{S}_{\rm BH} = \pi \left[ |\mathscr{Z}|^2 - 2i d_a \left( \frac{Y^a}{Y^0} \Upsilon - \frac{\bar{Y}^a}{\bar{Y}^0} \bar{\Upsilon}\right) - 2i \left( \Upsilon \frac{\partial G(Y^0, \Upsilon)}{\partial \Upsilon} - \bar{\Upsilon} \frac{\partial \bar{G}(\bar{Y}^0, \bar{\Upsilon})}{\partial \bar{\Upsilon}}\right)\right]\, .
\end{equation}
For future reference, we observe that if we parametrize $G(Y^0,\Upsilon)$ as
\begin{equation}
    G(Y^0, \Upsilon) = -\frac{i}{2 (2\pi)^3}\, \chi_E(X_3)\, (Y^0)^2 \mathcal{I}(\alpha) \,,
\end{equation}
and we isolate in \eqref{eq:BHentropylargevol} the terms depending only on $G(Y^0,\Upsilon)$ together with its derivatives, we obtain the simpler formula\footnote{Notice that, despite the piece $\mathcal{S}_{\rm BH}\big\rvert_{G = 0}$ in \eqref{eq:simplerentropyfn} not depending explicitly on $G(Y^0, \Upsilon)$, it still does implicitly via the stabilized value of $Y^0$, which also involves the higher-genus terms; cf. eq. \eqref{eq:q0equation}.}
\begin{equation}\label{eq:simplerentropyfn}
    \mathcal{S}_{\rm BH} = \mathcal{S}_{\rm BH}\bigg\rvert_{G = 0} + \frac{4\pi}{(2\pi)^3} \frac{\chi_E(X_3)}{2} |Y^0|^2 \, \text{Re}\left[\mathcal{I}(\alpha)-\text{Re}(\alpha) \frac{\partial \mathcal{I}(\alpha)}{\partial \alpha} \right] \,,
\end{equation}
where the first term in the  right-hand side corresponds to the entropy computed as if $G(Y^0,\Upsilon)$ were absent. In upcoming sections we will make frequent use of the above expressions, oftentimes particularizing to specific black hole systems that are well-suited for our purposes.

\section{Gluing Across Dimensions: Black Holes and EFT Transitions}\label{s:BHs&EFTtransitions}

Our aim in this section will be to study in detail the physics associated with the quantum corrections to the supersymmetric entropy. From the spacetime perspective, the latter are induced by an infinite number of higher-derivative F-terms that enter the 4d $\mathcal{N}=2$ effective action, cf. eq. \eqref{eq:superspacelagrangian}. To do so, we focus our attention on a particularly simple BPS black hole carrying D0-D2-D4 charges. This system belongs to the family of solutions specified in Section \ref{ss:entropyBPSBHs} and, as we will show, it can be used to describe all the relevant physical effects that we want to highlight here.

\medskip

Therefore, in Section \ref{sss:pertcorrectionsBHswithnoD6charge} we review the two-derivative solution and we discuss the leading-order quantum corrections to the entropy within the large volume approximation, which adopt the form of a perturbative power series. In particular, we show that the series expansion is governed by a real parameter $\alpha$ that is related to the ratio of the M-theory circle radius, $r_5$, and the horizon length-scale, $r_h$. As a consequence, for black holes with $r_h \lesssim r_5$ the perturbative expansion controlling the infinite set of local corrections to, e.g., its entropy appears to take over, thus leading to seemingly divergent results. Then the question arises as to how the higher-dimensional dual theory is able to resolve these issues and provide ultimately the correct physical quantities, given that such solutions are known to lift to 5d \emph{stable} supersymmetric configurations \cite{Gauntlett:2002nw}. Interestingly, it turns out that for this particular set-up one is able to resum analytically the non-local quantum effects induced by the full tower of (charged) Kaluza-Klein modes, and even compute the relevant non-perturbative prepotential at leading order in the large volume regime. 

\medskip

The key observation is that, despite the perturbative series having zero radius of convergence, it is possible to organize the latter into a Schwinger integral representation (cf. eq. \eqref{eq:generatingseries}), which splits into a perturbative contribution and a non-perturbative part. In Section \ref{ss:nonlocalcorrections}, we discuss the former, illustrating how the aforementioned non-localities allow us to `glue' explicitly two limiting EFT descriptions. Indeed, for $\alpha \ll 1$, i.e., when the black hole is large compared to the M-theory circle, the index is correctly reproduced by the four-dimensional theory. Instead, for $\alpha \gg 1$ limit, i.e., when the black hole radius is much smaller than the M-theory circle, the index corresponds to the one of the five-dimensional, uplifted, black string solution. Finally, in Section \ref{ss:nonlocalnonpertD0brane} we take into account further non-perturbative effects, which are seen to diverge along the 5d limit. Crucially, however, due to the purely imaginary phase associated to them, we are able to prove that they do not spoil the previous discussion. 

\subsection{Example 1: the D0-D2-D4 black hole}\label{ss:D0D2D4system}

\subsubsection{The two-derivative analysis}\label{sss:2derivativeD0D2D4}

Let us start by describing the D0-D2-D4 black hole system using first a purely two-derivative approach based on the leading-order cubic piece in the $\mathcal{N}=2$ prepotential \eqref{eq:prepotentialIIA} at large volume. This will already allow us to illustrate certain special features that the aforementioned system exhibits, without having to worry about the complications introduced by the higher-derivative expansion. We refer to \cite{Shmakova:1996nz,Behrndt:1996jn} for the original works on the subject.

\medskip

At the level of the attractor mechanism, the above restriction can be easily implemented by the substitutions 
\begin{equation}\label{eq:attractor2derivative}
	W^2 \to 0\, ,\qquad F(X^A, W^2) \to \mathcal{F} (X^A)\, ,
\end{equation}
which imply, in turn, that the quantities $\mathscr{K}$ and $\mathscr{Z}$ defined in eqs. \eqref{eq:generalizedcentralchargeBH} and \eqref{eq:generalizedkahlerpot} reduce to their two-derivative analogues, namely
\begin{equation}
	\mathscr{K} \to K\, ,\qquad \mathscr{Z} \to Z\, .
\end{equation}
From here it is straightforward to see that the stabilization equations adopt now the following simple form \cite{Ferrara:1995ih,Strominger:1996kf,Ferrara:1996dd,Ferrara:1996um}
\begin{equation}\label{eq:attractoreqs2derivative}
   ip^A = CX^A- \bar{C} \bar{X}^A\, , \qquad iq_A = C \mathcal{F}_{A}- \bar{C} \bar{\mathcal{F}}_{A}\, ,
\end{equation}
with $C= e^{K/2} \Bar{Z}$ some compensator field that ensures the symplectic and K\"ahler invariance of any solution to the attractor equations above (cf. discussion around \eqref{eq:normalizedgeneralizedprepotential}). Henceforth, we will concentrate on black holes characterized by having $p^0 = 0$, i.e., no D6-brane charge, as seen from the Type IIA perspective. Consequently, we deduce from \eqref{eq:attractoreqs2derivative} that the rescaled quantity $CX^0$ is purely real (and, e.g., non-negative), hence allowing us to completely solve the algebraic system as follows
\begin{equation}\label{eq:Y^anoD6@2derivative}
	CX^a = \frac{1}{6} CX^0 D^{ab} q_b + \frac{i}{2} p^a\, ,\qquad (CX^0)^2 = \frac{1}{4} \frac{D_{abc} p^a p^b p^c}{\hat{q}_0} \equiv (x^0)^2\, ,
\end{equation}
where $D^{ab}$ is the inverse matrix of $D_{ab} \equiv D_{abc} p^c$ ---which is assumed to exist,\footnote{\label{fnote:degattractors}This ensures that there exists a \emph{unique} solution to the algebraic system \eqref{eq:attractoreqs2derivative}, given precisely by \eqref{eq:Y^anoD6@2derivative}. 
Notice that $D_{ab}$ could be singular in special circumstances, such that $D^{ab}$ might not be well-defined  \cite{Marchesano:2023thx,Marchesano:2024tod,Castellano:2024gwi,CMP}.} and we defined $\hat{q}_0 \equiv q_0 + \frac{1}{12} D^{ab} q_a q_b$. Note that this latter shift may be interpreted as an induced D0-brane charge in the worldvolume of the D2- and D4-branes comprising the 4d black hole of the form $\delta q_0=q_a\, \text{Re}\, z^a-3p^a D_{abc}\, \text{Re}\, z^b\, \text{Re}\, z^c $.

\medskip

With these results at hand, we can now proceed and determine the relevant physical observables associated to the black hole solution under consideration. Thus, following the discussion in Section \ref{sss:BHsLargeVol} and using the restriction map \eqref{eq:attractor2derivative}, we determine the radius $r_h$ of the horizon in terms of the stabilized central charge
\begin{equation}\label{eq:ZnoD6simplified}
	\frac{r_h^2}{G_4}= |Z (q_A, p^B)|^2 = -\frac{D_{abc} p^a p^b p^c}{CX^0} = 2 \sqrt{ \frac16 |\hat{q}_0| \mathcal{K}_{abc}p^a p^b p^c}\, ,
\end{equation}
as well as the black hole entropy
\begin{equation}\label{eq:classicalentropynoD6}
	\mathcal{S}_{\rm BH} (q_A, p^B) = -4 \pi CX^0 \hat{q}_0 = 2 \pi \sqrt{ \frac16 |\hat{q}_0| \left( \mathcal{K}_{abc}p^a p^b p^c\right)}\, ,
\end{equation}
which indeed satisfies $\mathcal{S}_{\rm BH} = \pi r_h^2/G_4$, in perfect agreement with the Bekenstein-Hawking formula. 

\medskip

Lastly, in order to trust the validity of the present two-derivative solution, and given the fact that we have approximated the genus-0 prepotential by its leading-order cubic piece, we need to ensure that the non-trivial profile of every scalar field turned on by the black hole background belongs to the large volume approximation (cf. Section \ref{ss:LargevolAprox}) at every point outside the horizon. Luckily for us, the monotonicity properties of the BPS flow \cite{Ferrara:1995ih,Ferrara:1997tw} imply that this consistency condition is automatically satisfied if and only if \emph{i)} the boundary values measured at asymptotic infinity and \emph{ii)} the stabilized moduli at the attractor locus met those as well. In the following, we assume that the v.e.v.s $\braket{z^a}$ at infinity are such that the vacuum where we expand our black hole around indeed belongs to the large volume regime. Consequently, it is enough for us to check whether both the overall threefold volume as well as that associated to any individual holomorphic 2- and 4-cycle are large in string units, when evaluated at the horizon. The former may be easily computed to be
\begin{equation}\label{eq:vol@hor2derivative}
	\mathcal{V}_{\rm h} = \frac18  e^{-K}\, |X^0|^{-2} \bigg\rvert_{\rm hor} = \frac18\frac{|Z|^2}{|CX^0|^2} = \sqrt{\frac{6 |\hat{q}_0|^3}{ \mathcal{K}_{abc} p^a p^b p^c}}\, ,
\end{equation}
such that having $\mathcal{V}_{\rm h} \gg 1$ requires from imposing $|\hat{q}^0|^3 \gg \left| D_{abc} p^a p^b p^c \right|$. As for the latter, we only display here the saxionic part of the moduli fields
\begin{equation}\label{eq:Kahlermoduli2derivative}
	t^a_{\rm h} = \text{Im} \left( \frac{CX^a}{CX^0} \right)\bigg\rvert_{\rm hor} = p^a \sqrt{\frac{6 |\hat{q}_0|}{\mathcal{K}_{abc} p^a p^b p^c}}\, ,
\end{equation}
which can then be readily used to determine the volume of any minimal even-dimensional cycle in the internal geometry. Notice that, from eqs. \eqref{eq:ZnoD6simplified}-\eqref{eq:Kahlermoduli2derivative}, we deduce that in order for the aforementioned volumes to be positive definite and large we must have $p^a \gg \left| D_{abc} p^a p^b p^c/\hat{q}_0 \right|^{1/2} >0$ as well as $\hat{q}^0<0$. 

All in all, we conclude that the large volume approximation holds for our D0-D2-D4 black hole solutions if the following charge hierarchy is imposed 
\begin{align}\label{eq:chargehierarchy2derivative}
	(\hat{q}^0)^2,\, \left( p^a\right)^2 \gg \left| \frac{D_{abc} p^a p^b p^c}{\hat{q}_0} \right|\, ,
\end{align}
which can be easily attained by taking $|\hat{q}^0| \gg p^a$. Notice, however, that we have not specified the behavior of the quotient appearing in the r.h.s. of \eqref{eq:chargehierarchy2derivative} above, namely $(D_{abc} p^a p^b p^c/\hat{q}_0)^{1/2}$, that is moreover proportional to the quantity $x^0$ defined in eq. \eqref{eq:Y^anoD6@2derivative}. In any event, let us remark that if one insists on making sure that all subleading $\alpha'$ effects can be safely ignored, then we also need to ask for the individual charges $|\hat{q}^0|, p^a$ to be large, as we discuss next.

\subsection{Perturbative quantum corrections}
\label{sss:pertcorrectionsBHswithnoD6charge}

\subsubsection{Including higher-derivative corrections}
\label{sss:higherderivativeD0D2D4}

Up to now, the BPS black hole system under investigation has been described using the (purely bosonic) action displayed in eq. \eqref{eq:IIAaction4d}, where we moreover truncated the underlying $\mathcal{N}=2$ prepotential at leading cubic order, cf. eq. \eqref{eq:prepotentialIIA}. As it is clear, the latter indeed dominates the physics of the vector multiplets in the large volume approximation, but actually receives a plethora of perturbative and non-perturbative stringy $\alpha'$-corrections that can a priori modify these black hole solutions \cite{LopesCardoso:1999cv, LopesCardoso:1998tkj,LopesCardoso:1999xn}. Our aim in what follows will be to reconsider the two-derivative solution and embed it within the more general formalism that includes the relevant higher-derivative corrections for this work, as reviewed in Section \ref{ss:entropyBPSBHs}.

\medskip

Therefore, restricting ourselves again to the large radius regime, and following the discussion in Section \ref{sss:BHsLargeVol}, one concludes from \eqref{eq:qaequation} that the stabilized (rescaled) variables $Y^a$ are still of the form
\begin{equation}
	Y^a = \frac{1}{6} Y^0 D^{ab} q_b + \frac{i}{2} p^a\, ,
\end{equation}
and hence depend on the particular value of $Y^0$ that solves the attractor equation \eqref{eq:q0equation}. The latter, on the other hand, gets modified by the higher-order terms entering the generalized holomorphic prepotential, thus yielding the following implicit solution \cite{LopesCardoso:1999fsj}
\begin{equation}\label{eq:Y0noD6}
	(Y^0)^2= \frac{\frac14 D_{abc} p^a p^b p^c -d_a p^a \Upsilon}{\hat{q}_0 +i (G_0-\bar{G}_0)}\, ,\qquad \text{with}\ \ G_0 \equiv \frac{\partial G(Y^0,\Upsilon)}{\partial Y^0}\, .
\end{equation}
In general, however, given the particular form of the correction term \eqref{eq:Gfn}, it is not possible to solve \eqref{eq:Y0noD6} analytically. Nevertheless, one may hope to be able to perform some iterative procedure that provides the correct solution in terms of a series expansion depending on the real parameter $\alpha$ defined in \eqref{eq:Gfactors}. In fact, by comparing eqs. \eqref{eq:Y^anoD6@2derivative} and \eqref{eq:Y0noD6} it becomes evident that, in order to recover the results from the previous section, we need the following additional conditions to be satisfied
\begin{equation}\label{eq:q0>ImG_0}
	|\hat{q}_0| \gg p^a \gg 1\, ,\qquad |\hat{q}_0| \gg |i(G_0 - \bar{G}_0)|\, .
\end{equation}
This way, upon expanding around the two-derivative solution, one can explicitly solve for $Y^0$ in a power series whose correction terms are controlled by the quotient $\text{Im}\, G_0 (Y^0)/|\hat{q}_0|$, as follows
\begin{equation}\label{eq:implicitsolY^0}
	(Y^0)^2 = (y^0)^2 \left( 1 + \frac{i(G_0 (y^0, \Upsilon) - \bar{G}_0 (\bar{y}^0, \bar{\Upsilon}))}{|\hat{q}_0|}\, +\,  \ldots \right)\, ,
\end{equation}
where $(y^0)^2 = (x^0)^2 \left( 1-4d_a p^a \Upsilon/D_{bce} p^b p^c p^e\right)$.

\medskip

Regarding the generalized central charge and entropy, using the reality condition on $Y^0$ one deduces from eqs. \eqref{eq:Zequation} and \eqref{eq:BHentropylargevol} that they respectively reduce to
\begin{subequations}\label{eq:genZ&entropynoD6higherderivative}
   \begin{equation}\label{eq:genZnoD6}
	|\mathscr{Z}|^2 = -\frac{D_{abc} p^a p^b p^c - 2d_a p^a \Upsilon}{Y^0} + i Y^0 \left( G_0-\bar{G}_0\right)\, ,
   \end{equation}
   \begin{equation}\label{eq:BHentropynoD6}
	\mathcal{S}_{\rm BH} = -4 \pi Y^0 \hat{q}_0 - i \pi \left( 3Y^0 G_0 + 2 \Upsilon G_{\Upsilon} - \text{h.c.}\right)\, ,
   \end{equation}
\end{subequations}
which can be expressed solely in terms of the black hole charges once we have solved for $Y^0$. Indeed, upon inserting the leading-order solution \eqref{eq:implicitsolY^0}, the latter read as
\begin{subequations}\label{eq:genZ&entropynoD6}
	\begin{equation}\label{eq:genZnoD6simplified}
		|\mathscr{Z}|^2 = 2 \sqrt{ \frac16 |\hat{q}_0| \mathcal{K}_{abc}p^a p^b p^c} + \ldots\, ,
	\end{equation}
	\begin{equation}\label{eq:approxentropynoD6}
		\mathcal{S}_{\rm BH} = 2 \pi \sqrt{ \frac16 |\hat{q}_0| \left( \mathcal{K}_{abc}p^a p^b p^c + c_{2,a}\, p^a\right)} - 2 \pi i \left( G (y^0, \Upsilon) - \bar{G} (\bar{y}^0, \bar{\Upsilon})\right)+ \ldots\, ,
	\end{equation}
\end{subequations}
whose resemblance with those shown in \eqref{eq:ZnoD6simplified}-\eqref{eq:classicalentropynoD6} is manifest. Let us also remark that, as already noticed in earlier works (see, e.g., \cite{Behrndt:1998eq, LopesCardoso:1998tkj}), in order to reproduce the quantity within the square root in the black hole entropy above it is crucial to take into account \emph{both} the deviations from the area law in \eqref{eq:entropy} as well as the correction to the horizon radius itself, namely to the generalized central charge $\mathscr{Z}$. 

\medskip

Finally, let us try to understand the extra constraints imposed by the hierarchy \eqref{eq:q0>ImG_0}. The first one is required so as to ensure that the corrections to the attractor solution and black hole entropy due to the genus-1 term in \eqref{eq:holomorphicprepotential@largevol} are in fact subleading with respect to the two-derivative results. The second condition, on the other hand, is more interesting, since its net effect is to suppress the higher-genus terms as well.\footnote{Actually, it suppresses the full tower of quantum corrections $\delta \mathcal{F}_g$ associated to D0-brane states, as captured by $G(Y^0, \Upsilon)$, which also includes a genus-0 contribution.} Furthermore, it can be translated into an equivalent mathematical statement on the value for $y^0$ (equivalently $x^0$ in \eqref{eq:Y^anoD6@2derivative}), whose precise growth with the gauge charges has been ignored so far. To see this, we should first compute the imaginary part of $G_0(Y^0, \Upsilon)$, which from the local 4d EFT perspective is given by the following asymptotic series
\begin{align}
	\label{eq:G0}
	G_0(Y^0, \Upsilon) = -\frac{i}{2 (2\pi)^3}\, \chi_E(X_3)\, Y^0 \sum_{g=0, 2, 3, \ldots} (2-2g)\, c^3_{g-1}\, \alpha^{2g} + \ldots\, ,
\end{align}
such that 
\begin{equation}\label{eq:ImG0}
	\begin{aligned}
		i\left(\bar{G}_0 - G_0\right) &= - \frac{\chi_E(X_3)}{(2\pi)^3}\, Y^0 \sum_{g=0, 2, 3, \ldots} (2-2g)\,c^3_{g-1}\, \alpha^{2g} + \ldots\\
		&= -\frac{\chi_E(X_3)}{8(2\pi)^3}\, |\Upsilon|^{1/2} \sum_{g=0, 2, 3, \ldots} (2-2g)\,c^3_{g-1}\, \alpha^{2g-1} + \ldots\, .
	\end{aligned}
\end{equation}
Notice that for sufficiently small expansion parameter $\alpha$, the quantity $i(\bar{G}_0 - G_0)$ grows like $1/\alpha \sim Y^0$. Therefore, imposing the condition $|\hat{q}_0| \gg |i(G_0 - \bar{G}_0)|$ amounts to having $Y^0 \sim y^0 \gg 1$ at the attractor locus, since we also have that $|\hat{q}_0| \gg y^0$, as per eq. \eqref{eq:chargehierarchy2derivative}. If this is so, then the iterative procedure followed to arrive at the solution \eqref{eq:implicitsolY^0} ---as well as the physical quantities derived thereafter--- is self-consistent. Thus, we find that the 4d EFT supplemented with the higher-derivative F-terms displayed in \eqref{eq:superspacelagrangian}, correctly accounts for the physical properties of the D0-D2-D4 black hole system if the following refined charge hierarchy is attained
\begin{align}\label{eq:finalchargehierarchy}
	(\hat{q}^0)^2,\, \left( p^a\right)^2 \gg \left| \frac{D_{abc} p^a p^b p^c}{\hat{q}_0} \right| \gg 1\, .
\end{align}
From here it is moreover straightforward to determine explicitly the contribution of the $G$-dependent terms to the entropy \eqref{eq:BHentropynoD6}, yielding a final answer of the form
\begin{equation}
\label{eq:asymptoticentropy4dregimenoD6}
	\mathcal{S}_{\rm BH} = 2 \pi \sqrt{\frac{1}{6} |\hat{q}_0| \left( \mathcal{K}_{abc} p^a p^b p^c + c_{2,a}\, p^a \right)} -\frac{\chi_E(X_3)}{4\pi^2}\, \sum_g c^3_{g-1} (y^0)^{2-2g}\, +\, \ldots\, ,
\end{equation}
where the ellipsis are meant to capture further subleading terms in $1/|\hat{q}^0|$, cf. eq. \eqref{eq:implicitsolY^0}. Hence, as advertised, the hierarchy \eqref{eq:finalchargehierarchy} precisely ensures that every quantum-induced contribution to the entropy indeed becomes negligible with respect to the quantity already calculated in \eqref{eq:classicalentropynoD6}.

\medskip

For future reference, let us show here how one should compute the corrected volumes in the internal Calabi--Yau geometry at the attractor (i.e., horizon) locus, once the higher-derivative effects have been properly taken into account. For instance, the overall threefold volume reads now as
	\begin{equation}\label{eq:vol@hor}
		\mathcal{V}_{\rm h} = \frac18  e^{-\mathscr{K}}\, |X^0|^{-2} \bigg\rvert_{\rm hor} =\; \frac18\frac{|\mathscr{Z}|^2}{|Y^0|^2} = \sqrt{\frac{6 |\hat{q}_0|^3}{ \mathcal{K}_{abc}p^a p^b p^c}} + \ldots\, ,
	\end{equation}
whereas those of the remaining even-dimensional cycles can be deduced upon appropriately contracting the stabilized K\"ahler coordinates
\begin{equation}\label{eq:Kahlermoduli}
	t^a_{\rm h} = \text{Im}\, \left( \frac{Y^a}{Y^0}\right)\bigg\rvert_{\rm hor} = p^a \sqrt{\frac{\hat{q}_0}{D_{abc}p^a p^b p^c}} + \ldots\, .
\end{equation}
Notice that, once again, the hierarchy \eqref{eq:finalchargehierarchy} is enough so as to ensure that the large volume approximation holds herein.

\subsubsection{The transition regime}
\label{sss:SmallBHsD0D2D4}

The analysis presented in Sections \ref{ss:D0D2D4system} and \ref{sss:higherderivativeD0D2D4} is very instructive and teaches us that, in those cases where our description can be reliably embedded into the large volume patch, the four-dimensional EFT provides sensible answers for the relevant physical properties of the supersymmetric black hole solutions considered therein. However, we noticed that the additional contributions arising from higher-derivative terms included in the half-superspace integral \eqref{eq:superspacelagrangian} lead to a certain tower of 4d local operators involving the (anti-sefl-dual part of the) graviton and graviphoton field strengths, which correct all these quantities via some perturbative series depending on a real-valued expansion parameter $\alpha = 1/Y^0$, cf. eq. \eqref{eq:G0}. Our main concern in what follows will be to understand both its physical significance as well as whether one could reach some pathological regime where the series would seem to break down, thus invalidating the four-dimensional effective description of the BPS system. 

\medskip

Let us start by considering the explicit definition of the parameter $\alpha$, which we recall here for the comfort of the reader
\begin{equation}\label{eq:alphadef}
	\begin{aligned}
		\alpha &= \frac{1}{8}\, \frac{|\Upsilon|^{1/2}}{Y^0}= \frac{1}{8}\, \frac{|\Upsilon|^{1/2}}{X^0 e^{\mathscr{K}/2} \bar{\mathscr{Z}}}\, ,
	\end{aligned}
\end{equation}
where we have substituted above the defining equation for $Y^0$. Therefore, upon taking its absolute value and using eq. \eqref{eq:vol@hor}, one arrives at
\begin{align}\label{eq:alpha=r_5/rh}
	|\alpha| = \frac{1}{8}\, \frac{|\Upsilon|^{1/2}}{|X^0| e^{\mathscr{K}/2} |\mathscr{Z}|} \stackrel{\eqref{eq:attractoreqs}}{=} \frac{\sqrt{8\mathcal{V}_{\rm h}}}{|\mathscr{Z}|} = \frac{r_5}{r_h}\, ,
\end{align}
where $r_5$ is the physical radius of the dual M-theory circle evaluated at the horizon ---as computed from the D0 mass, and $r_h$ denotes that of the black hole. Note that the last identity in \eqref{eq:alpha=r_5/rh} readily follows from the identification $r_h= |\mathscr{Z}| \kappa_4/\sqrt{8\pi}$ (cf. eq. \eqref{eq:nearHorizonmetric}), as well as the fact that the internal $\mathbf{S}^1$ radius is captured by the characteristic Compton wavelength of the Kaluza-Klein replica, which in the present context correspond to the D0-brane states. The latter have a (running) mass that is easily determined to be $m_{\rm D0}= \sqrt{8\pi} |X^0| e^{\mathscr{K}/2}/\kappa_4$. Hence, when evaluated at the attractor point, the (absolute value of the) expansion parameter precisely captures the relative size between the black hole horizon and M-theory circle.

\begin{figure}[t!]
	\begin{center}
        \begin{overpic}[scale=0.60,trim = 1cm 4cm 1cm 2cm, clip]{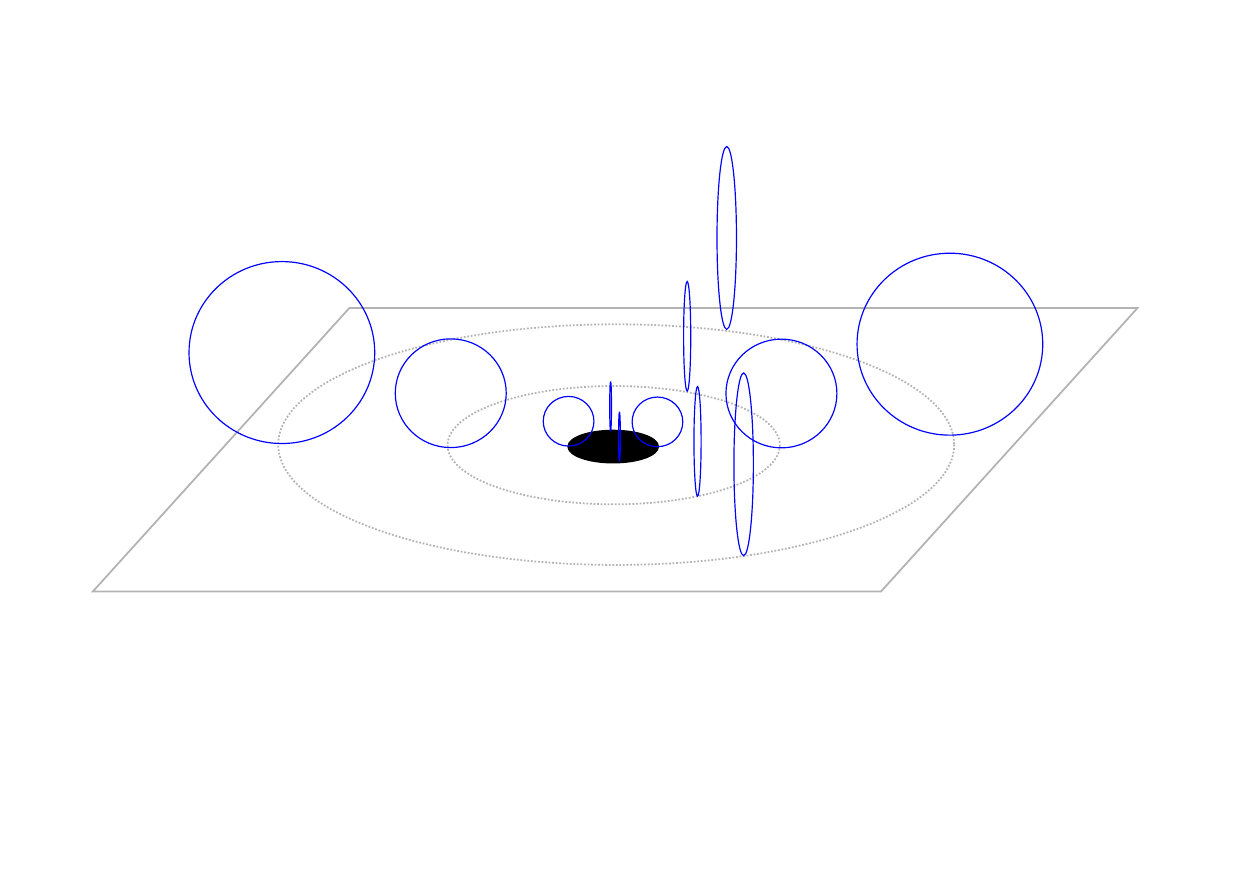} 
		  \put(82,36){$\mathbf{S}^1$} 
            \put(82,8){$\mathbb{R}^3$} 
		\end{overpic}
		\caption{\small Schematic depiction of the spatial profile for the M-theory circle (blue) in the 4d supersymmetric black hole background. Its asymptotic size is determined by the vacuum expectation value of the Calabi--Yau volume (in string units), and it evolves smoothly according to the attractor flow towards its fixed value at the horizon. If the size of the black hole is comparable to that of the extra circle, KK gravitons can be easily excited, yielding large corrections to e.g., the entropy.} 
		\label{fig:BH&Mthycirclefibration}
	\end{center}
\end{figure}

\medskip

Therefore, recall from our discussion in Section \ref{sss:higherderivativeD0D2D4} that the regime where the 4d EFT seemed to organize itself in a perturbative and well-behaved way so as to correctly reproduce the physical properties of the D0-D2-D4 black hole system, occurred precisely when $\alpha \ll 1$. When this is the case, the black hole becomes much bigger than the size of the extra dimension, such that a four-dimensional effective description must be able to describe the relevant physics. On the other hand, if we now consider the opposite situation where both radii are close to each other ---namely when $r_5 \gtrsim r_h$, then a putative 5d picture seems to be required. This corresponds to black hole solutions with $\alpha \gtrsim \mathcal{O}(1)$ at the horizon, and for those something interesting must be going on, since the series controlling $G (\alpha, \Upsilon)$ breaks down very quickly. This stems from the fact that the perturbative expansion capturing the quantum deformations of the black hole solutions exhibit numerical coefficients that grow in a factorial way, namely $c_{g-1}^3 \sim (2g-3)!$\,. Consequently, the series expansion
\begin{equation}\label{eq:absvalueGfn}
	G(Y^0, \Upsilon)\, \sim\, -\frac{i}{2 (2\pi)^3}\, \chi_E(X_3)\, (Y^0)^2 \sum_{g=0}^{\infty} c^3_{g-1}\, \alpha^{2g}\, ,
\end{equation}
can only provide an \emph{asymptotic} approximation \cite{Shenker:1990uf,Pasquetti:2010bps} to the exact result which is valid for $|\alpha| \ll 1$ (see Appendix \ref{ap:Asymptotic&Borel} for details), since it has formally zero radius of convergence. In mathematical terms, this means that for any order $N$ in the sum \eqref{eq:absvalueGfn}, the truncated series up to and including $k=N$, gives a better estimate for $G(\alpha, \Upsilon)$ the smaller $\alpha$ is.\footnote{In fact, as demonstrated in Appendix \ref{ss:D0nonpertcontribution}, the optimal truncation for \eqref{eq:absvalueGfn} occurs when we cut off the series at $g_{\star} \sim \frac12 \left(1+\frac{4\pi^2}{|\alpha|} \right)$.} Similarly, the larger we take $\alpha$, the more it deviates from the correct resummed result, leading ultimately to a seemingly divergent behavior. 

\medskip

Notice that the above discussion can be easily extended so as to accommodate other 4d BPS black hole systems which do not necessarily exhibit a real-valued expansion parameter $\alpha$. This rests on two important facts. First of all, note that what determines the ratio between the relevant scales is the absolute value of the expansion parameter $\alpha$, cf. eq. \eqref{eq:alpha=r_5/rh}, which does not care about its complex phase. And secondly, one can argue that the exact same considerations regarding the asymptotic properties of the series defining $G(Y^0, \Upsilon)$ also apply when the variable is complex instead of real-valued, see Appendix \ref{ap:Asymptotic&Borel}. In fact, an explicit black hole system where the aforementioned quantity is purely imaginary will be presented in Section \ref{s:other4dBHs} below.

\medskip

All in all, we conclude that the point $|\alpha| = \mathcal{O}(1)$ marks some sharp transition \cite{Calderon-Infante:2025ldq} where the 4d EFT ---including the tower of higher-curvature and higher-derivative local operators derived from \eqref{eq:superspacelagrangian}--- provides unphysical results, at least with regard to certain black hole properties. However, the interpretation of this failure 
is perfectly reasonable according to our discussion above. Indeed, the four-dimensional supergravity theory starts giving misleading predictions for the physics associated to certain BPS black holes precisely when these attain sizes which are of the order of the extra compact dimension (in the dual M-theory picture). At this point, one can no longer view the internal circle to be adiabatically fibered over the spatial $\mathbb{R}^3$ external to the horizon (see Figure \ref{fig:BH&Mthycirclefibration}), and in fact local fluctuations in the black hole geometry can easily excite KK modes (i.e., D0-branes). Hence, one should not expect a purely 4d description to be able to capture the physical properties associated to these systems, since they already belong to the five-dimensional realm. On the other hand, the higher-dimensional EFT should be able to cure somehow this pathology upon including highly-non local effects (when seen from the 4d perspective), which is what we will devote all our efforts to in the upcoming sections.

\subsection{The non-local resolution and the EFT transition}\label{ss:nonlocalcorrections}

\subsubsection{A resummed entropy formula}
\label{sss:nonlocalresummation}

In order to obtain a well-defined expression (of the universal piece) of the quantum-corrected generalized prepotential \eqref{eq:holomorphicprepotential@largevol} beyond the asymptotic series expansion \eqref{eq:Gfn}, we should evaluate more carefully the one-loop calculation associated to the D0-brane tower, which as already explained yields \cite{LopesCardoso:1999fsj, Gopakumar:1998ii}
\begin{align}\label{eq:G&I(alpha)}
	G(Y^0, \Upsilon) =\frac{i}{2 (2\pi)^3}\, \chi_E(X_3)\, (Y^0)^2\, \mathcal{I}(\alpha)\, ,
\end{align}
where we have defined\footnote{\label{fnote:tau&svars}Note that we have performed the change of variable $\tau=is$ in the proper time integral when going from \eqref{eq:generatingseries} to eq. \eqref{eq:I(alpha)} above.}
\begin{align}\label{eq:I(alpha)}
	\mathcal{I}(\alpha)\, =\, \frac{\alpha^2}{4} \sum_{n \in \mathbb{Z}}\int_{0^+}^{\infty} \frac{\text{d}s}{s}\, \frac{1}{\sinh^2\left( \pi n \alpha  s\right)}\, e^{-4\pi^2 n^2 i s} \,=\, \mathcal{I}^{(p)}(\alpha)\, +\, \mathcal{I}^{(np)}(\alpha)\, .
\end{align}
In the following, we will only be concerned with the perturbative contribution, $\mathcal{I}^{(p)}(\alpha)$, to the above integral, and we defer to Section \ref{ss:nonlocalnonpertD0brane} the discussion about the non-perturbative corrections, denoted here by $\mathcal{I}^{(np)} (\alpha)$. Hence, if we insist on substituting the Laurent series
\begin{equation}\label{eq:expansionsinh^-2(x)}
	\frac{1}{\sinh^{2}(x)} = \sum_{n=0}^{\infty} \frac{2^{2n}(1-2n)}{(2n)!} B_{2n} x^{2n-2}\, ,
\end{equation}
and subsequently exchange the order of the summation and integration in \eqref{eq:I(alpha)}, we recover the asymptotic approximation \eqref{eq:Gfn} for the perturbative part, $\mathcal{I}^{(p)}(\alpha)$. Alternatively, one may evaluate directly the above integral upon using the mathematical identity ${\sum_{n \in \mathbb{Z}} e^{2\pi i n \theta} = \sum_{k \in \mathbb{Z}} \delta (\theta-k)}$, which rather gives
\begin{figure}[t!]
	\begin{center}
		\subfigure[\label{sfig:RealpartIcurve}]{\includegraphics[width=0.48\textwidth]{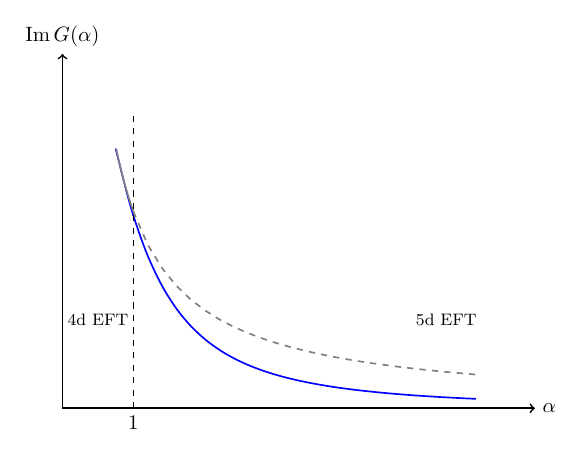}}
		\quad
		\subfigure[\label{sfig:ImaginarypartIcurve}]{\includegraphics[width=0.48\textwidth]{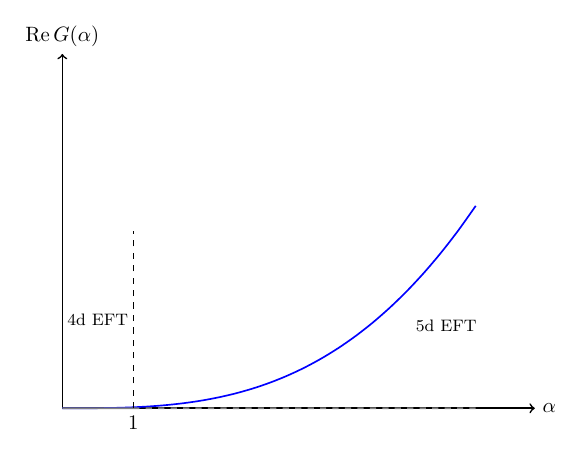}}
		\caption{\small Relevant $\alpha'$-corrections to the generalized holomorphic prepotential in the D0-D2-D4 black hole background close to the large volume regime. In M-theory language, the real part shown in $\textbf{(a)}$ arises from perturbative contributions associated to D0-brane states, whilst those displayed in $\textbf{(b)}$ account for non-perturbative Schwinger-like corrections.}
		\label{fig:Icurve}
	\end{center}
\end{figure}
%
\begin{align}\label{eq:I(alpha)pert}
	\mathcal{I}^{(p)}(\alpha) = \frac{\alpha^2}{4} \sum_{n \in \mathbb{Z}}\int_{0^+}^{\infty} \frac{\text{d}s}{s}\, \frac{1}{\sinh^2\left( \frac{\alpha s}{2}\right)}\, e^{-2\pi i n s}\, =\, \frac{\alpha^2}{4}\, \sum_{k=1}^{\infty} \frac{1}{k \sinh^2 \left( \frac{\alpha k}{2}\right)}\, .
\end{align}
The previous expression can be further massaged by expanding the denominator in \eqref{eq:I(alpha)pert} and performing the summation over the index $k$, thus arriving at
\begin{align}
	\label{eq:D0towernonlocal}
	G^{(p)}(Y^0, \Upsilon) = -\frac{i}{2 (2\pi)^3}\, \chi_E(X_3)\, (Y^0)^2\, \alpha^2\, \sum_{n=1}^{\infty} n \log \left( 1-e^{-\alpha n}\right)\, .
\end{align}
Notice that this function is non-analytic around $\alpha=0$, and this is in fact the reason why the series expansion in terms of purely four-dimensional operators displayed in eq. \eqref{eq:Gfn} crucially necessitates from further \emph{non-local} contributions. These corrections are nevertheless highly suppressed precisely when the black hole is large compared to the M-theory circle, namely when $\alpha \ll 1$, such that the quantities derived from the asymptotic series \eqref{eq:G0} match their ultra-violet completion obtained from \eqref{eq:D0towernonlocal}, which indeed verifies
\begin{align}
	G^{(p)}(Y^0, \Upsilon)\, \sim\, \frac{i}{2 (2\pi)^3}\, \chi_E(X_3)\, (Y^0)^2\, \zeta (3)\, =\, \frac{i}{2 (2\pi)^3}\, \chi_E(X_3)\, \zeta(3) \left( \frac{-\Upsilon}{64}\right) \alpha^{-2}\, , \quad \text{as}\ \ \alpha \to 0\, .
\end{align}
However, as soon as we get close to $\alpha \gtrsim \mathcal{O}(1)$, the aforementioned non-localities become essential and must be incorporated into the analysis so as to obtain a well-behaved, finite answer. Furthermore, it is easy to check that the function $G^{(p)}(Y^0, \Upsilon)$ shown in \eqref{eq:D0towernonlocal} is monotonic (see Figure \ref{sfig:RealpartIcurve}), and in fact satisfies
\begin{align}
	G^{(p)}(Y^0, \Upsilon)\, \sim\, \frac{i}{2 (2\pi)^3}\, \chi_E(X_3)\, (Y^0)^2\, \frac{\alpha^2 \csch^2 (\alpha/2)}{4} \to 0\, , \qquad \text{as}\ \ \alpha \to \infty\, .
\end{align}
This means, in turn, that once the 4d black holes have passed the 5d threshold they can be effectively described using a genuine M-theoretic analysis in five non-compact dimensions, see Section \ref{sss:gluing5d} for details. Incidentally, let us mention that the monotonicity properties of the resummed version of $G(\alpha, \Upsilon)$ (and derivatives thereof, cf. eq. \eqref{eq:resummedImG0} below) ensure that the iterative solution described around \eqref{eq:implicitsolY^0} is well-defined for \emph{all} values $\alpha>0$.

\medskip

For completeness, we compute here the resummed quantity controlling the quantum deformations of the black hole solutions
\begin{align}\label{eq:resummedImG0}
	i\left(\bar{G}_0 - G_0\right) =\frac{\chi_E(X_3)}{(2\pi)^3}\, \alpha^2\, \sum_{n=1}^{\infty} \frac{n^2\, e^{-\alpha n}}{1-e^{-\alpha n}}\, ,
\end{align}
which, as can be readily checked, tends to zero as well when $\alpha\to\infty$. This allows us to write down the full\footnote{We stress one more time that in this work we are only keeping track of the universal quantum correction arising from constant worldsheet maps into the target Calabi--Yau threefold \cite{Bershadsky:1993cx}.} quantum-corrected black hole entropy
\begin{equation}\label{eq:finalresummedentropynoD6}
	\begin{aligned}
		\mathcal{S}_{\text{BH}}\, =\ & 2 \pi \sqrt{\frac{1}{6} |\hat{q}_0| \left( \mathcal{K}_{abc} p^a p^b p^c + c_{2,a}\, p^a \right)} \left( 1 -  \frac{\chi_E(X_3)\,Y^0\, \alpha^2}{(2\pi)^3 |\hat{q}_0|}\, \sum_{n=1}^{\infty} \frac{n^2\, e^{-\alpha n}}{1-e^{-\alpha n}}\right)^{-1/2}\\
		&-\frac{\chi_E(X_3)}{4\pi^2} (Y^0)^2 \alpha^2 \left( \sum_{n=1}^{\infty} n \log\left(1 - e^{-\alpha n}\right) - (Y^0)^{-1} \sum_{n=1}^{\infty} \frac{n^2 e^{-\alpha n}}{1 - e^{-\alpha n}} \right)\, ,
	\end{aligned}
\end{equation}
where one must substitute the particular value of $Y^0$ that solves the attractor equation displayed in \eqref{eq:Y0noD6}. This expression should be moreover understood as the resummed version of eq. \eqref{eq:asymptoticentropy4dregimenoD6} above.

\medskip

Let us also note, in passing, that the present analysis reinforces the idea that the smallest possible black hole size is attained when the linear term in the generalized prepotential \eqref{eq:holomorphicprepotential@largevol} becomes of the same order as (or even dominates over) the classical cubic piece. This happens whenever the magnetic charges $p^a$ are all of order one, and in this case, eq. \eqref{eq:finalresummedentropynoD6} precisely accounts for the minimal black hole entropy. The latter provides an $\mathcal{O}(\Mpf^2/\LQG^2)$ number, with $\LQG$ denoting here the quantum gravity cut-off, i.e., the 5d Planck scale) \cite{Cribiori:2022nke,Calderon-Infante:2023uhz,Calderon-Infante:2025pls, 4dBHs}.

\subsubsection{Explicit gluing with the 5d solution}
\label{sss:gluing5d}

In this section we want to clarify our interpretation of  (\ref{eq:finalresummedentropynoD6}). We claim that it provides the full quantum-corrected black hole entropy of the D0-D2-D4 system when evaluated \emph{at} the large volume point. In the next section, we will explicitly show that the non-perturbative corrections encoded in the Schwinger integral do not contribute to the entropy formula, despite not vanishing. Here, we describe and check the consistency of the entropy, namely that the latter must be well-defined even when going beyond the regime of validity of our starting EFT and entering a new (possibly very different) one, thus effectively gluing the two complementary descriptions.

\medskip

As is well known, the four-dimensional EFT considered so far (cf. Section \ref{ss:4dsugrahigherderivatives}) has a natural embedding into a 5d theory with one direction compactified on a circle \cite{Witten:1995zh,Cadavid:1995bk}. At the classical ---i.e., two-derivative--- level, the supersymmetric black hole solutions studied in this section can be uplifted to five-dimensional, supersymmetric black strings wrapping the internal periodic direction \cite{Maldacena:1996ky}. Therefore, in the limit of infinite compactification radius they would appear to extend indefinitely. In string theory, this scenario is realized by uplifting Type IIA compactified on a Calabi--Yau threefold to M-theory reduced on the same compact space times a circle, and subsequently taking the large $\mathbf{S}^1$ limit. These solutions exist classically for every value of the black hole and the (asymptotic) compactification circle radii \cite{Gauntlett:2002nw}. However, one could naively wonder whether quantum corrections could spoil them. Importantly, notice that these two pictures must be regarded as complementary, limiting descriptions of the same physical object, since they naturally arise upon using two different EFTs. 
In particular, the 4d black hole is valid as a four-dimensional EFT solution as long as the quantum corrections are (highly) suppressed, which is equivalent to being in the regime where $\alpha \ll 1$ at the horizon. On the other hand, for $\alpha \gg 1$ the correct EFT description is the one of a five-dimensional black string wrapped on the M-theory circle, with the horizon transverse to and much smaller than the latter. Finally, for $\alpha = \mathcal{O}(1)$ the physical object still requires a 5d EFT description where the quantum corrections associated to the compactification circle are not necessarily small and need to be properly taken into account. 

Crucially, however, despite the difficulties in correctly describing the transition regime, 
the physical object still exists. Therefore, if a full quantum-corrected entropy in four dimensions is available, it might be well defined even when $\alpha \gtrsim \mathcal{O}(1)$. Indeed, the fact that it a priori accounts for both non-local and non-perturbative effects could potentially enable us to extrapolate certain physical properties beyond the failure of the EFT itself. In this regard, a simple but highly non-trivial test for the resummed entropy \eqref{eq:finalresummedentropynoD6} is that we can actually take the decompactification limit and reproduce the entropy density of a 5d black string.\footnote{We treat the system as an infinite five-dimensional string with an infra-red regularization that renders its total length finite (and equal to the volume of the extra circle). The entropy density is then defined in units of the infra-red regulator.} Thus, the aforementioned corrected 4d entropy is capable of correctly cross the EFT transition point. We dedicate the rest of this section to prove such an important result.

\medskip

The entropy of the BPS black holes of interest is exactly known in certain regimes. Given that we are considering a decompactification limit to five dimensions, various strategies can be followed. 
We focus first on the results based on microstate counting. In \cite{Maldacena:1997de}, and following the approach of \cite{Strominger:1996sh}, they computed the leading-order contribution to the microstate degeneracy of the 4d $\mathcal{N}=2$ black holes with $p^0=0$ (cf. eqs. \eqref{eq:Y0noD6}-\eqref{eq:BHentropynoD6})
\begin{equation}\label{eq:entropymicro}
     \mathcal{S}_{\text{micro}} = 2 \pi \sqrt{\frac{|\hat{q}_0| c_L}{6}} + \ldots\, ,
\end{equation}
where $c_L$ is the (left) central charge\footnote{For unitary conformal theories, the central charge must satisfy $c_L \geq 0$ \cite{Cappelli:1986hf,DiFrancesco:1997nk}. From the present, geometrical perspective, this is ensured by the fact that \eqref{eq:centralcharge} equals the second Chern number of the 4-cycle class $P\subset X_3$ wrapped by the M5-branes, which is non-negative for nef divisors \cite{Miyaoka1987,kanazawa2013}} of the sigma model associated to the moduli space of the worldvolume theory of the branes whose gravitational backreation generates the black hole. The central charge can be evaluated to be  \cite{Maldacena:1997de, Vafa:1997gr, Harvey:1998bx}
\begin{equation}\label{eq:centralcharge}
    c_L =  \mathcal{K}_{abc} p^a p^b p^c + c_{2,a}\, p^a\, .
\end{equation}
Strictly speaking, though, the above microscopic entropy formula (\ref{eq:entropymicro}) is only valid for $|\hat{q}_0| \gg c_L $  and $\mathcal{K}_{abc} p^a p^b p^c \ll \mathcal{V}_{X_3}$. But this is exactly equivalent to impose $\alpha \gg 1$ and thus to consider black hole solutions that are not anymore weakly curved, i.e., in the `small' black hole regime (cf. Section \ref{sss:SmallBHsD0D2D4}). Hence, taking the aforementioned limit in eq. \eqref{eq:finalresummedentropynoD6} yields 
\begin{equation}\label{eq:limitentropy}
	\mathcal{S}_{\text{BH}} \xrightarrow{\alpha \rightarrow \infty} 2 \pi \sqrt{\frac{1}{6} |\hat{q}_0| \left( \mathcal{K}_{abc} p^a p^b p^c + c_{2,a}\, p^a \right)} \,.
\end{equation}
which precisely reproduces (\ref{eq:entropymicro}).

\medskip

Let us consider now the infinitely extended black string uplift in five non-compact dimensions. A fundamental feature of these solutions is that they admit a near-horizon geometry of the form AdS$_3\times \mathbf{S}^2$. Treating the AdS$_3$ throat as a boundary, we can then compute the entropy of the configuration by evaluating the Cardy formula for the associated two-dimensional dual conformal field theory \cite{Maldacena:1997de, Ooguri:2004zv}. The resulting entropy is nothing but (\ref{eq:entropymicro}) with the central charge taken to be precisely (\ref{eq:centralcharge}). 

Interestingly, and in contrast to the four-dimensional case, in five dimensions one can actually prove that the entropy has the structure (\ref{eq:entropymicro}) also with a macroscopic computation. We start by clarifying what is the interpretation of the macroscopic entropy of a black string and, in general, of any extended black $p$-brane. For simplicity, we discuss this point in the two-derivative approximation. Clearly, if the entropy were simply the analog of the Bekenstein--Hawking area law we would obtain that all extended black-branes would have infinite entropy. Therefore, the way to obtain meaningful thermodynamics for such objects is to introduce regularized quantities in the form of worldvolume densities (see, for instance, \cite{deAntonioMartin:2012bi, Meessen:2012su}). Reinstating Newton's constant, we can conveniently define the entropy density $s$ for a $p$-brane living in $d$ spacetime dimensions as
\begin{equation}
    s = \frac{A_{\text{hor}}}{4 G_d}\,,
\end{equation}
where $A_{\text{hor}}$ is the $(d-p-2)$-dimensional black-brane transverse horizon, and $G_d$ is the $d$-dimensional Newton constant. Hence, for a five-dimensional black string one obtains 
\begin{equation}\label{eq:entropydensity}
    s =  \frac{A_{\text{hor}}/(2\pi R_5)}{4 G_4} \,,
\end{equation}
where $R_5$ has the role of a regulator measuring the total string length and we used the relation
\begin{equation}
G_5 = G_4 \, 2 \pi R_5 \,.
\end{equation}
Following Wald's prescription, one can easily show that (\ref{eq:entropydensity}) satisfies the first law of thermodynamics (more details can be found in \cite{Gomez-Fayren:2023wxk}). Notice that the above definition is such that, when compactified down to four dimensions, we recover the area law for the associated four-dimensional black hole. From now on, we will simply refer to the entropy of the black string as the entropy density.  

With this, we are finally in good position to proceed with the actual evaluation of the macroscopic entropy. Let us note that the supersymmetric black string is in fact a very special configuration. Thanks to its near-horizon structure AdS$_3\times \mathbf{S}^2$, one can easily use Wald's formula. We first dimensionally reduce the 5d $\mathcal{N}=1$ supergravity theory on the compact part of the near-horizon metric with constant matter fields, thus obtaining a three-dimensional effective Lagrangian 
\begin{equation}
    S = \frac{1}{16 \pi G_3} \int d^3x \sqrt{-g}\, \mathcal{L}_3 + S_{\text{bndy}} \,.
\end{equation}
Then, applying Wald's prescription on the resulting 3d action, one obtains precisely (\ref{eq:entropymicro}). The quantity $c$ playing the role of the central charge is now the action integral evaluated on an AdS$_3\times \mathbf{S}^2$ background, whose radii are fixed by a certain extremization procedure \cite{Sen:2005wa,Kraus:2005vz}
\begin{equation} \label{eq:centralchargemacro}
    c = \frac{\ell_{\rm AdS}}{2 G_3} g_{\mu\nu} \frac
    {\partial \mathcal{L}_3}{\partial \mathcal{R}_{\mu\nu}} \,,
\end{equation}
where $\mathcal{R}_{\mu\nu}$ is the Ricci tensor. Equation (\ref{eq:centralchargemacro}) is not completely determined by the near-horizon geometry and to evaluate it we must know the precise structure of $\mathcal{L}_3$. Interestingly, this computation correctly reproduces the central charge (\ref{eq:centralcharge}) upon considering just the standard, two-derivative 5d $\mathcal{N}=1$ supergravity action supplemented with the known four-derivative corrections \cite{Castro:2007sd}. Consequently, the macroscopic computation can be regarded as one-loop exact. On top of that, this confirms that (\ref{eq:limitentropy}) gives not only the leading term in the decompactification limit, but actually the \emph{exact} entropy of a five-dimensional black string extended along an infinitely long compact circle of volume equal to $\mathcal{V}_{\rm h}$ in 5d Planck units. 

\medskip

To sum up, let us recall that the entropy (\ref{eq:finalresummedentropynoD6}) was derived via a macroscopic computation in four dimensions, and the fact that it interpolates between the 4d and 5d regimes clarifies what is happening here. Non-perturbative contributions should not enter in the entropy of a BPS black string corresponding to a 4d $\mathcal{N}=2$ D0-D2-D4 BPS black hole system, whereas all non-local, higher-genus contributions must be suppressed along the $\alpha\gg 1$ limit. The only surviving corrections are those associated to the one-loop piece of the prepotential, which directly descends from the $t_8 t_8 \mathcal{R}^4$ term in 11d supergravity \cite{Antoniadis:1997eg}.

\subsection{Including non-perturbative effects}
\label{ss:nonlocalnonpertD0brane}

Notice that, from the perspective of the auxiliary topological string theory that can be used to compute certain terms within the generalized holomorphic prepotential \eqref{eq:generalizedprepotential}, we have restricted ourselves so far to the \emph{perturbative} sector of the theory. Hence, since it is well-known that one should actually expect further \emph{non-perturbative} contributions to arise (see, e.g., \cite{Marino:2024tbx} for a recent review on the topic), it is thus natural to wonder whether and how these additional effects could affect our previous analysis. 

\medskip

Our aim in the following will be to reconsider this point and show that, in fact, the main conclusions drawn from last section are left unchanged. To do so, we take two alternative routes that ultimately lead to the same answer. The first one proceeds, as explained in Section \ref{sss:directcomputationSchwinger}, by carefully evaluating the Schwinger determinant ---in complex proper time--- separately for each state in the D0-brane tower. Conversely, in Section \ref{sss:alternativeCauchy}, we derive an equivalent prescription by resumming the full tower of one-loop contributions, thereby treating both perturbative and non-perturbative corrections on an equal footing. We also briefly elaborate on the limitations associated to this second approach, which is very reminiscent of the recent proposal for computing the non-perturbative topological string partition function put forward in \cite{Hattab:2024ewk,Hattab:2024chf,Hattab:2024yol,Hattab:2024ssg}. For a lengthier discussion see Section \ref{sss:problemsCauchyformulation} below.

\subsubsection{Direct evaluation of the Schwinger integral}\label{sss:directcomputationSchwinger}

Our first strategy to obtain a non-perturbative definition of the leading-order prepotential at large volume proceeds similarly as we did in Section \ref{sss:nonlocalresummation}. There, following \cite{Gopakumar:1998ii,Gopakumar:1998jq}, we showed explicitly how by resorting to the dual M-theory description one is able to rewrite the relevant asymptotic series in an integral Schwinger-like form, cf. eqs. \eqref{eq:G&I(alpha)} and \eqref{eq:I(alpha)}. Next, we performed the integration using an appropriate change of variable and a certain Fourier transform, which led us directly to the \emph{resummed} perturbative expression \eqref{eq:D0towernonlocal}. However, in doing so we were not concerned with some subtleties related to both the state-dependent change of integration variable (cf. footnote \ref{fnote:subtletychangevar}), as well as to possible singularities that could arise within the complex $s$-plane. In fact, it is easy to realize that, when relabeling $s \to s/2\pi n$ so as to reach the l.h.s. of \eqref{eq:I(alpha)pert}, we must separate between states with $n \geq 0$ and $n<0$, since their contours cannot be simply deformed into one another due to the (double) poles arising from the hyperbolic sine in the denominator of the integrand. Hence, taking this into account leads to the following two distinct contributions within $\mathcal{I}(\alpha)$
\begin{subequations}\label{eq:fullnonpertI(alpha)}
	\begin{equation}\label{eq:n>=0D0branes}
		\mathcal{I}_{n\geq 0}\,(\alpha) = \frac{\alpha^2}{4} \sum_{n \geq 0} \int_{0^+}^{\infty} \frac{\text{d}s}{s}\, \frac{ e^{-2\pi i n s}}{\sinh^2\left( \frac{\alpha s}{2}\right)}\, ,
	\end{equation}
	\begin{equation}\label{eq:n<0D0branes}
		\begin{aligned} 
			\mathcal{I}_{n <0}\,(\alpha) = \frac{\alpha^2}{4} \sum_{n \geq 1} \int_{0^-}^{-\infty} \frac{\text{d}s}{s}\, \frac{e^{2\pi i n s}}{\sinh^2\left( \frac{\alpha s}{2}\right)}\, .
		\end{aligned}
	\end{equation}
\end{subequations}
%
\begin{figure}[t!]
	\begin{center}
		\includegraphics[scale=0.8]{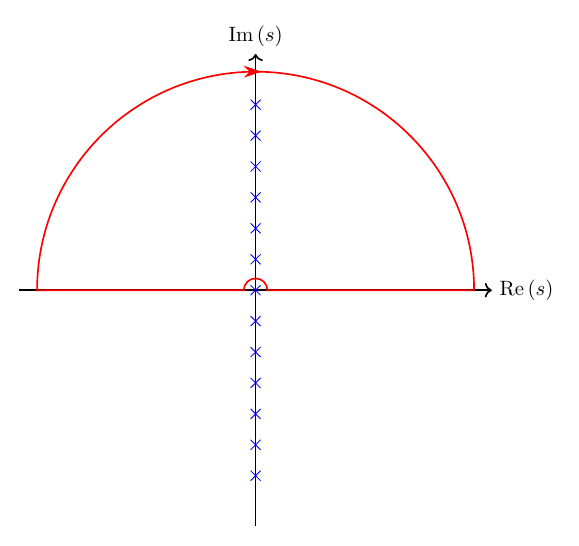}
		\caption{\small Integral contour in the complex $s$-plane that allow us to deform the one-loop determinant \eqref{eq:n<0D0branes} associated to D0-brane states with $n<0$ from the negative to the positive real axis. The singularities (blue crosses) located along the imaginary axis are associated to non-perturbative pair supergraviton production, and give rise to a non-trivial imaginary part for $\mathcal{I}(\alpha)$, as defined in \eqref{eq:I(alpha)}. Note that causality  ---as well as unitarity, in the form of the $i \epsilon$ prescription, fixes how the latter should be encircled \cite{Chadha:1977my}.} 
		\label{fig:D0branepoles}
	\end{center}
\end{figure}
As a next step, and in order to be able to perform the Poisson resummation leading to the r.h.s. of \eqref{eq:I(alpha)pert}, we need to deform the contour integral \eqref{eq:n<0D0branes} within the upper half plane so that it coincides with the positive real axis, thereby implying that we should pick up the residues of the infinitely many poles located $s=\frac{2\pi i k}{\alpha}$ for each $k \in \mathbb{N}$ (see Figure \ref{fig:D0branepoles}). The latter give rise to a non-perturbative contribution of the form\footnote{Notice that \eqref{eq:Inonpertalpha1stmethod} can be rewritten in terms of a single function $\varrho (\alpha)= i \frac{\alpha^2}{2\pi}\, \sum_{k=1}^{\infty} k^{-2} \left(1 - e^{4\pi^2k/\alpha}\right)^{-1}$ as follows
\begin{equation}
    \mathcal{I}^{(np)}(\alpha) = \alpha^2\frac{d}{d\alpha}\left( \frac{1}{\alpha}\,\varrho (\alpha)\right)\, .\notag
\end{equation}}
\begin{equation}\label{eq:Inonpertalpha1stmethod}
	\begin{aligned}
		\mathcal{I}^{(np)}(\alpha)\, &=\, -2 \pi i \alpha \sum_{n, k=1}^{\infty} \frac{n}{k}\, e^{-\frac{4\pi^2 k n}{\alpha}} \left( 1+ \frac{\alpha}{4\pi^2 kn}\right)\\
        &=\, -2 \pi i \alpha\, \sum_{k=1}^{\infty} \frac{e^{4 \pi^2 k/\alpha}}{k \left( e^{4\pi^2k/\alpha}-1\right)^2} - \frac{\alpha^2}{2\pi i}\, \sum_{k=1}^{\infty} \frac{1}{k^2 \left(1 - e^{4\pi^2k/\alpha}\right)}\, ,
	\end{aligned}
\end{equation}
where one can reach the second equality after summing over the index $n$. The above expression may be moreover expanded in the two limiting regimes which are most relevant for this work, namely when the corresponding 4d black hole is much bigger than the dual KK scale ($\alpha \ll 1$)
\begin{equation}\label{eq:Inonpertalpha<<1}
	\begin{aligned}
		\mathcal{I}^{(np)}(\alpha)\, &\sim\, -2 \pi i \alpha \sum_{n, k=1}^{\infty} \frac{n}{k}\, e^{-4\pi^2 k n/\alpha}\, \sim\,  -\frac{2\pi i \alpha}{\left(2 \sinh(2\pi^2/\alpha)\right)^2}\, ,
	\end{aligned}
\end{equation}
or alternatively when it belongs to the parent 5d theory ($\alpha \gg 1$), thus obtaining instead
\begin{equation}\label{eq:Inonpertalpha>>1}
	\mathcal{I}^{(np)}(\alpha)\, \sim\, -4 \zeta(3) \left(\frac{\alpha}{2\pi i}\right)^3\, .
\end{equation}
For illustrative purposes, we have depicted the exact non-perturbative contribution, computed from \eqref{eq:Inonpertalpha1stmethod}, in Figure \ref{sfig:ImaginarypartIcurve}. Notice, in particular, the polynomial dependence with the expansion parameter $\alpha$ that arises in the deep five-dimensional regime, whose physical origin may seem surprising from the perspective of the auxiliary topological string theory. Furthermore, one might object that such divergent behavior could potentially undermine the discussion presented in Section \ref{sss:gluing5d}, where it was argued that in the $\alpha \to \infty$ limit the entropy formula should match the one computed directly within the uplifted 5d supergravity theory. Very remarkably, we observe that this additional non-perturbative correction in $G(Y^0, \Upsilon)$ does \emph{not} modify the attractor equations nor the entropy function associated to the black hole system we are interested in here, since only the real part of $\mathcal{I}(\alpha)$ contributes to those, cf. eqs. \eqref{eq:Y0noD6}-\eqref{eq:genZ&entropynoD6higherderivative}. Let us stress here that the absence of this kind of corrections is intimately related to the non-perturbative stability of the BPS black hole background, a property that is ensured by supersymmetry. In any event, it is interesting to see explicitly how this expectation is borne out in the present set-up (see Section \ref{s:other4dBHs} for further evidence). 

\medskip

We also note, in passing, that the present analysis is in agreement with recent results obtained in \cite{Lin:2024jug}, where it was shown that extremal Reissner-N\"ordstrom black holes exhibit a non-trivial spatial profile for their decay rate induced by non-perturbative emission of Swchinger pairs due to charged particles already existing in the theory, unless the latter are also extremal. Therefore, given that both the black hole solutions considered herein and the charged D0-branes fulfill this condition \cite{Gendler:2020dfp,Heidenreich:2020upe,Heidenreich:2024dmr}, it makes perfect sense that such a decay channel does not exist in this case. 

\subsubsection{Alternative computation of the Schwinger determinant}\label{sss:alternativeCauchy}

Let us now present a different method so as to compute the non-perturbative corrections to the generalized holomorphic prepotential $F(Y^a, \Upsilon)$ due to the infinite tower of D0-branes. The emphasis will be placed on Cauchy's residue theorem, which allows us to obtain both perturbative and non-perturbative contributions from the singularity structure of a single, resummed Schwinger integral. 

\begin{figure}[t!]
	\begin{center}
		\includegraphics[scale=0.8]{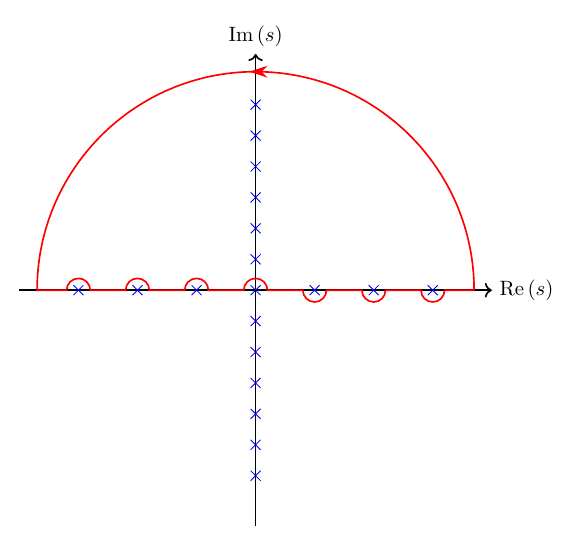}
		\caption{\small Integral contour in the complex $s$-plane employed to evaluate the integral \eqref{eq:CauchyD0}. The singularities located along the imaginary axis are associated to non-perturbative D0-brane effects, whereas the real poles correspond to the perturbative piece \eqref{eq:I(alpha)pert}.} 
		\label{fig:contourD0}
	\end{center}
\end{figure}

Hence, after repeating the same steps outlined at the beginning of Section \ref{sss:directcomputationSchwinger}, we arrive at two different integrals for the sector of positive (respectively negative) charged D0-brane bound states. Next, taking advantage of the fact that the non-perturbative poles are all located along the imaginary axis, we can slightly deform the integration ray for each separate integral towards/away the vertical axis.\footnote{Notice that the poles along the positive and negative axes must be shifted in opposite directions. Specifically, performing the shift starting from \eqref{eq:I(alpha)} and subsequently changing coordinates results in opposite shifts for the positive and negative modes. The direction of the shift is, in turn, determined by the requirement that the exponent of the exponential has a negative real part when evaluated along the real axis.} This allows us to resum the geometric series in $e^{2\pi i n s}$, such that eqs. \eqref{eq:n>=0D0branes} and \eqref{eq:n<0D0branes} reduce to
\begin{subequations}\label{eq:apfullnonpertI(alpha)}
	\begin{equation} 
		 \mathcal{I}_{n\geq 0}\,(\alpha) \,  \frac{4}{\alpha^2}  =  \int_{0^+}^{\infty} \frac{\text{d}s}{s}\, \frac{\sum_{n \geq 0} e^{-2\pi i n (s-i 0^+)}}{\sinh^2\left( \frac{\alpha s}{2}\right)} =  \int_{0^+}^{\infty} \frac{\text{d}s}{s}\, \frac{1}{1-e^{-2\pi i (s-i 0^+)}}\frac{1}{\sinh^2\left( \frac{\alpha s}{2}\right)}\, , \label{eq:integrandA}
	\end{equation}
	\begin{equation}
		\begin{aligned} 
		 \mathcal{I}_{n< 0}\,(\alpha) \, \frac{4}{\alpha^2} =  \int_{0^-}^{-\infty} \frac{\text{d}s}{s}\, \frac{\sum_{n \geq 1} e^{2\pi i n (s+i 0^+)}}{\sinh^2\left( \frac{\alpha s}{2}\right)} = \int_{-\infty}^{0^-} \frac{\text{d}s}{s}\, \frac{1}{1-e^{-2\pi i (s+i 0^+)}}\frac{1}{\sinh^2\left( \frac{\alpha s}{2}\right)}\, .
		\end{aligned}
	\end{equation}
\end{subequations}
Subsequently, we can add to the integration contour the semi-circle at infinity in the upper half plane since it does not contribute to the complex integral.\footnote{\label{fnote:realpartalpha}This is true, in general, only if $\text{Re}\, \alpha \ne 0 $. See Section \ref{sss:problemsCauchyformulation} for details on this point.} Finally, by  gluing the integrals in \eqref{eq:apfullnonpertI(alpha)} to avoid the pole at the origin, we construct a closed path in the complex $s$-plane (see Figure \ref{fig:contourD0}), thereby enabling us to rewrite \eqref{eq:I(alpha)} as follows\footnote{We emphasize that a similar approach to obtaining the full contribution of a given D0-D2 bound state to the generalized prepotential was recently proposed in \cite{Hattab:2024ewk}.} 

%
\begin{align}\label{eq:CauchyD0}
	\mathcal{I}(\alpha) = \frac{\alpha^2}{4} \oint \frac{\text{d}s}{s}\, \frac{1}{1-e^{-2\pi i s}}\frac{1}{\sinh^2\left( \frac{\alpha s}{2}\right)}\, ,
\end{align}
which can be finally evaluated upon using the residue theorem. Interestingly, there are two kinds of poles that contribute to the integral \eqref{eq:CauchyD0}. On the one hand, those occurring along the real axis $s= k\in \mathbb{Z}$ provide for the perturbative piece already computed in \eqref{eq:I(alpha)pert}. On the other hand, the poles located at $s=\frac{2\pi i n}{\alpha}$ account for non-perturbative D0-brane corrections. Hence, as a final result one obtains
\begin{subequations}\label{eq:residues}
    \begin{align}
    \mathcal{I}^{(p)}(\alpha)\, & = \, \alpha^2 \sum_{k=1}^{\infty} \frac{1}{4  k \sinh^2\left(\frac{k \alpha}{2}\right)}\,, \\
     \mathcal{I}^{(np)}(\alpha)\,& =\, -2 \pi i \alpha \, \sum_{n=1}^{\infty}  \, \frac{4n\pi^2 + 2 \alpha  \sinh\left(\frac{2 n \pi^2}{\alpha}\right) e^{-\frac{2 n \pi^2}{\alpha}}}{16 \pi^2 n^2 \sinh^2\left(\frac{2 n \pi^2}{\alpha}\right)} \,, \label{eq:I(alpha)nonpert}
    \end{align}
\end{subequations}
thus reproducing our previous expressions for the perturbative \eqref{eq:I(alpha)pert} and non-perturbative contributions \eqref{eq:Inonpertalpha1stmethod}, respectively. 

\medskip

Let us take the opportunity here to stress that the fact that both prescriptions to compute the perturbative and non-perturbative quantum contributions to the generalized prepotential due to the D0-brane tower agree rests, at the end of the day, on us being able to deform the contours for positive/negatively charged states, as well as to add the arc at infinity shown in Figure \ref{fig:contourD0} without any additional cost. Importantly, though, this may not always be necessarily the case, which ultimately depends on the complex phase that the expansion parameter $\alpha$ exhibits (cf. footnote \ref{fnote:realpartalpha}). We will elaborate further on this topic later on in Section \ref{sss:problemsCauchyformulation}.

\section{The Fate of Other BPS Black Hole Systems}\label{s:other4dBHs}

In Section \ref{s:BHs&EFTtransitions}, we have illustrated how certain supersymmetric black hole solutions are able to probe the ultra-violet cut-off scale of the theory that is used to describe both its geometry and physical properties. To do so, we focused on a particular family of BPS systems pertaining to the large volume regime, and studied in detail the convergence properties of the most relevant quantum corrections that deform their physical observables, such as the entropy. However, along the course of our investigation, several interesting comments were raised that we believe hold more generally, since the argumentation proceeded oftentimes in a rather solution-independent way (see, for instance, Section \ref{sss:SmallBHsD0D2D4}). Consequently, our aim in this section will be to see whether (and to what extent) these considerations apply to other BPS black holes in four spacetime dimensions.

\medskip

To accomplish this, we analyze in Section \ref{ss:4dBHsas5dBHs} another BPS system involving D2- and D6-brane charge. The reason for selecting this family of solutions will become clear along the way. Therefore, following the same strategy as in the previous chapter, we first describe these black holes from the perspective of the two-derivative supergravity theory. Subsequently, in Section \ref{sss:quantumcorrected5dBH}, we repeat the analysis taking into account the effect of the higher-derivative F-terms introduced around eq. \eqref{eq:superspacelagrangian}. A key difference between this configuration and the D0-D2-D4 black hole system is that the expansion parameter controlling the quantum deformations of the theory is now purely imaginary. However, as we argue in Section \ref{sss:SmallBHsD2D6}, by appropriately choosing the gauge charges one is able to probe the pathological regime $|\alpha| = \mathcal{O}(1)$ here as well. Nevertheless, in Section \ref{ss:nonlocal&nonpert5dBHs}, we show that it is still possible to include highly non-local effects due to the tower of D0-branes which allow us to resum their Schwinger contribution in an exact (i.e., non asymptotic) way. Crucially, we also find that non-perturbative effects are absent in this class of backgrounds, and hence do not modify the attractor solutions nor the entropy. This nicely matches with the observations made in previous sections.

\medskip

Before proceeding with our discussion, let us briefly summarize here our findings. First of all, we observe that, as soon as we turn on the D6-brane charge in the system, the solution can no longer explore the genuine 5d regime, which we recall corresponds to the $|\alpha| \to \infty$ limit. This can be readily checked from the attractor mechanism, even at two-derivative level. The latter imposes an upper bound on the stabilized value for the (absolute value of) expansion parameter $\alpha$, whenever $p^0 \neq 0$. Alternatively, from the M-theory perspective, one can argue that having non-trivial D6-brane charge is equivalent to introducing a background Taub-NUT geometry whose center coincides with that of the black hole (see Appendix \ref{ap:5dspinningBH} for details). This implies, among other things, that in the attractor locus at most ${r_h \sim r_5}$ can occur, thus preventing us from performing a clean matching with the putative 5d index, similarly to what we did in Section \ref{sss:gluing5d}. Still, by analyzing a representative example introduced in Section \ref{ss:4dBHsas5dBHs}, we confirm that an EFT transition must happen when $|\alpha|$ becomes of order one, which is signaled by an apparent singular behavior exhibited by the asymptotic series of corrections to, e.g., the entropy. This problem can be cured by resorting to the uplifted 5d theory, thereby including highly non-local effects involving the extra circular direction. In any event, very remarkably, we find that non-perturbative contributions in the Scwhinger integral also decouple from this kind of solutions, as it was the case in the simpler D0-D2-D4 system. Finally, in Section \ref{sss:problemsCauchyformulation} we point out that a simple alternative approach to compute the non-perturbative pair-production-like effects based on Cauchy (cf. Section \ref{sss:alternativeCauchy}) seems difficult to apply herein, thus requiring from a special treatment.

\subsection{Example 2: The D2-D6 black hole}\label{ss:4dBHsas5dBHs}

\subsubsection{The two-derivative analysis}\label{sss:2derivativeD2D6}

The family of black hole solutions with $p^0 = 0$ introduced in Section \ref{ss:D0D2D4system} is rather special, since they exhibit an explicit attractor for any combination of the remaining quantized charges (see, however, footnote \ref{fnote:degattractors}). On the other hand, in the most general situation with non-trivial $p^0, q_0$ and arbitrary $(q_a, p^b)$-charges, the system is characterized instead by a set of algebraic quadratic equations which may or may not have a real solution, even at the classical level of approximation \cite{Shmakova:1996nz}. Therefore, in order to provide yet another instance where quantum corrections to the entropy $\mathcal{S}_{\rm BH}$ can be determined and subsequently studied, we consider in what follows the restricted case of 4d black holes with $p^a=0$, i.e., with no D4-brane charge. The reason for this choice is twofold. First, such solutions ---which are shown to exist for any relative value of the horizon and \emph{asymptotic} M-theory radii \cite{Gauntlett:2002nw,Gaiotto:2005gf}--- can be more easily analyzed than their most general counterparts. Second, they uplift to five-dimensional spinning BPS black holes at the center of a Taub-NUT geometry (cf. Appendix \ref{ap:5dspinningBH}), in contrast to the D0-D2-D4 configuration, which instead corresponds to 5d black strings extended along the compactification circle.

\medskip

Let us start then by describing the system at leading order, namely without the higher-derivative corrections. Having $p^a=0$ means that the rescaled moduli take the following simple form at the horizon locus (cf. eq. \eqref{eq:attractoreqs2derivative})
\begin{align}\label{eq:YmodulinoD4}
	CX^0= \text{Re}\; CX^0 + i\, \frac{p^0}{2}\, , \qquad CX^a= \bar{C} \bar{X}^a =  \text{Re}\; CX^a\, .
\end{align}
Furthermore, from the attractor equations \eqref{eq:qaequation}-\eqref{eq:q0equation}, and using the restriction map \eqref{eq:attractor2derivative}, we also deduce that
\begin{subequations}\label{eq:qaq0eqsnoD4}
	\begin{equation}\label{eq:qaeqsnoD4}
	D_{abc}\, (CX^b) (CX^c) = - \frac{q_a}{3p^0} |CX^0|^2\, ,
 	\end{equation}
 	\begin{equation}\label{eq:q0eqsnoD4}
	q_0 = \frac{2\, p^0\, \text{Re}\; CX^0 \left(D_{abc}\, (CX^a) (CX^b) (CX^c)\right)}{|CX^0|^4} = -\frac{2\,\, \text{Re}\; CX^0\, (q_a CX^a)}{3|CX^0|^2}\, , 
 	\end{equation}
\end{subequations}
where $|CX^0|^2 = \left( \text{Re}\; CX^0\right)^2 + \frac{(p^0)^2}{4}$. These can be slightly simplified upon defining the following collection of real-valued variables \cite{Shmakova:1996nz}
\begin{align}\label{eq:xvariables}
	x^A= \text{Re}\; CX^A \sqrt{\frac{3}{|CX^0|^2}}\, .
\end{align}
In terms of those, the set of equations \eqref{eq:qaq0eqsnoD4} read as
\begin{equation}\label{eq:qaq0eqsnoD4xvar}
	D_{abc} x^b x^c = - \frac{q_a}{p^0}\, ,\qquad q_0 = -\frac{2}{9}q_a x^a x^0\, , 
\end{equation}
whereas\footnote{\label{fnote:consistencycond5d}Note that the existence of a real attractor solution requires having $(q_a x^a)^2 \geq \frac{27}{4} (q_0)^2$. This is a 4d manifestation of the five-dimensional inequality $|Z_{\rm 5d}|^3 \geq J_{\rm L}^2$. Here, $|Z_{\rm 5d}|^{3/2} = D_{abc} L^a L^b L^c$ ---with $L^a$ verifying $3 D_{abc} L^b L^c = -q_a^{\rm 5d}$--- is related to the central charge of the 5d black hole,  whereas $J_{\rm L} = (p^0)^2 q_0/2$ measures its (left-)angular momentum \cite{Kallosh:1996vy}. The former is determined, in turn, by the electric charges $q_a^{\rm 5d}=p^0 q_a$ \cite{Larsen:2006xm}.}
\begin{align}\label{eq:|Y^0|}
	|CX^0|^2= \frac{(p^0)^2 (q_a x^a)^2}{4 (q_a x^a)^2-27(q_0)^2}\, .
\end{align}
Hence, we arrive at an algebraic set of $h^{1,1} (X_3)+1$ real quadratic equations. The latter must be mutually compatible and admit a physical solution for us to claim the existence of a BPS configuration associated with the corresponding vector of (quantized) charges. From this, one can readily compute the absolute value of the central charge 
\begin{equation}\label{eq:classicalcentralchargenoD4}
	\begin{aligned}
		|Z|^2 &= D_{abc}\, (CX^a) (CX^b) (CX^c) \left[ \frac{3p^0}{|CX^0|^2} - \frac{p^0}{|CX^0|^4} \left( 3(\text{Re}\; CX^0)^2- \frac{(p^0)^2}{4}\right)\right]\\
		&=- (q_a CX^a) \left[1 - \frac{1}{|CX^0|^2}\left( (\text{Re}\; CX^0)^2- \frac{(p^0)^2}{12}\right)\right]\, ,
	\end{aligned}
\end{equation}
as well as the leading-order (i.e., classical) entropy, which is given, as usual, by $\mathcal{S}_{\rm BH} = \pi |Z|^2$. In order to simplify things even further, let us assume that there is no D0-brane charge in the system, i.e., $q_0=0$. In the dual five-dimensional theory, this translates into having zero angular momentum, and from \eqref{eq:qaq0eqsnoD4xvar} we conclude that $x^0 =0$ as well, such that $CX^0$ ends up being now purely imaginary. In addition, it is easy to show that in this case we have
\begin{equation}\label{eq:classicalcentralchargenoD4noD0}
	|Z|^2 = - \frac{4}{3} (q_a CX^a)\, ,
\end{equation}
in perfect agreement with the results of \cite{Shmakova:1996nz}. In order to elucidate the charge hierarchy needed for the solution to be well-behaved and within the large volume regime, we should study again the size of the different, relevant cycles evaluated at the horizon. On the one hand, we find
\begin{align}\label{eq:volnoD4noD0}
	\mathcal{V}_{\rm h} & =\; \frac18\frac{|Z|^2}{|CX^0|^2} = \frac23 \frac{(-q_a CX^a)}{(p^0)^2} = \frac{D_{abc} CX^a CX^b CX^c}{i \left(CX^0 \right)^3}\, ,
\end{align}
for the overall threefold volume, whilst
\begin{align}\label{eq:KahlermodulinoD4noD0}
	t^a_{\rm h} = \text{Im}\, \left( \frac{CX^a}{CX^0}\right)\bigg\rvert_{\rm hor} = -2\, \frac{CX^a}{p^0} = -\frac12\frac{p^0 CX^a}{|CX^0|^2}\, .
\end{align}
determines the attractor values for the K\"ahler moduli. We take, without loss of generality, $p^0>0$ in what follows. Therefore, from eqs. \eqref{eq:volnoD4noD0}-\eqref{eq:KahlermodulinoD4noD0} we conclude that $x^a$ (equivalently $CX^a$) must be negative definite, which also requires $q_a > 0$, as per \eqref{eq:qaq0eqsnoD4xvar}. Furthermore, asking for large volumes (in string units) at the attractor point translates into having
\begin{align}\label{eq:chargereqsnoD4noD0}
	\left| \frac{CX^a}{CX^0}\right| \gg 1\, ,
\end{align}
a condition that can be easily attained upon imposing the hierarchy $q_a \gg p^0$.\footnote{Notice that, given a solution $\{ x^a\}$ of \eqref{eq:qaq0eqsnoD4xvar}, one may obtain similar ones upon rescaling $q_a \to k^\beta q_a$ and ${p^0 \to k^\gamma p^0}$, which results into $x^a \to k^{\frac{\beta - \gamma}{2}} x^a$. Hence, by taking $\beta >\gamma$ and $k\gg 1$, one can easily achieve \eqref{eq:chargereqsnoD4noD0}.} Notice that, similarly to what happened with the D0-D2-D4 system (cf. \eqref{eq:chargehierarchy2derivative}), at this level of approximation we do not need to specify the asymptotics of $CX^0$, which in the present case is entirely determined by the D6-brane charge $p^0$. The latter turns out to control the importance of the relevant, perturbative quantum corrections, as we discuss next.

\subsection{Perturbative quantum corrections}\label{sss:quantumcorrected5dBH}

\subsubsection{Including higher-derivative corrections}
\label{ssss:solutionD2D6}

We consider in the following the quantum deformations induced by higher-derivative, protected terms derived from \eqref{eq:superspacelagrangian}. In this case, the classical attractor equations displayed in \eqref{eq:qaq0eqsnoD4} get modified as follows (cf. eq. \eqref{eq:rescaledvars})
\begin{subequations}\label{eq:qaq0eqsnoD4quantum}
	\begin{equation}\label{eq:qaeqsnoD4quantum}
		3D_{abc} Y^b Y^c = - \frac{q_a}{p^0} |Y^0|^2 - d_a \Upsilon\, ,
	\end{equation}
	\begin{equation}\label{eq:q0eqsnoD4quantum}
		q_0 = \frac{2\, p^0\, \text{Re}\, Y^0 \left(D_{abc} Y^a Y^b Y^c + d_a Y^a \Upsilon\right)}{|Y^0|^4} -i \left( G_0 - \bar{G}_0\right)\, .
	\end{equation}
\end{subequations}
Notice, in particular, the second (correction) term in the r.h.s. of eq. \eqref{eq:q0eqsnoD4quantum} due to the $g>1$ Gopakumar-Vafa operators. Given this structure, it is natural to ask ourselves whether the solution described in the previous section will survive once we take into account the aforementioned higher-order contributions. Namely, one would like to know if declaring ${q_0 = \text{Re}\, Y^0=0}$ is consistent with \eqref{eq:qaq0eqsnoD4quantum} above. To show that this is indeed the case, we only need to focus on the quantity $\text{Im}\, G_0 (Y^0, \Upsilon)$. Hence, upon taking into account that $Y^0$ is purely imaginary, it is straightforward to compute both $G(Y^0, \Upsilon)$ and $G_0(Y^0, \Upsilon)$ directly, which now read
\begin{subequations}\label{eq:locG&G0noD4D0}
	\begin{equation}\label{eq:locGnoD4D0}
		G(Y^0, \Upsilon) = \frac{i}{2 (2\pi)^3}\, \chi_E(X_3)\, |Y^0|^2 \sum_{g=0, 2, 3, \ldots} (-1)^g c^3_{g-1}\, |\alpha|^{2g}\, ,
	\end{equation}
	\begin{equation}\label{eq:locG0noD4D0}
		\frac{\partial G(Y^0, \Upsilon)}{\partial Y^0} = \frac{\chi_E(X_3)}{2 (2\pi)^3} |Y^0| \sum_{g=0, 2, 3, \ldots} (-1)^g (2-2g) c^3_{g-1}\, |\alpha|^{2g}\, .
	\end{equation}
\end{subequations}
Crucially, the reality condition on $G_0(Y^0, \Upsilon)$ implies that the dangerous term appearing in \eqref{eq:q0eqsnoD4quantum} vanishes identically. This, in turn, allows us to conclude that the classical solution presented in Section \ref{sss:2derivativeD2D6} still survives after including all relevant, \emph{perturbative} corrections. 

\medskip

For completeness, we also show here the generalized central charge, which is given by
\begin{equation}\label{eq:gencentralchargenoD4}
	\begin{aligned}
		|\mathscr{Z}|^2 &=- \frac{4}{3} Y^a \left(q_a - \frac{1}{12 p^0} c_{2,a} \right) + p^0\, \text{Re}\, G_0\, ,
	\end{aligned}
\end{equation}
with $\text{Re}\, G_0 = G_0$, as well as the quantum-corrected black hole entropy
\begin{equation}\label{eq:entropynoD6higherderivative}
	\mathcal{S}_{\rm BH} = - \frac{4}{3}\pi Y^a \left(q_a + \frac{1}{6 p^0} c_{2,a} \right) + \frac{ \chi_E(X_3) (p^0)^2}{4 (2\pi)^2} \sum_{g=0, 2, 3, \ldots} (-1)^g c^3_{g-1}\, |\alpha|^{2g}\, .
\end{equation}
where one should substitute in the above pair of equations the solution for $Y^a$ to the implicit equation \eqref{eq:qaeqsnoD4quantum}, as well as $|\alpha|=2/p^0$. Finally, note that in order to recover the results obtained from the previous two-derivative approach (plus a series of small corrections), we need to have $|Y^0| \gg 1$ at the horizon, which fixes the asymptotics of the magnetic charge $p^0$. 

\subsubsection{The transition regime} \label{sss:SmallBHsD2D6}

\begin{figure}[t!]
\begin{minipage}[t]{0.3\textwidth}
    \centering 
    \raisebox{1.5cm}{
    \begin{overpic}[scale=0.25,trim = 6cm 5cm 7cm 5cm, clip]{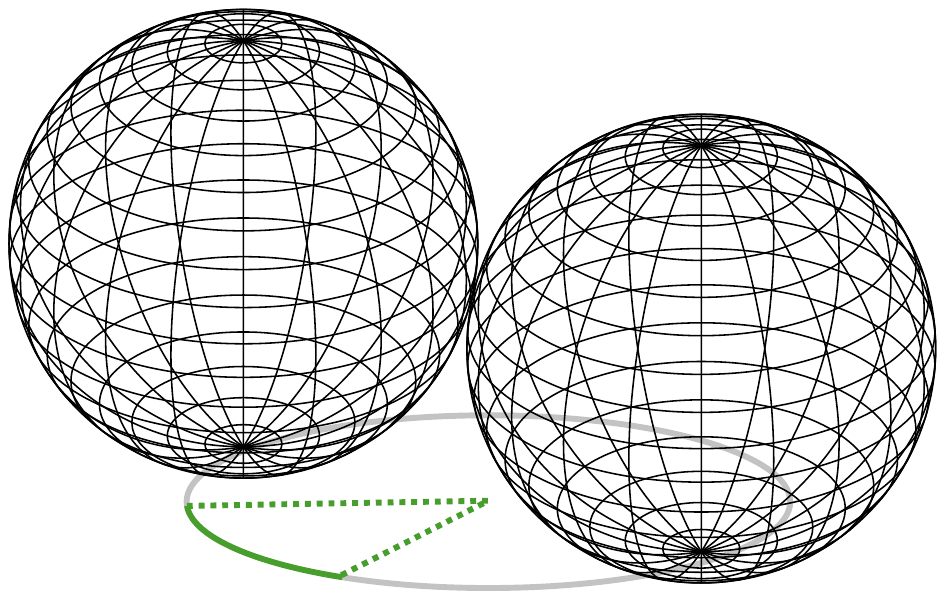} 
    \put(10, -2){$r_h/p^0$} 
    \put(46, 7){$r_h$} 
    \end{overpic}
    }
\end{minipage}%
\hfill
\begin{minipage}[t]{0.7\textwidth}
    \centering
   \begin{overpic}[scale=0.35,trim = 1cm 4cm 1cm 2cm, clip]{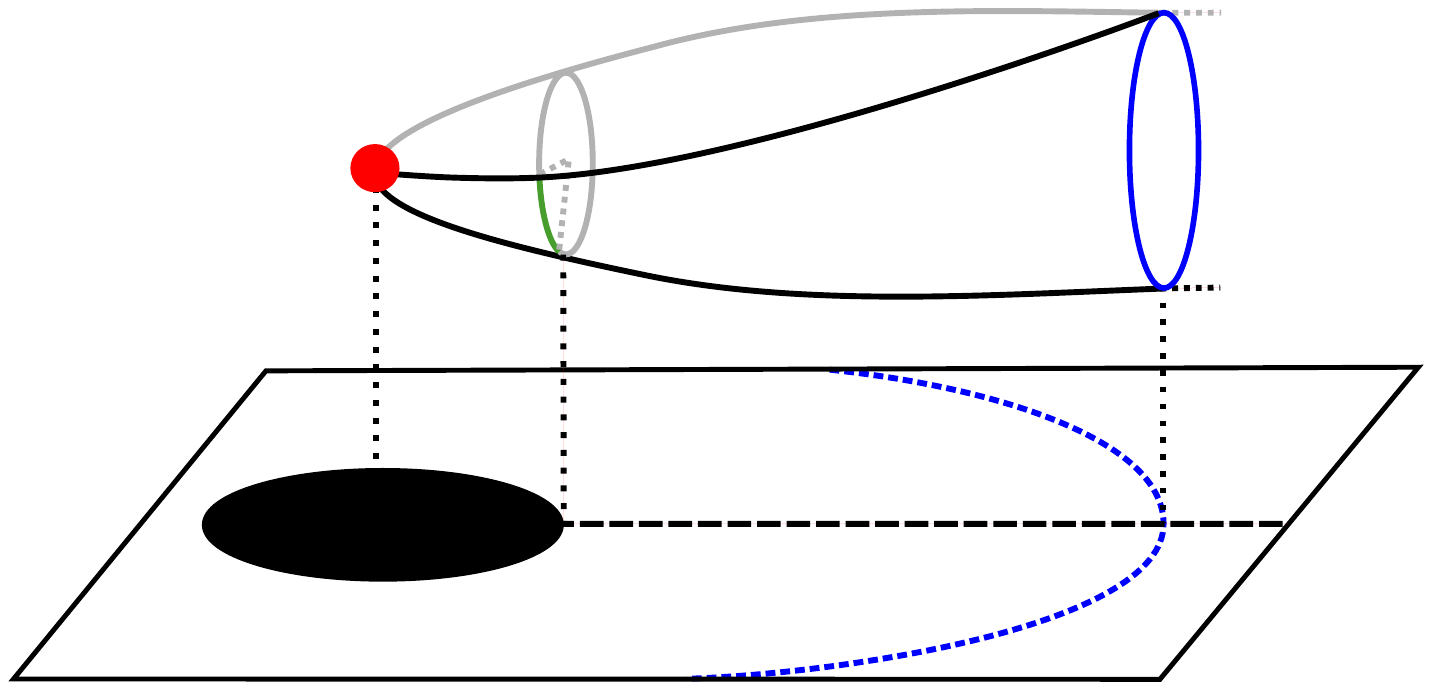} 
		  \put(83,36){$\mathbf{S}^1$} 
            \put(84,8){$\mathbb{R}^3$} 
	\end{overpic}
\end{minipage}
\caption{\small Schematic depiction of the induced profile for the extra compact direction (blue) in the 5d supersymmetric black hole background. The 5d black hole carries Taub-NUT charge and the spacetime has the geometry of a $\mathbf{S}^1$ circle fibration over the $\mathbf{S}^2$ component of $\mathbb{R}^3$. Asymptotically, the circle radius is finite and much smaller than the one associated to the 2-sphere (right). Close to the black hole horizon, the Taub-NUT geometry sourced by the D6-brane charge $p^0$ relates the scales of the $\mathbf{S}^2$ and the $\mathbf{S}^1$. In particular, the black hole horizon has the form of lens space $\mathbf{S}^3/\mathbb{Z}_{p^0}$ (left) with radius $r_h$.}
\label{fig:LensSpceandCigarBH}
\end{figure}

As discussed in Section \ref{sss:SmallBHsD0D2D4}, most of the considerations presented therein regarding the validity of the black hole solutions and the asymptotic behavior of the series of quantum corrections, should equally apply here as well. Nevertheless, there are, in fact, various important differences which are worth emphasizing. We start by focusing on the parameter $\alpha$, and highlight some of the properties which emerged during the construction of the BPS black hole solution. The latter is still defined as follows
\begin{equation}
    \alpha^2 = - \frac{1}{64} \frac{\Upsilon}{(Y^0{})^2}\,.
\end{equation}
Recall that in the D0-D2-D4 case, the attractor equations fixed $Y^0$ to be purely real. On the other hand, for the present D2-D6 configuration we have instead $Y^0 = i p^0/2$, which now implies that $\alpha$, when evaluated at the attractor point, becomes purely imaginary
\begin{equation} \label{eq:alphaD2D6}
    \alpha = -i \,|\alpha| \,, \hspace{1cm} |\alpha| = \frac{2}{p^0} \,.
\end{equation}
Regardless, the physical interpretation of the parameter $|\alpha|$ remains unchanged. Namely, once we sit at the horizon, it determines the ratio between $r_5$, i.e., the physical size of the dual M-theory circle ---as computed from the D0 mass, and the black hole radius $r_h$
\begin{equation}
    |\alpha| \stackrel{\eqref{eq:attractoreqs}}{=} \frac{\sqrt{8\mathcal{V}_{\rm h}}}{|\mathscr{Z}|} = \frac{r_5}{r_h}\,,
\end{equation}
where we used 
\begin{equation} \label{eq:radii}
    \mathcal{V}_{\rm h} =\; \frac18\frac{|\mathscr{Z}|^2}{|Y^0|^2}  \,,\qquad  r_5  = \frac{\kappa_4}{\sqrt{8\pi}} \frac{|{\mathscr{Z}}|}{|Y^0|}  \,, \qquad r_h= |\mathscr{Z}| \frac{\kappa_4}{\sqrt{8\pi}} \,.
\end{equation}
Crucially, and in contrast to the D0-D2-D4 example, the accessible range of $|\alpha|$ in this case appears to be upper bounded. This is a direct consequence of the relationship between the latter quantity and $p^0$ (cf. \eqref{eq:alphaD2D6}), which represents the amount of D6-brane charge in the system and is, as such, quantized, i.e., $p^0 \in \mathbb{Z}$. Hence, since our solution to the attractor equations \eqref{eq:qaq0eqsnoD4quantum} is well-defined as long as $p^0 > 0$, we find that $|\alpha| \in (0, 2]$, with $|\alpha| = 2$ corresponding to the particular choice $p^0 = 1$, whereas $\alpha = 0$ can be rather identified with the formal limit $p^0 \gg 1$. Consequently, a putative higher-dimensional regime with $r_5 \gg r_h$ can never be achieved, and the four-dimensional, quantum corrections to the entropy described in Section \ref{ssss:solutionD2D6} do not get diluted, unlike the case of 4d black holes originating from 5d wrapped black strings. The black hole radius is lower bounded by the physical radius of the M-theory circle at the horizon
\begin{equation}\label{eq:boundhh}
    r_h \gtrsim r_5 \,,
\end{equation}
so that we can only explore the purely 4d regime $|\alpha| \ll 1$, as well as the transition region $|\alpha| \sim 1$. From the point of view of the 5d embedding theory, such obstruction can be intuitively understood by recalling that the present solutions do uplift to black holes with Taub-NUT charge, where both centers coincide in spacetime. As a consequence, even if we try to make the radius of the five-dimensional black hole small compared to the size of the asymptotic circle, in the near-horizon geometry the latter becomes just one angular coordinate ---in an orbifold of $\mathbf{S}^3$--- whose radius behaves like that of the horizon itself (cf. Figure \ref{fig:LensSpceandCigarBH}). Note that, in general, the 5d and 4d black hole radii differ from each other. However, once the proper relation between them is established, it is straightforward to verify that a bound on the 5d radius implies a corresponding bound on $|\alpha|$. (See Appendix \ref{ss:5dspinningBHs&Taub-NUT} for details.) 

Let us also remark that the absence of an asymptotic five-dimensional regime prevents the existence of a genuine 5d configuration in the decompactified theory that could be used to `glue' with the D2-D6 black hole across the transition regime. This observation was already clear from the perspective of the classical, two-derivative theory, given that the KK monopoles sourcing the Taub-NUT charge are topological solitons that exist only in presence of compact directions, see Appendix \ref{ss:Taub-NUTgeneral}. Furthermore, we can confirm now that the previous conclusion is not modified upon including the relevant set of quantum corrections (perturbative and non-perturbative). Therefore, we cannot test our quantum-corrected entropy in the same way as in Section \ref{s:BHs&EFTtransitions} for the D0-D2-D4 black hole, where we ultimately recovered the (regularized) entropy of the 5d black string carrying M2, M5 and KK charges by taking the $\alpha \to \infty$ limit.

\medskip

Finally, we want to emphasize that the possibility of having a BPS solution with D2- and D6-brane charges depends, after properly accounting for the relevant perturbative quantum corrections, on the particular complex phase exhibited by the latter (cf. discussion around eq. \eqref{eq:locG&G0noD4D0}). Hence, in order to claim that the solution described in Section \ref{ssss:solutionD2D6} is indeed consistent\footnote{This amounts to being able to argue that setting $q_0 = \text{Re}\, Y^0=0$ in \eqref{eq:attractoreqs} is ultimately justified.} ---with $\alpha$ given by (\ref{eq:alphaD2D6}), one needs to analyze how the non-local and non-perturbative corrections (if any) behave in the present set-up. This is what we turn to next. 

\subsection{Non-local and non-perturbative effects}\label{ss:nonlocal&nonpert5dBHs} 

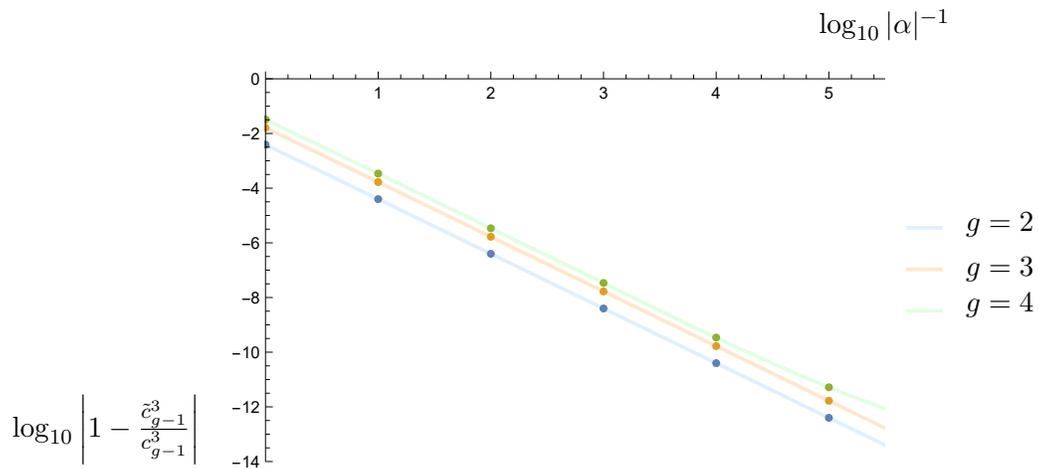
\begin{figure}[t!]
	\begin{center}
        \hspace{2cm}
		\begin{overpic}[scale = 0.95]{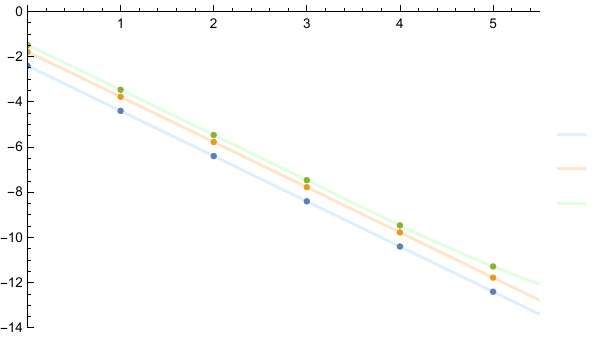}
                \put(80,60){$\log_{10}|\alpha|^{-1}$} 
                \put(-30,5){$\log_{10}\left|1- \frac{\tilde{c}_{g-1}^3}{ c_{g-1}^3}\right|$} 
                \put(100,33){$g=2$}
                \put(100,27){$g=3$}
                \put(100,22){$g=4$}
		\end{overpic}
		\caption{\small Numerical approximation of the coefficients in the series expansion of $\mathcal{I}(|\alpha|)$ when $|\alpha| \ll 1$. Along the  horizontal axis we denote the values of $\log_{10} |\alpha|^{-1}$ inserted in \eqref{eq:I(alpha)selfdualmagneticintegraldef}, 
        whilst the vertical axys measures how close the approximated coefficients $\tilde{c}^3_{g-1}$ turn out being with respect to the exact ones $c^3_{g-1}$, cf. eq. \eqref{eq:locGnoD4D0}. The dots represent the actual values of $\alpha$ that were employed. The colored lines have been extracted by interpolating the data, and correspond to different orders $g$ of the expansion.}
        \label{fig:errors}
	\end{center}
\end{figure}

The aim of this section will be to study in detail the non-local and non-perturbative effects lurking in the one-loop determinant \eqref{eq:generatingseries} associated to the full tower of D0-brane states, when evaluated in the D2-D6 black hole background. Thus, we proceed as in Section \ref{sss:nonlocalresummation} by focusing on the dominant quantum deformations of the generalized holomorphic prepotential. These are given by
\begin{align}
	G(Y^0, \Upsilon) = -\frac{i}{2 (2\pi)^3}\, \chi_E(X_3)\, |Y^0|^2\, \mathcal{I}(|\alpha|)\, ,
\end{align}
with
\begin{align}\label{eq:I(alpha)selfdualmagnetic}
	\mathcal{I}(|\alpha|)\, =\, \frac{|\alpha|^2}{4} \sum_{n \in \mathbb{Z}}\int_{0^+}^{\infty} \frac{\text{d}s}{s}\, \frac{1}{\sin^2\left( \pi n |\alpha|  s\right)}\, e^{-4\pi^2 n^2 i s}\, ,
\end{align}
and where we made use of the purely imaginary nature of the expansion parameter $\alpha$, cf. eqs. \eqref{eq:G&I(alpha)} and \eqref{eq:I(alpha)}. Crucially, and in contrast to what happened in the D0-D2-D4 system, we observe that the poles in the Schwinger integral are now real, such that we can freely deform the integration contour towards the imaginary axis without encountering any singularity that could account for some additional non-perturbative effect. Upon doing so, we obtain
\begin{equation}
\label{eq:I(alpha)selfdualmagneticintegraldef}
\begin{aligned}
    \mathcal{I}(|\alpha|)\, &=\, \zeta(3)\, -\frac{|\alpha|^2}{2} \sum_{n>0}\int_{0}^{\infty} \frac{\text{d}\tau}{\tau}\, e^{-4\pi^2 n^2 \tau} \left( \frac{1}{\sinh^2\left( \pi n |\alpha|  \tau\right)}- \frac{1}{(\pi n |\alpha|  \tau)^2} +  \frac13\right)\\
    &= \zeta(3)\, -\, \frac{|\alpha|^2}{2} \int_{0}^{\infty} \frac{\text{d}s}{s}\, \frac{e^{-\frac{4\pi s}{|\alpha|}}}{1-e^{-\frac{4\pi s}{|\alpha|}}} \left( \frac{1}{\sinh^2\left( s\right)} - \frac{1}{s^2}+\frac13\right)\, ,
\end{aligned}
\end{equation}
%
\begin{figure}[t!]
	\begin{center}
        \hspace{-1cm}
		\begin{overpic}[scale = 1]{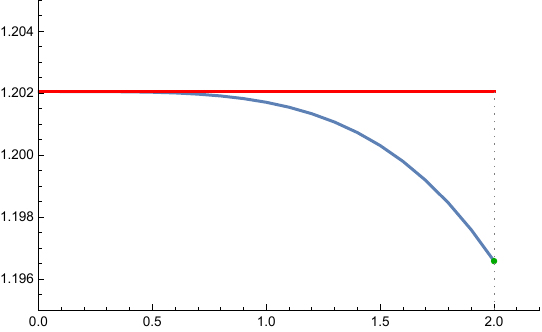}
		    \put(10,47){$\scriptstyle \mathcal{I}(0) = \zeta(3)$} 
                \put(60,30){$\scriptstyle \mathcal{I}(|\alpha|)$} 
                \put(95,10){$\scriptstyle \mathcal{I}(2) = 1.19658... $} 
                \put(95,-1){$|\alpha|$} 
		\end{overpic}
		\caption{\small Numerical plot of $\mathcal{I}(|\alpha|)$ as a function of the (purely imaginary) expansion parameter $|\alpha|$. The actual computation corresponds to the `renormalized' expression displayed in \eqref{eq:I(alpha)selfdualmagneticintegraldef}. Notice that the integral is convergent and finite for every $|\alpha| \in [0,2)$ within its physical domain (see footnote \ref{fnote:alphalargeD2D6}).}
        \label{fig:convergenceintegral}
	\end{center}
\end{figure}
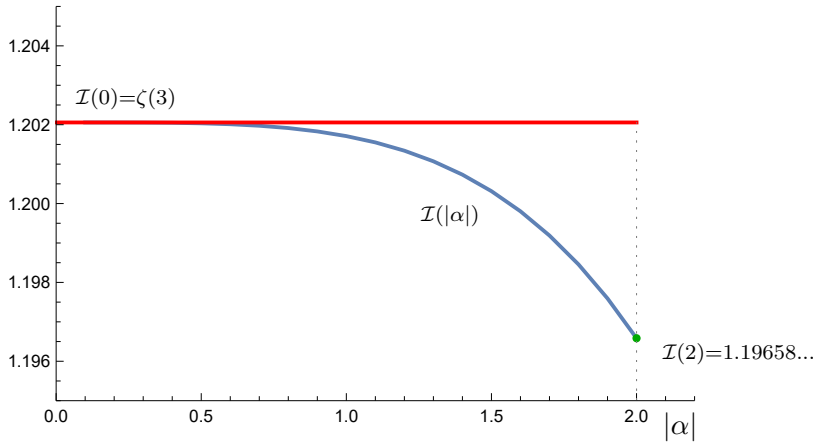
%
where we have explicitly separated the contributions for $g\leq 1$ and $g>1$, subsequently introduced the integration parameter $s = \pi n |\alpha| \tau$ and finally performed the summation of the geometric series in $e^{-4\pi s/|\alpha|}$. The terms subtracted in the parenthesis correspond to the regularization of the pole at the origin, which allow us to safely remove the cutoff $0^+$ in \eqref{eq:I(alpha)selfdualmagnetic} (see, for instance, \cite{Schwartz:2014sze}). We note, in particular, that the integral above can be easily checked to be convergent (cf. Figure \ref{fig:convergenceintegral}) and moreover defines the exact, resummed version of the asymptotic series \eqref{eq:locGnoD4D0}. To see this, one may insert back the Laurent series for $\text{csch}^2(x)$ at $x=0$ (cf. eq. \eqref{eq:expansionsinh^-2(x)}) in \eqref{eq:I(alpha)selfdualmagneticintegraldef}, subsequently exchange the order of summation and integration, and finally perform the integral for each term independently. For completeness, we show in Figure \ref{fig:errors} a numerical evaluation\footnote{The authors would like to thank Alessandro Lenoci for useful explanations on how to perform the high-precision numerical evaluation.} of the error made by the asymptotic approximation to the one-loop determinant when $|\alpha| \ll 1$.\footnote{\label{fnote:alphalargeD2D6}As a side note, let us point out that the integral \eqref{eq:I(alpha)selfdualmagneticintegraldef} is well-behaved and monotonic for $|\alpha| >2$ as well. It becomes negative around $|\alpha| \simeq 8$, and moreover behaves like $-|\alpha|^3$ for $|\alpha|\gg 1$, similarly to what happened with the non-perturbative contribution $\mathcal{I}^{(np)}(\alpha)$ in the D0-D2-D4 case, cf. eq. \eqref{eq:Inonpertalpha>>1}.} Let us also mention that the absence of additional non-perturbative terms in $G(Y^0, \Upsilon)$ can be understood as well from the fact that the series \eqref{eq:locG&G0noD4D0} are, in this case, \emph{alternating} and thus Borel summable (see Appendix \ref{ap:Asymptotic&Borel} for details), contrary to what happened in the wrapped black string background, cf. eq. \eqref{eq:G0}.

\medskip

Notice how the main two black hole systems described in this work differ in various crucial aspects of the physics. First, as already discussed, due to the range value of $\alpha$ they should be regarded as either purely four-dimensional objects (i.e., the D2-D6-brane case), or rather as a BPS configuration that is able to smoothly interpolate between the 4d and 5d regimes (i.e., the D0-D2-D4 system). Secondly, they exhibit either a real or purely imaginary expansion parameter $\alpha$, which is moreover associated with the presence of non-trivial or vanishing non-perturbative corrections (induced by the D0-brane tower) to the generalized prepotential $\eqref{eq:holomorphicprepotential@largevol}$, respectively. This fact is actually familiar from quantum electro-dynamics (QED), where the occurrence of non-perturbative pair production depends on whether a purely electric or magnetic constant field strength is applied (see, e.g., \cite{Kim:2003qp}). Relatedly, for the specific case of (anti-)self-dual backgrounds,\footnote{In QED, the anti-self-duality condition on the field strength $F_{\mu \nu}$ implies that $\boldsymbol{E} = i \boldsymbol{B}$. This restriction admits two different solutions, namely $\boldsymbol{B}$ real and $\boldsymbol{E}$ imaginary, or viceversa. They are usually referred to as (self-dual) \emph{magnetic} and \emph{electric}, respectively, thus exhibiting very different non-perturbative properties \cite{Dunne:2002qg}.} one finds that depending on the dimensionless ratio $\gamma = 2eF_-/m^2$ being real or imaginary, it may be possible to create real Schwinger pairs \cite{Dunne:2001pp,Dunne:2002qf,Dunne:2002qg,Kim:2003qp}. We could therefore regard these two systems as the 4d $\mathcal{N}=2$ gravitational analogues.

\subsubsection{Challenges and Obstructions in the Cauchy formulation}\label{sss:problemsCauchyformulation}

The derivation of equation \eqref{eq:CauchyD0} we outlined in Section \ref{sss:alternativeCauchy} appeared to be quite general and thus one might wonder if one could repeat the same steps with a generic complex-valued $\alpha$ so as to extend its validity to other cases as well. In the following, we will argue that this is indeed the case except for certain special choices of the aforementioned parameter, where the Cauchy prescription seems to fail ---in a dramatic fashion. In particular, as we will see, the complex integral is not well-defined if $\text{Re}\, \alpha=0$. This is precisely the case of interest for us in the present section, and the upshot will be that we cannot use equation \eqref{eq:CauchyD0} for the D2-D6 black hole configuration. Notice that the presence of a pathological behavior in the Gopakumar-Vafa prescription whenever the topological string coupling ---related to our parameter $\alpha$ here--- becomes purely imaginary has been already pointed out elsewhere in the literature (see, e.g., \cite{Marino:2024tbx} and references therein).

Let us consider then the naive analytic extension of \eqref{eq:CauchyD0} to all complex values of $\alpha$
\begin{align}\label{eq:complexCauchyD0}
	\mathcal{I}(\alpha) = \frac{\alpha^2}{4} \oint \frac{\text{d}s}{s}\, \frac{1}{1-e^{-2\pi i s}}\frac{1}{\sinh^2\left( \frac{\alpha s}{2}\right)}\, , \qquad \alpha = |\alpha| e^{i\theta_\alpha} \in \mathbb{C}\, .
\end{align}
First, note that since the integral is an even function of $\alpha$, we may, without loss of generality, restrict our analysis to the case $\text{Re}(\alpha) \ge 0$, i.e., we take $\theta_\alpha \in (-\pi/2, \pi/2]$ in what follows. The main effect of having a non-zero phase $\theta_\alpha$ in (\ref{eq:complexCauchyD0}) is that now the non-perturbative poles appear to be rotated in the $s$-plane (see Figure \ref{fig:RotatedContour} below). More concretely, they are located at $s =\frac{2 \pi n}{|\alpha|}\, \exp{(i\pi/2 - i\theta_\alpha)} $, with $n \in \mathbb{Z}$,  such that they do not lie anymore along the imaginary axis. Furthermore, in the limiting case where $\alpha$ exhibits no real part, the non-perturbative singularities fall onto the real axis, and they are given accordingly by $s = \frac{2 \pi n}{|\alpha|} $. In fact, for certain values of $|\alpha|$, some (or even all) of the poles might even coincide. Nevertheless, in all the cases that are relevant for us, $|\alpha|$ is fixed by the attractor mechanism to be a \emph{rational} number (cf. eq. \eqref{eq:attractoreqs}). Consequently, we always find a tower of simple poles at $s = k$ and an analogous infinite set of double poles at $s =\frac{2 \pi i n}{\alpha}$,\footnote{To be precise, this statement holds for non-zero $k$ and $n$, since the pole at $s=0$ is actually of fourth order.} with their residues still specified by \eqref{eq:residues}. Moreover, their asymptotic behavior for $\text{Re}\, \alpha > 0$ is found to be
\begin{subequations}
    \begin{align}
    \mathcal{I}^{(p)}(\alpha)\, & \sim  \, \alpha^2 \sum_{k=1}^{\infty}  \frac{1}{k}e^{- k \alpha}  \,, \\
     \mathcal{I}^{(np)}(\alpha)\,& \sim \, - 2 \pi i \alpha \sum_{n=1}^{\infty}     \frac{1}{n} e^{- \frac{4 n \pi^2}{\alpha}} \,,
    \end{align}
\end{subequations}
with the series being indeed convergent. However, when $\text{Re}\, \alpha = 0$, the asymptotics change abruptly, and we instead obtain
\begin{subequations}\label{eq:alphaimaginaryasymptotics}
    \begin{align}
    \mathcal{I}^{(p)}(\alpha)\, & \sim  \, - \alpha^2 \sum_{k=1}^{\infty} \frac{1}{4 k \sin^2\left(\frac{k |\alpha|}{2}\right)}  \,, \\
    \mathcal{I}^{(np)}(\alpha)\,& \sim \, 2 \pi i \alpha \sum_{n=1}^{\infty} \frac{1}{4 n \sin^2\left(\frac{2 n \pi^2}{|\alpha|}\right)} \,.
    \end{align}
\end{subequations}
Note that both series are now badly divergent, as well as their sum.\footnote{One might have hoped that for $\alpha = i |\alpha|$, where the two sums have opposite signs, they would cancel each other out. However, this never actually occurs, even if we set $|\alpha|=2\pi$, as can be easily verified from \eqref{eq:residues}.} Furthermore, their behavior is not only bounded from below by that of the harmonic series ---which is known to diverge, but they are in fact dominated by the contribution of terms with $n,k \in \mathbb{Z}$ which render the argument of the sine close to $\pi \mathbb{N}$. Being slightly more precise, one can argue that for every irrational number $\gamma$ there are infinitely many integer pairs $(p_\gamma,q_\gamma)$ satisfying
\begin{equation}\label{eq:Dirichlet}
    0 < \left| \gamma - \frac{p_\gamma}{q_\gamma}\right| < \frac{1}{q_\gamma^2} \,, 
\end{equation}
with $q_\gamma$ arbitrarily large, as per Dirichlet's approximation theorem \cite{apostol2012modular}. Hence, for any given such pair, one may establish the following lower bound
\begin{equation}
   \frac{1}{\sin^2(p_\pi)}\, =\, \frac{1}{\sin^2(\pi q_\pi + (p_\pi-\pi q_\pi))}\, \sim\, \frac{1}{|\pi q_\pi - p_\pi|^{2}}\, >\, q_\pi^2\, \sim\, \frac{p_\pi^2}{\pi^2} \,.
\end{equation}
Therefore, in the simple case where we set $|\alpha| = 2$, the perturbative series seem to be dominated by terms with $k = p_\pi$, whereas for the non-perturbative sum, the dominant contribution arises from terms with $n = p_{1/\pi}$. This two sets of `quasi-poles' do not match, and hence the partial sums grow in an oscillatory manner, as one may readily check.

\begin{figure}[t!]
	\begin{center}
        \hspace{2cm}
		\begin{overpic}[scale = 0.8] {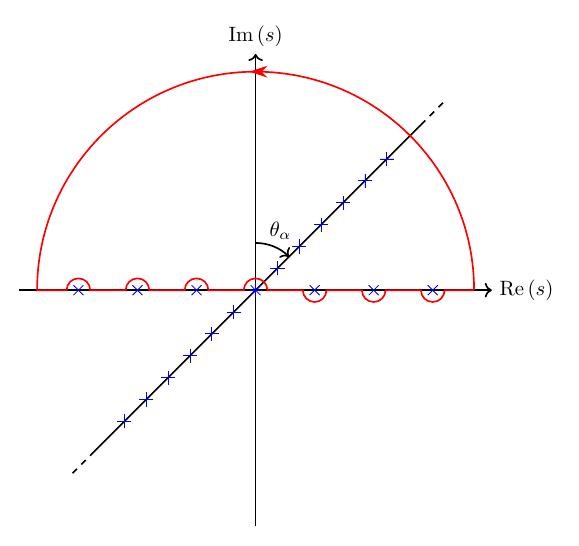}
		\end{overpic}
		\caption{\small Integral contour in the complex $s$-plane employed to evaluate the one-loop integral \eqref{eq:complexCauchyD0}. The non-perturbative singularities no longer lie along the imaginary axes if $\alpha$ has a non-vanishing complex phase $\theta_\alpha$. The real poles still correspond to the perturbative piece. In the limit of $\text{Re}\, \alpha \to 0$, all the poles therefore become real. However, for the rational values of $\alpha$ enforced by the attractor mechanism, perturbative and non-perturbative singularities do not coincide.}
        \label{fig:RotatedContour}
	\end{center}
\end{figure}

\medskip

The divergences we are encountering in the case of $\text{Re}(\alpha)=0$ signal that we should not be allowed to use anymore the residue theorem when evaluating \eqref{eq:complexCauchyD0}. Indeed, in its standard formulation (see, e.g., \cite{Ahlfors1966}), one considers a finite arc as well as finitely many poles. One can then formally extend the computation to the case of infinitely many isolated singularities by considering a discrete family of contours $\{\mathcal{C}_N\}$ which enclose $N$ such poles, and subsequently take the limit $N \to \infty$. This process is well-defined if and only if the limit exists, i.e., provided the series of residues converges. For $\alpha$ imaginary this does not happen, as we just discussed, and thus we cannot use the residue theorem to evaluate the integral. Notice that this oscillatory behavior can be ultimately traced back to the contributions associated to the infinitesimal semi-circles surrounding each pole along the real axis. In essence, what happens is that, since both sets of singularities are dense with respect to each other, the small arcs around the `quasi-poles' get an enhancement ---due to the closeness to the nearest singularity--- that grows as we go towards real-infinity, as per \eqref{eq:Dirichlet}.

In addition, it is worth highlighting that, from this perspective, one can also understand the origin of the harmonic-like behavior exhibited by the series of residues \eqref{eq:alphaimaginaryasymptotics}. Recall that the starting point was eq. \eqref{eq:I(alpha)}, which does require special care when $\alpha$ is purely imaginary. Whenever this happens, what one can do is to consider a shift so as to move the non-perturbative singularities infinitesimally away from the real axis, as we did in \eqref{eq:apfullnonpertI(alpha)}. While this deformation may seem sufficient to eliminate the divergences caused by going exactly through poles, it is not enough to fully regularize the expression. To see this, let us consider the integrand of \eqref{eq:integrandA} with $\alpha = i |\alpha|$, and where the non-perturbative poles properly are shifted as explained before, namely
\begin{equation}
     \frac{1}{s}\, \frac{1}{1-e^{-2\pi i (s-i 0^+)}}\,\frac{1}{\sinh^2\left( \frac{\alpha (s-i 0^+)}{2}\right)}\,.
\end{equation}
The introduction of a cutoff $0^+$ bounds the norm of the factors appearing in the integrand according to
\begin{subequations}
\begin{align}
    2 \pi 0^+ \lesssim   & \left| 1-e^{-2\pi i (s-i 0^+)} \right|  \lesssim  2 - 2 \pi 0^+  \,, \\ 
      \left(\frac{|\alpha| 0^+}{2}\right)^2 \lesssim &
      \left|\sinh^2\left( \frac{\alpha (s-i 0^+)}{2}\right) \right| \lesssim 1 + \left(\frac{|\alpha| 0^+}{2}\right)^2 \,,
\end{align}
\end{subequations}
and is thus a priori able to remove the divergences caused by the singularities. However, at the same time it also introduces an upper bound for the hyperbolic sine in the denominator, which does not suppress anymore the $1/s$ factor for large values of $s$. This explains, in turn, the logarithmic sub-divergence exhibited by the series \eqref{eq:alphaimaginaryasymptotics}.

All these considerations lead us conclude that simply deforming the contour of integration in e.g., \eqref{eq:apfullnonpertI(alpha)} so as to avoid the infinitely many isolated poles is not sufficient to completely regularize the integral, and, in fact, it becomes crucial to rotate the contour from the real towards the imaginary axis. Indeed, it is easy to see that upon doing so one arrives at
\begin{align}\label{eq:complexCauchyD0alphaimaginary}
	\mathcal{I}(|\alpha|) = -\frac{|\alpha|^2}{4} \int_{-\infty}^{\infty} \frac{\text{d}\tau}{\tau}\, \frac{1}{1-e^{-2\pi \tau}}\frac{1}{\sinh^2\left( \frac{|\alpha| \tau}{2}\right)}\, ,
\end{align}
where we substituted $\alpha= i |\alpha|$ and we defined $\tau= is$, cf. footnote \ref{fnote:tau&svars}. Note that the above expression is clearly reminiscent of the exact result derived in eq. \eqref{eq:I(alpha)selfdualmagneticintegraldef}, and it is moreover well-defined and convergent (except for the pole at $\tau =0$, which must be carefully dealt with as in the rest of this work). Furthermore, from this point of view it becomes clear that the pathologies we referred to in our previous discussion would be absent in \eqref{eq:complexCauchyD0alphaimaginary} and hence only appear if we close the contour by adding the arc at infinity, whereby picking up the residues displayed in \eqref{eq:residues}.
	
\section{Conclusions and Outlook}
\label{s:conclusions}

In this work, we have revisited the role played by macroscopic quantum corrections to the black hole entropy in 4d $\mathcal{N}=2$ supersymmetric effective field theories arising from Type IIA string theory compactified on Calabi–Yau threefolds. More precisely, by considering an infinite series of higher-derivative F-terms, we examined the most relevant perturbative and non-perturbative contributions to the supersymmetric black hole index close to the large volume point, emphasizing the interplay between different limiting (dual) descriptions in four and five spacetime dimensions. In particular, we obtained an explicit and well-defined, analytic expression for the entropy which allows us to track the underlying physical system across the transition between the two different field-theoretic descriptions. As a byproduct, this study highlights the importance of EFT transitions in understanding quantum-gravitational aspects of black hole thermodynamics.

\medskip

One of the main conclusions of our investigation is that the aforementioned quantum corrections to the entropy exhibit distinct behaviors as the characteristic size of the black hole approaches the Kaluza-Klein scale, where the effects of the extra dimensions become relevant. In particular, we identified a `transition regime' (cf. Section \ref{sss:SmallBHsD0D2D4}) where an infinite number of local higher-curvature and higher-derivative operators seem to induce pathological contributions to the macroscopic black hole entropy, signaling the failure of the 4d $\mathcal{N}= 2$ EFT to correctly describe such configurations. Indeed, the corrections organize into an asymptotic series with expansion parameter $\alpha$ which can be interpreted as the ratio of the inverse D0-brane mass to the black hole radius. We showed that one can regularize the series trough a resummation procedure and how the highly non-local perturbative effects induced by the infinite tower of Kaluza-Klein states (identified herein as D0 bound states) plays a crucial role in regulating certain ultra-violet divergences naively exhibited by the entropy function. This was explicitly analyzed for two BPS configurations, namely the D0-D2-D4 (Section \ref{s:BHs&EFTtransitions}) and the D2-D6 (Section \ref{s:other4dBHs}) black holes. Furthermore, in a sense, these two systems are somewhat complementary. From a more technical point of view, the resummation procedure that was used for the former black hole can be readily extended for more general charge configurations. Crucially, though, it fails precisely when we turn off D0- and D4-brane charges, which is equivalent to the regime of purely imaginary $\alpha$, see discussion in Section \ref{sss:problemsCauchyformulation}. Therefore, in Section \ref{s:other4dBHs}, we not only studied a second example but also clarified how to extend the resummation prescription employed in Section \ref{s:BHs&EFTtransitions} to this case as well. Moreover, as discussed in Section \ref{sss:gluing5d}, by lifting the solution to five dimensions and enforcing a parametrically small horizon radius compared to the size of the internal circle (whenever possible), one finds perfect agreement with the microscopic counting \cite{Maldacena:1997de,Vafa:1997gr,Harvey:1998bx} and one-loop exact computations in the dual M-theory description \cite{Castro:2007sd}. The transition regime can be crossed completely by taking some appropriate limit only for the D0-D2-D4 system. Indeed, the aforementioned black hole becomes a 5d black string wrapped along the extra compact direction, which grows indefinitely in the decompactification limit. Instead, the D2-D6 solutions uplifts to a five-dimensional black hole with Kaluza--Klein monopole charge, and it exists as long as the extra dimension is strictly compact. In this latter case, we can explore the transition regime; however, a topological obstruction prevents us from taking the full decompactification limit for this class of configurations. Remarkably, we also found that additional non-perturbative corrections ---which are oftentimes present and may be related to non-trivial pair production rate of Kaluza-Klein gravitons (and superpartners thereof) in the anti-self-dual graviphoton constant background close to the black hole horizon--- seem to not modify the most relevant physical properties of the solutions considered in this work, in particular their associated entropy (cf. Sections \ref{ss:nonlocalnonpertD0brane} and \ref{ss:nonlocal&nonpert5dBHs}). On the one hand, this further supports the consistency of the analysis carried out here. On the other hand, it raises the question of whether the same phenomenon might also occur for more general BPS black holes that can be constructed in the 4d theory. A more detailed investigation of these issues is left for future work \cite{Castellano:2025yur}.

\medskip

Our results may open up several promising avenues for future research. For instance, we restricted ourselves throughout this work to the large volume regime, where the dominant set of quantum correction to the generalized prepotential is universal and adopts a rather simple form (see Section \ref{sss:largevolprepotential} for details). Thus, it would be interesting to incorporate additional worldsheet instanton effects and see whether our conclusions are modified, if at all. Similarly, it would be valuable to extend this analysis to other singularities within the vector multiplet moduli space, which is also known to encode certain dualities with six-dimensional supergavity theories (obtained from F-theory compactified on a Calabi--Yau threefold), as well as with four-dimensional heterotic or Type II string compactifications \cite{Lee:2019wij}. The crucial difference with respect to our analysis has to do with the fact that, in those cases, the relevant corrections to the macroscopic entropy would be interpreted as quantum effects associated to massive particles other than Kaluza-Klein replica, or even exhibiting different spin statistics. Results along these lines will be reported in an upcoming work \cite{4dBHs}.

On another note, one may hope to be able to obtain from this perspective further insights into the non-perturbative behavior of certain topological string theories, which are known to capture the same prepotential controlling the higher-derivative corrections to the entropy \cite{Ooguri:2004zv}. In fact, we saw that in the D0-D2-D4 black hole background, namely the one associated to the unique system that is able to explore the genuine 5d regime, there should be a priori certain non-perturbative contributions to the generalized prepotential. These can be equivalently determined via a careful study of a one-loop integral associated to BPS states in M-theory \cite{Gopakumar:1998ii,Gopakumar:1998jq,Pasquetti:2010bps} (see also \cite{Hattab:2024ewk,Hattab:2024ssg} for recent related results). Hence, even though the latter did not ultimately affect any of the black hole observables we actually cared about in this work, they certainly exhibited interesting behaviors, particularly so along the decompactification limit. Interestingly, this regime can be analogously understood as the strong coupling limit of the auxiliary topological string theory. Consequently, we believe that a proper identification with the Kaluza-Klein production in a dual gravity theory (see \cite{Gabriel_2000,Friedmann:2002gx,Russo:2009ga} for earlier works) may offer key insights into the subject. 

Equally interesting is the fate of small black hole systems, namely those which seem to have vanishing Bekenstein-Hawking entropy at leading order in the charges (see e.g., \cite{Cano:2018hut} and references therein), in the presence of this kind of quantum corrections. Thus, it would be important to elucidate how these configurations (as well as their relevant thermodynamic properties) may be modified upon taking into account the full set of higher-derivative contributions to the supersymmetric entropy considered herein.

\medskip

We hope that our work serves to encourage further investigations into these and related exciting research directions.

\section*{Acknowledgements}
	
We are indebted to José Calderón-Infante, Matilda Delgado, Álvaro Herráez, Luis Ibáñez, Dieter L\"ust, Sameer Murthy, Tomás Ortín, Eran Palti, Alejandro Ruipérez, Savdeep Sethi and Ángel Uranga for illuminating discussions and useful comments on the manuscript. We also acknowledge valuable conversations with Alek Bedroya, Jake McNamara, Cumrun Vafa and Max Wiesner. A.C. would like to thank José Calderón-Infante and Álvaro Herráez for collaboration on related topics. The authors thank IFT-Madrid for hospitality and support during the different stages of this work. A.C. acknowledges the hospitality of the Department of Physics at Harvard University during the early stages of this work. The work of A.C. is supported by a Kadanoff and an Associate KICP fellowships, as well as through the NSF grants PHY-2014195 and PHY-2412985. A.C. and M.Z. are also grateful to Teresa Lobo and Miriam Gori for their continuous encouragement and support.
	

\appendix

\section{Asymptotic Series, Borel Resummation and Resurgence}
\label{ap:Asymptotic&Borel}

In this appendix we provide a brief overview on the mathematical theory of asymptotic series and resurgence, with an eye to direct applications in quantum field theory \cite{Dyson:1952tj} and string theory \cite{Shenker:1990uf}. Therefore, in Section \ref{ss:generalities} we first introduce and define these objects, paying special attention to their regime of validity. Subsequently, we comment on how the large order expansion of the aforementioned series contains relevant information for reconstructing the exact non-perturbative answer. Finally, in Section \ref{ss:D0nonpertcontribution} we illustrate all these matters in the most relevant example for this work, namely the (universal piece of the) non-perturbative corrections to the generalized holomorphic prepotential in Type IIA string theory at large volume due to D0-brane states in the Gopakumar-Vafa prescription \cite{Gopakumar:1998ii,Gopakumar:1998jq}.

\subsection{Asymptotic expansions and optimal truncation}\label{ss:generalities}

Mathematically, we say that a $\mathbb{R}$-valued function $f(x)$ has an \emph{asymptotic series expansion} around some point $x_0$,\footnote{\label{fnote:asymptoticseriesinf}It is also possible to define asymptotic series around infinity as follows
\begin{align}
   f(x) \sim \sum_{\ell=0}^{\infty} a_{\ell}\, x^{-\ell}\, , \qquad \text{as}\ \ x \to \infty \notag\, ,
\end{align}
for which the analogue of \eqref{eq:asymptoticseriesdef} becomes instead
\begin{align}
   \lim_{x \to \infty} \frac{f(x)-\sum_{\ell=0}^{N} a_{\ell}\, x^{-\ell}}{x^{-N}} =0 \notag\, .
\end{align}
} denoted here by
\begin{align}\label{eq:asymptoticseriesnotation}
   f(x)\, \sim\, \sum_{\ell=0}^{\infty} a_{\ell} \left( x-x_0\right)^{\ell}\, , \qquad \text{as}\ \ x \to x_0\, ,
\end{align}
if for any fixed order $N\geq 0$ in the sum, the difference between the truncated series and the exact value $f(x)$ is of $\mathcal{O}\left((x-x_0)^{N+1} \right)$. This condition can be written formally as
\begin{align}\label{eq:asymptoticseriesdef}
   \lim_{x \to x_0} \frac{f(x)-\sum_{\ell=0}^{N} a_{\ell} \left( x-x_0\right)^{\ell}}{\left( x-x_0\right)^N} =0\, .
\end{align}
More generally, one may also accommodate here the possibility of $f(x)$ behaving asymptotically as another mathematical expression $g(x)$ ---comprised perhaps by more elementary functions, upon declaring
\begin{align}
   \frac{f(x)}{g(x)}\, \sim\, \sum_{\ell=0}^{\infty} a_{\ell} \left( x-x_0\right)^{\ell}\, , \qquad \text{as}\ \ x \to x_0\, .
\end{align}
Notice that the above definition resembles ---but is actually different than--- that corresponding to convergent power series (i.e., Taylor/Laurent expansions). Hence, the infinite series specified by \eqref{eq:asymptoticseriesnotation} might be non-convergent, but nonetheless it must be that condition \eqref{eq:asymptoticseriesdef} holds for any of its finite order truncations. In fact, this concept can be readily extended to include complex functions as well, even though in that case one usually needs to be slightly more careful about the validity regime of the approximation due to e.g., Stokes' phenomena \cite{Dorigoni:2014hea} (see also the discussion around eq. \eqref{eq:Stokesdiscontinuity} below). Furthermore, it is easy to show that a function can have at most one asymptotic expansion around some point $x_0$ (or infinity, see footnote \ref{fnote:asymptoticseriesinf}), but the reverse statement is not true, i.e., two different functions $f(x)$ and $h(x)$ may share the same asymptotic series at a given point within their domain of definition.

\subsubsection{Optimal truncation and best approximation}\label{sss:optimaltruncation}

In the rest of this appendix we will focus on those asymptotic series which can only be interpreted as formal expansions, since they do not converge for any value of their argument. The latter are usually of the form 
\begin{align}\label{eq:Gevrey1}
\varphi (z)\, =\, \sum_{\ell=0}^{\infty} a_{\ell}\, z^{\ell}\, , \qquad \text{with}\ \ a_{\ell} \sim (\beta \ell)!\, ,
\end{align}
where we have defined $z=x-x_0$ in eq. \eqref{eq:asymptoticseriesnotation} above. Notice the factorial growth exhibited by the coefficients defining $\varphi (z)$, which in fact is responsible for the latter to be non-convergent. Indeed, regardless of how close to the origin we choose to evaluate the series, the prefactors $a_{\ell}$ eventually dominate and make the sum diverge in an unbounded exponential fashion. However, upon truncating the sum, the expression \eqref{eq:Gevrey1} provides for a sequence of approximations that become more accurate as we take $z \to 0$. Consequently, this implies that the identification $f(z) \sim \varphi (z)$ is not uniform, and a natural question that arises then is how to choose the \emph{optimal truncation} that provides the best approximation to the exact value of $f(z)$. Of course, one can always define the former as the particular order $\ell = N_{\star}$ for which the difference between the partial sums $\varphi_{N_{\star}}(z) =\sum_{\ell=0}^{N_{\star}} a_{\ell}\, z^{\ell}$ and the exact result is minimized. 
In practice, however, it is oftentimes the case that we do not have access to the function $f(z)$, so that we need to resort to any other useful definition that only depends on the asymptotic expansion $\varphi (z)$. Interestingly, even though there exists as of today no formal proof in the mathematical literature, it has been experimentally observed \cite{white2010asymptotic} that the optimal truncation for any asymptotic series of the form \eqref{eq:Gevrey1} seems to be attained for the maximum order $\ell$ such that 
\begin{align}\label{eq:optimaltruncationdef}
\left|\frac{a_{\ell}\, z^{\ell}}{a_{\ell+1}\, z^{\ell+1}}\right| > 1\, ,
\end{align}
remains true.\footnote{Note that if some of the expansion coefficients are vanishing, one should then compare pairs of consecutive non-zero terms in the series \eqref{eq:Gevrey1}.} This, in turn, is equivalent to ask for the value of $\ell=N_{\star}+1$ that minimizes $|a_{\ell}\, z^{\ell}|$, which as already stressed, will depend in general on the argument $z$.

\medskip

Finally, let us briefly comment on the regime of validity of any asymptotic expansion of the form specified by \eqref{eq:asymptoticseriesnotation}. In general, it is difficult to sharply and unambiguously define the value of $z$ where the series $\varphi(z)$ stops giving an accurate approximation to the exact function $f(z)$, for any of its finite order truncations. However, one can still estimate the breaking of the series by asking at which point the \emph{recessive} of the optimal truncation becomes comparable to the best approximation itself. Namely, suppose we declare that
\begin{align}
f (z)\, \sim\, \sum_{\ell=0}^{\infty} a_{\ell}\, z^{\ell}\, , \qquad \text{as}\ \ z \to 0\, ,
\end{align}
and we define the recessive $\mathscr{R}(z)$ as follows
\begin{align}
f(z)\, =\, \sum_{\ell=0}^{N_{\star}} a_{\ell}\, z^{\ell}\, +\, \mathscr{R}(z)\, ,
\end{align}
which is an exact (i.e., not asymptotic) relation. Then, we say that the asymptotic approximation breaks down at a \emph{sector boundary}, i.e., whenever $|\mathscr{R}(z)| \gtrsim \left|\sum_{\ell=0}^{N_{\star}} a_{\ell}\, z^{\ell} \right|$ holds, since from that point on the dominant and recessive contributions get exchanged.\footnote{Whenever we are deep within the regime of validity of a given asymptotic approximation, it is usually the case that the quantity $\mathscr{R}(z)$ becomes exponentially suppressed in $1/z$.} Notice that this definition would of course require from knowing the exact function $f(z)$, but it is good enough for our purposes herein. In any event, one can roughly estimate this happening whenever the optimal truncation becomes just the first term within the series \eqref{eq:Gevrey1}, as we will illustrate in a concrete example in Section \ref{ss:D0nonpertcontribution} below.

\subsection{Borel resummation and resurgent structures}\label{ss:resurgence}

One of the most surprising and interesting facts about quantum mechanics and quantum field theory \cite{Bender:1971gu,Bender:1973rz,Chadha:1977my,Collins:1977dw, Zinn-Justin:1980oco,LeGuillou:1990nq} concerns the observation that, oftentimes, the large-order behavior of a given perturbative asymptotic series secretly contains non-trivial (partial) information about its non-perturbative completion. This subtle connection is the object of study of the mathematical theory of \emph{resurgence} \cite{ecalle1981fonctions,Dorigoni:2014hea}. Here we would like to review some useful concepts and results that will allow us to better understand the discussion presented in Sections \ref{ss:nonlocalcorrections} and \ref{ss:nonlocal&nonpert5dBHs} of this work. Our treatment follows closely that of refs. \cite{Pasquetti:2010bps,Gu:2023mgf}.

\medskip

Therefore, let us assume that we are handed an asymptotic series of the form \eqref{eq:Gevrey1}. In order to study its resurgent properties, we can first perform a Borel transform as follows
\begin{align}\label{eq:boreltransformdef}
\mathcal{B} [\varphi] (\zeta)\, =\, \sum_{\ell=0}^{\infty} \frac{a_{\ell}}{(\beta \ell)!}\, \zeta^{\ell}\, ,
\end{align}
which removes by hand the problematic growth in the expansion coefficients of the original series $\varphi(z)$. In general, however, the resulting function will have singularities located within the Borel complex $\zeta$-plane, and it is a crucial task for us to find those. The reason being that, in fact, one can define the Borel sum 
\begin{align}\label{eq:Borelsumdef}
    \hat{\varphi} (z) &= \int_0^{\infty} \text{d}s\, e^{-s}\, \mathcal{B} [\varphi] (z s^{\beta})\, ,
\end{align}
which by construction has the same asymptotic expansion than the starting series $\varphi(z)$,\footnote{This can be easily shown upon using the definition of the $\Gamma$-function, namely $\Gamma (x) = \int_0^{\infty} \text{d}s\, s^{x-1}\, e^{-s}$.} 
such that if the Borel transform does not present any pole along the positive real line, one can then Borel resum the series \eqref{eq:Gevrey1} so as to obtain a finite, unambiguous result. Whenever this is the case, we say that the latter is Borel summable. This happens, for instance, with the leading-order (within the large volume patch) quantum corrections to the generalized holomorphic prepotential in the D2-D6 black hole background described in Section \ref{ss:4dBHsas5dBHs}. If, on the contrary, there exist some singularities along the domain of integration, one needs to specify a contour within the Borel plane so as to avoid them, which typically introduces certain (non-perturbative) ambiguities in the process of resummation. In what follows, we assume that there exist possibly infinitely many such singularities that we label by $\zeta_{\omega}$, which are moreover logarithmic branch cuts, such that near $\zeta = \zeta_\omega$ we have
\begin{align}\label{eq:expansionaroundsingus}
\mathcal{B} [\varphi] (\zeta_\omega + \chi)\, =\, - \frac{\mathsf{S}_\omega}{2\pi} \log (\chi)\, \mathcal{B} [\varphi_\omega] (\chi)\, +\, \ldots\, ,
\end{align}
where the complex numbers $\mathsf{S}_\omega$ are denoted Stokes constants, whilst the ellipsis is meant to indicate further regular terms in the variable $\chi$. Notice that we have also introduced an additional series $\mathcal{B} [\varphi_\omega] (\chi)$ in \eqref{eq:expansionaroundsingus} of the form
\begin{align}
 \mathcal{B} [\varphi_\omega] (\zeta) = \sum_{\ell=0}^{\infty} \frac{b_n}{(\beta \ell)!}\, \zeta^{\ell}\, ,
\end{align}
which has finite convergence radius and should be actually regarded as the Borel transform of
\begin{align}
\varphi_\omega (z) = \sum_{\ell=0}^{\infty} b_n\, z^{\ell}\, .
\end{align}
The above collection of data is what is usually referred to as the resurgent structure \cite{Gu:2021ize}, from which one can introduce the formal quantities
\begin{align}
\Phi_\omega (z) = e^{-\zeta_\omega/z^{1/\beta}}\varphi_\omega (z)\, ,
\end{align}
that are called \emph{trans-series} (see, e.g., \cite{Dorigoni:2014hea} and references therein). The physical relevance of these objects lies on the fact that they typically capture certain non-perturbative sectors of a given physical theory, such as instanton corrections \cite{Marino:2015yie}. Indeed, upon deforming the contour of integration in \eqref{eq:Borelsumdef} so as to cross any such singularity, one finds a discontinuity in the Borel transform given by
\begin{align}\label{eq:Stokesdiscontinuity}
\hat{\varphi}_+(z)-\hat{\varphi}_-(z) = i \mathsf{S}_{\omega}\, e^{-\zeta_\omega/z^{1/\beta}} \hat{\varphi}_-(z)\, ,
\end{align}
where the subscript $\pm$ indicates whether the ray of integration ---starting from the origin--- lies above/below the singularity.

\medskip

Let us finish this section by emphasizing that sometimes one can unambiguously define the Borel resummation of the original asymptotic series, even in the presence of multiple singularities within the Borel plane. This happens for instance if there exists a physical Schwinger-like representation for the function of interest \cite{Chadha:1977my}. In that case, it may be possible to define the integral in a way that avoids the aforementioned singularities, thus allowing for a physical interpretation of the non-perturbative corrections and the Stokes phenomenon displayed in \eqref{eq:Stokesdiscontinuity}.

\subsection{D0-brane contribution to $F(X, W^2)$ close to the large volume point}\label{ss:D0nonpertcontribution}

As stressed in Sections \ref{sss:SmallBHsD0D2D4} and \ref{sss:SmallBHsD2D6}, the main object of study in this work, i.e., the generalized holomorphic prepotential, exhibits an asymptotic-like behavior when evaluated close to the large volume point. This stems from the fact that the expansion coefficients in the perturbative series defining $G(Y^0, \Upsilon)$ close to the large radius point grow as $c_{g-1}^3 \sim \Gamma(2g-2)$, cf. eq. \eqref{eq:Fg>1LCS}. Consequently, all such corrections ---together with their contribution to the generalized central charge and black hole entropy--- have, strictly speaking, zero radius of convergence and should be regarded as approximate expressions valid for $|\alpha| \ll 1$. Our aim in the following will be to illustrate the different concepts introduced in this appendix within the present, four-dimensional set-up. We focus on the most relevant cases of BPS black holes with vanishing D6-brane charge (Section \ref{sss:nonalternating}), or zero D0- and D4-charge (Section \ref{sss:alternating}).

\subsubsection{The non-alternating case}\label{sss:nonalternating}

Let us consider first the D0-D2-D4 system analyzed in Section \ref{s:BHs&EFTtransitions}. Hereafter, we focus on the quantity that controls the quantum corrections to both the stabilized central charge and indexed entropy, namely $\text{Im}\, G_0$. The latter was computed in \eqref{eq:ImG0}, which we recall here for the comfort of the reader
\begin{equation}\label{eq:apImG0}
    i\left(\bar{G}_0 - G_0\right) = -\frac{\chi_E(X_3)}{8(2\pi)^3}\, |\Upsilon|^{1/2} \sum_{g=0, 2, 3, \ldots} (2-2g)\,c^3_{g-1}\, \alpha^{2g-1} + \ldots\, ,
\end{equation}
with $\alpha \geq 0$ and $\Upsilon = -64$, when evaluated at the attractor point. Indeed, the above series can be recast in the form \eqref{eq:Gevrey1}, as follows 
\begin{align}\label{eq:defphi(alpha)}
   \varphi (\alpha) \equiv \frac{i (\bar{G}_0 - G_0) (4\pi)^3}{|\Upsilon|^{1/2} \chi_E(X_3)} - \frac{2 \zeta (3)}{\alpha} = \sum_{g=2}^{\infty} \frac{4 \zeta(2g) \zeta(2g-2) (2g-1)!}{(2\pi)^{4g-2}}\, \alpha^{2g-1}\, ,
\end{align}
where we have subtracted by hand the genus zero and one terms, as well as substituted explicitly the numerical dependence of the expansion coefficients $c^3_{g-1}$ in \eqref{eq:apImG0}. Notice that the terms in $\varphi (\alpha)$ are \emph{non-alternating}, which will have important consequences when trying to extract its associated resurgent structure, cf. discussion around \eqref{eq:Borelsum}.

\subsubsection*{The optimal truncation}

As explained at the beginning of the appendix, the accuracy and convergence properties of the successive truncations that one may consider for $\varphi (\alpha)$ defined in eq. \eqref{eq:defphi(alpha)} above will depend on the value of the expansion parameter $\alpha$. Hence, it becomes important to estimate in a precise way where we must cut off the series, depending on the latter. The answer to this question is provided by the optimal truncation technique, which seeks to find the best approximation to the exact result by truncating the series so that the error that is made gets minimized. Thus, according to the discussion presented in Section \ref{sss:optimaltruncation}, we should proceed by looking for the maximum value of $g$ such that \eqref{eq:optimaltruncationdef} still holds, which in the case at hand is easily determined by the following minimization condition
\begin{align}\label{eq:optimal}
    \frac{d}{d(2g)} \log \left(a_g\, \alpha^{2g-1} \right) = \frac{d}{d(2g)} \log \left( \frac{4 \zeta(2g) \zeta(2g-2) (2g-1)!}{(2\pi)^{4g-2}}\, \alpha^{2g-1}\right) =0\, .
\end{align}
Moreover, upon assuming momentarily that the extremum is attained for large values of $g$ and using Stirling's formula $k! \sim (k/e)^{k}$, we find
\begin{align}\label{eq:gstar}
    \frac{d}{d(2g)} \log \left(a_g\, \alpha^{2g-1} \right)\, \sim\, \log \left[ \frac{(2g-1) \alpha}{4\pi^2} \right] =0 \Longleftrightarrow2g_{\star}-1 \sim \frac{4\pi^2}{\alpha}\, ,
\end{align}
which is indeed consistent with our original assumption as long as $\alpha \ll 1$. What this means, in practice, is that whenever we have perturbative control, it is enough to include just a few contributions within the sum in order to get an accurate result, and the smaller the expansion parameter is, the later one encounters significant deviations from the exact value. For instance, in the particular case where $\alpha = 1/20$, the optimal truncation happens for $g_{\star} \sim 395$ and in fact the series starts deviating ---in a sharp exponential way--- from the exact resummed result \eqref{eq:resummedImG0} when including terms with $g\gtrsim 1070$.

\subsubsection*{Borel resummation}

Furthermore, according to the theory of resurgence, it may happen that the previous asymptotic series contains non-trivial information about further non-perturbative physics that are not visible at any order in perturbation theory. Here we will show that this is indeed the case, using the machinery of Borel resummation reviewed in Section \ref{ss:resurgence}.

Hence, let us first compute the Borel transform of the series \eqref{eq:defphi(alpha)}. This yields
\begin{align}\label{eq:Boreltransform}
    \mathcal{B} [\varphi] (\zeta) = \sum_{g=2}^{\infty} \frac{a_{g}}{(2g-1)!}\, \zeta^{2g-1} = \sum_{g=2}^{\infty} 4 \zeta(2g) \zeta(2g-2) \left( \frac{\zeta}{4 \pi^2} \right)^{2g-1}\, .
\end{align}
Subsequently, we perform the Borel sum
\begin{align}\label{eq:Borelsum}
    \hat{\varphi} (\alpha) &= \int_0^{\infty} \text{d}s\, e^{-s}\, \mathcal{B} [\varphi] (s \alpha) = 4 \int_0^{\infty} \text{d}s\, e^{-s}\, \sum_{k, n=1}^{\infty} \sum_{g=2}^{\infty} k^{-2g} n^{2-2g}  \left( \frac{s \alpha}{4 \pi^2} \right)^{2g-1}\notag\\
    &=4 \int_0^{\infty} \text{d}s\, \sum_{k, n=1}^{\infty} \frac{n}{k}\, \frac{e^{-s} \left( \frac{s \alpha}{4 \pi^2 k n} \right)^{3}}{1-\left( \frac{s \alpha}{4 \pi^2 k n} \right)^2}\, ,
\end{align}
where to arrive at the second equality we inserted the definition of the $\zeta$-function
\begin{align}
    \zeta(x) = \sum_{k=1}^{\infty} k^{-x}\, ,
\end{align}
which is convergent when $\text{Re}\, (x)>1$, and in the last step we carried out the sum over the free index $g$. Let us note that \eqref{eq:Borelsum} can be written more suggestively as follows
\begin{equation}\label{eq:Borelsumcot}
\begin{aligned}
    \hat{\varphi} (\alpha) &= \sum_{n=1}^{\infty} \frac{(4\pi n)^2}{\alpha} \int_0^{\infty} \text{d}s\, e^{-\frac{4\pi n s}{\alpha}} \sum_{k=1}^{\infty} \frac{1}{k^4}\, \frac{s^{3}}{1- \frac{s^2}{k^2 \pi^2}}\\
    &= -\frac12 \sum_{n=1}^{\infty} \frac{(4\pi n)^2}{\alpha} \int_0^{\infty} \text{d}s\, e^{-\frac{4\pi n s}{\alpha}} \left( \cot s -\frac1s + \frac{s}{3}\right)\, ,
\end{aligned}
\end{equation}
where we used the following mathematical identity 
\begin{equation}\label{eq:seriescotx}
    \cot x -\frac1x + \frac{x}{3} = -\frac{2}{\pi^4}\sum_{k=1}^{\infty} \frac{1}{k^4}\, \frac{x^{3}}{1- \frac{x^2}{k^2 \pi^2}}\, ,
\end{equation}
so as to obtain the final expression. Therefore, it becomes clear that the integral \eqref{eq:Borelsumcot} (equivalently \eqref{eq:Borelsum}) exhibits an infinite number of poles located at $s= \frac{4 \pi^2 k n}{\alpha}$ for any pair of positive integers $( k, n)$. This poles can be conveniently arranged into infinitely many overlapping BPS rays, corresponding to $k$ worldline windings of a bound state of $n$ D0-branes. This implies that one should expect a non-perturbative completion of $\varphi(\alpha)$ to include corrections of order $e^{-4\pi^2 kn/\alpha}$. Notably, this is precisely confirmed by the exact computation  performed in the main text (cf. eq. \eqref{eq:Inonpertalpha1stmethod}). 

\subsubsection{The alternating case}\label{sss:alternating}

Finally, and for illustrative purposes, let us repeat the above exercise now specializing to the D2-D6 system analyzed in Section \ref{s:other4dBHs}. Notice that the main difference with respect to the analysis performed in Section \ref{sss:nonalternating} is that, in this case, the expansion parameter $\alpha$ controlling the asymptotic series $G(Y^0, \Upsilon)$ is purely imaginary. This means that the latter becomes now \emph{alternating}, which has non-trivial implications, as we argue in the following.

We start from the quantity
\begin{equation}\label{eq:apReG0}
    \bar{G}_0 + G_0 = \frac{\chi_E(X_3)}{8(2\pi)^3}\, |\Upsilon|^{1/2} \sum_{g=0, 2, 3, \ldots} (-1)^g (2-2g)\,c^3_{g-1}\, |\alpha|^{2g-1} + \ldots\, ,
\end{equation}
which is the analogue of eq. \eqref{eq:apImG0}, and subsequently define the asymptotic series 
\begin{align}\label{eq:defphi(alpha)D2D6}
   \varphi (|\alpha|) \equiv -\frac{(\bar{G}_0 + G_0) (4\pi)^3}{|\Upsilon|^{1/2} \chi_E(X_3)} - \frac{2 \zeta (3)}{|\alpha|} = \sum_{g=2}^{\infty} (-1)^{g}\, \frac{4 \zeta(2g) \zeta(2g-2) (2g-1)!}{(2\pi)^{4g-2}}\, |\alpha|^{2g-1}\, .
\end{align}
On the one hand, regarding the optimal truncation and best approximation, for any given value of $|\alpha|$, it is easy to see from the definition \eqref{eq:optimaltruncationdef} that the situation remains the same as compared to the D0-D2-D4 system. Therefore, the exact same argument as before leads to the estimate $2g_\star -1 = \frac{4\pi^2}{|\alpha|^2}$, which is strictly valid as long as $|\alpha| \ll 1$.

On the other hand, the resurgent structure of the series gets crucially modified due to the fact that the expansion coefficients $a_g$ have now alternating sign. Indeed, upon performing the Borel sum of \eqref{eq:defphi(alpha)D2D6} one finds
\begin{align}\label{eq:BorelsumD2D6}
    \hat{\varphi} (|\alpha|) = 4 \int_0^{\infty} \text{d}s\, e^{-s}\, \sum_{k, n=1}^{\infty} \sum_{g=2}^{\infty} (-1)^g\, k^{-2g} n^{2-2g}  \left( \frac{s |\alpha|}{4 \pi^2} \right)^{2g-1} = 4 \int_0^{\infty} \text{d}s\, \sum_{k, n=1}^{\infty} \frac{n}{k}\, \frac{e^{-s} \left( \frac{s |\alpha|}{4 \pi^2 k n} \right)^{3}}{1+\left( \frac{s |\alpha|}{4 \pi^2 k n} \right)}\, ,
\end{align}
which does not present poles along the real line and is thus Borel summable. In fact, using the following series expansion for $\coth x$
\begin{equation}\label{eq:seriescothx}
    \coth x -\frac1x - \frac{x}{3} = -\frac{2}{\pi^4}\sum_{k=1}^{\infty} \frac{1}{k^4}\, \frac{x^{3}}{1 + \frac{x^2}{k^2 \pi^2}}\, ,
\end{equation}
one may write \eqref{eq:BorelsumD2D6} as follows
\begin{equation}\label{eq:Borelsumcoth}
\begin{aligned}
    \hat{\varphi} (|\alpha|) = -\frac12 \sum_{n=1}^{\infty} \frac{(4\pi n)^2}{|\alpha|} \int_0^{\infty} \text{d}s\, e^{-\frac{4\pi n s}{|\alpha|}} \left( \coth s -\frac1s - \frac{s}{3}\right)\, ,
\end{aligned}
\end{equation}
Note that the absence of singularities along the integration domain is nothing but a reflection of the fact that the D2-D6 system should not exhibit any further non-perturbative contribution to $G(Y^0, \Upsilon)$, contrary to what happened in the previous case. 
	
\section{5d BPS Black Holes and Taub-NUT Geometries}\label{ap:5dspinningBH}

The purpose of this appendix is to understand in simple physical terms why the black hole solutions described in Section \ref{s:other4dBHs} exhibit an upper bound on the parameter $|\alpha|$, when evaluated at the attractor point. Recall from our discussion in Section \ref{sss:SmallBHsD2D6} (cf. in particular eq. \eqref{eq:alpha=r_5/rh}) that the aforementioned quantity determines the relative size of the M-theory circle compared to that of the BPS black hole, with the former measured at the horizon locus as the inverse of the D0 mass. In fact, this is the main reason why we can reproduce the 4d entropy from a purely five-dimensional perspective only for black holes having $p^0=0$, as these are the ones able to probe the parametric regime $r_5/r_h \to \infty$. We refer to Section \ref{sss:gluing5d} for details on this. 

\subsection{The Taub-NUT geometry}\label{ss:Taub-NUTgeneral}

Let us start by briefly reviewing the Taub-NUT solution \cite{Taub:1950ez,Newman:1963yy}, since it will play a major role in our subsequent analysis. This configuration may be regarded as a gravitational instanton \cite{Hawking:1976jb} with finite energy-momentum that solves the (euclidean version of) the Einstein field equations in $\mathbb{R}^4$. Its line element reads as
\begin{equation}\label{eq:TaubNUTgeometry}
    ds_{\rm TN}^2=  \frac14 \frac{\rho+R}{\rho-R}d\rho^2 + \frac{\rho-R}{\rho+R} R^2 \sigma_3^2 + \frac14 \left( \rho^2-R^2\right) \left( \sigma_1^2 + \sigma_2^2\right)\, ,
\end{equation}
where $\rho \geq R$ denotes the radial coordinate, and 
\begin{equation}
    \begin{aligned}\label{eq:leftinv1forms}
        \sigma_1 &= -\sin \psi d\theta +\cos \psi \sin \theta d\phi\, ,\\
        \sigma_2 &= \cos \psi d\theta +\sin \psi \sin \theta d\phi\, ,\\
        \sigma_3 &= d\psi +\cos \theta d\phi\, ,
    \end{aligned}
\end{equation}
%
\begin{figure}[t!]
	\begin{center}
		\begin{overpic}[scale=0.35,trim=70 120 70 100, clip]{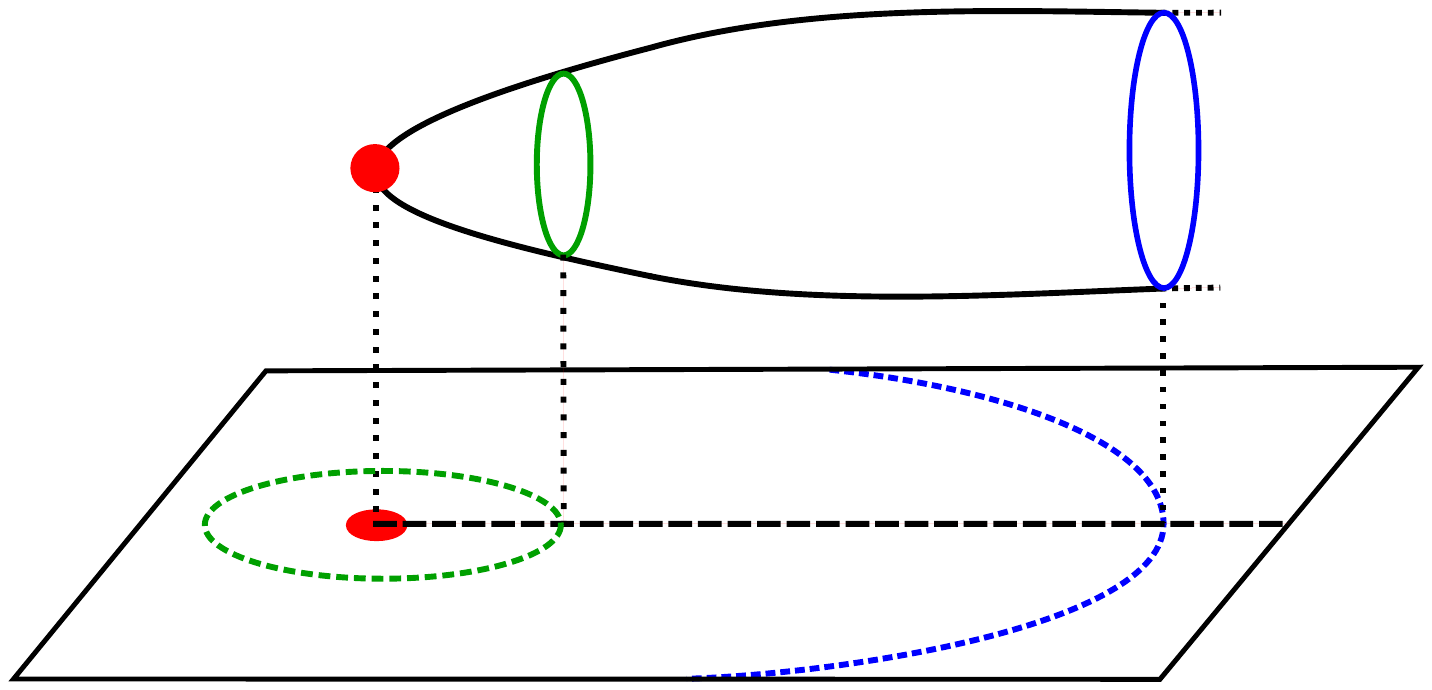}
        \put(88, 45){$\mathbf{S}^1$}
        \put(74, 40){$R$}
        \put(77, 18){$r$}
        \put(44, 40){$R_\epsilon$}
        \put(43, 18){$\epsilon$}
        \put(10, 9){$\mathbb{R}^3$}
		\end{overpic}
		\caption{\small Schematic depiction of the Taub-NUT geometry for a single Kaluza-Klein (KK) monopole. It can be regarded as a gravitational instanton interpolating between an asymptotic spacetime with local topology $\mathbb{R}^3\times \mathbf{S}^1$ and a smooth configuration at its core locally of the form $\mathbb{R}^4$. Asymptotically, the circle radius $R$ is much smaller than the radial coordinate $r \gg R$. Close to the core, the circle radius $R_\epsilon$ scales as the radial coordinate $R_\epsilon \sim \epsilon$ and the circle fibration of $\mathbf{S}^1$ on the $\mathbf{S}^2 \subset \mathbb{R}^3$ has the topological structure of an $\mathbf{S}^3$. For $N$ KK monopoles the sphere is replaced by the lens space $\mathbf{S}^3/\mathbb{Z}_N$.}
		\label{fig:TaubNUT}
	\end{center}
\end{figure}
are left-invariant\footnote{The right-invariant set may be obtained directly from \eqref{eq:leftinv1forms} upon interchanging $\psi \leftrightarrow \phi$.} 1-forms of $SU(2) \cong \mathbf{S}^3$ defined in terms of the Euler angles, namely $0 \leq \theta < \pi,\, 0 \leq \phi < 2\pi,\, 0 \leq \psi < 4\pi$. These covectors moreover satisfy the algebra relations
\begin{equation}
    d\sigma_i = \frac12 \epsilon_{ijk} d\sigma_j \wedge d\sigma_k\, .
\end{equation}
Geometrically, the above four-dimensional manifold can be seen to smoothly interpolate between a (fibered) compactified space at infinity of local topology $\mathbb{R}^3\times \mathbf{S}^1$
\begin{equation}\label{eq:farawaygeometry}
    ds_{\rm TN}^2\, \sim\,  \frac14 \left( d\rho^2 + \rho^2 d\Omega_2^2\right) + R^2 \left( d\psi +\cos \theta d\phi\right)^2\, ,\qquad \text{when}\ \ \rho \gg R\, ,
\end{equation}
with the circle having an asymptotic radius of size given by $R$, and a \emph{non-singular} $\mathbb{R}^4$ near the tip $\rho=R$ (see Figure \ref{fig:TaubNUT}), namely
\begin{equation}\label{eq:nearhorizongeometry}
    ds_{\rm TN}^2\, \sim\,  \frac{R}{2 \varepsilon}\left( d\varepsilon^2 + 4\varepsilon^2 d\Omega_3^2 \right)\, ,\qquad \text{for}\ \ \rho = R + \varepsilon\, ,
\end{equation}
where $d\Omega_3^2 = \frac14 (d\Omega_2^2 + \left( d\psi +\cos \theta d\phi\right)^2)$ denotes the metric of the unit $\mathbf{S}^3$, written here in a manifestly left-invariant way. Note that the regularity of the near-horizon geometry is readily seen by performing the change of radial coordinate $u= \sqrt{2R\varepsilon}$, which brings the metric \eqref{eq:nearhorizongeometry} to the standard form $ds_{\mathbb{R}^4}^2 = du^2 + u^2 d\Omega_3^2$.

\medskip

The physical significance of this solution lies on the fact that, when embedded in theories living in $d > 4$ spacetimes with dynamical gravity, it describes a Kaluza-Klein monopole \cite{Sorkin:1983ns,Gross:1983hb}, where the compactification circle becomes parametrized by the angular coordinate $\psi$. Interestingly, it turns out that one can easily generalize \eqref{eq:TaubNUTgeometry} to the case where there are $N$ distinct KK monopoles at various locations, which may or may not coincide \cite{Hawking:1976jb,Sen:1997js}. This is what we review next. 

\medskip

Let us first rewrite the metric \eqref{eq:TaubNUTgeometry} using isotropic coordinates
\begin{equation}\label{eq:TaubNUTgeometryisotropic}
    ds_{\rm TN}^2=  \frac14 \left(1+ \frac{2R}{u} \right) \left(du^2 + u^2 d\Omega_2^2\right)+ R^2 \left(1+ \frac{2R}{u} \right)^{-1} \left(d\psi +\cos \theta d\phi \right)^2\, ,
\end{equation}
where we have defined $u =\rho-R \geq 0$. Introducing now $r = u/2$, the previous line element can be recast in the following convenient form
\begin{equation}\label{eq:HawkingTaubNUT}
    ds_{\rm TN}^2=  f(r) \left(\delta_{ij} dx^i dx^j\right)+ f(r)^{-1} (dx^4 + \omega_i dx^i)^2\, ,
\end{equation}
with
\begin{equation}\label{eq:TaubNUTfunctions}
    f(r)= 1+ \frac{R}{r}\, ,\qquad \omega= R \cos \theta d\phi\, ,
\end{equation}
and where $x^4 \sim x^4 + 4\pi R$. Notice that the above quantities satisfy the relation (in differential form notation)
\begin{equation}\label{eq:TaubNUTconstraint}
    df= \star_3 d \omega\, ,
\end{equation}
where the Hodge dual is taken with respect to the flat metric in $\mathbb{R}^3$. In fact, since $f(r)$ satisfies a Poisson equation with a $\delta$-like source, one might regard the 1-form $\omega$ precisely as the Kaluza-Klein photon, whose non-trivial (magnetic) background follows from eq. \eqref{eq:TaubNUTconstraint}. Incidentally, one can show that Einstein field equations are indeed satisfied by metrics of the form \eqref{eq:HawkingTaubNUT} if and only if the aforementioned two conditions are verified. Therefore, a straightforward generalization of the single KK monopole configuration is readily obtained by considering multi-centered solutions, with\footnote{The fact that one can a priori superimpose various Taub-NUT geometries can be traced back to a gravitational no-force condition \cite{Hawking:1976jb} in euclidean four-dimensional space, similarly to what happens with multi-centered Reissner-N\"ordstrom black holes \cite{Majumdar:1947eu,Papaetrou:1947ib,Hartle:1972ya} in 4d Minkowski.} 
\begin{equation}\label{eq:multicenteredTaubNUT}
    f(r)= 1+ \sum_{k=1}^N \frac{R}{|\mathbf{x}-\mathbf{x_k}|}\, ,\qquad \omega =\sum_{k=1}^N \omega_k\, ,
\end{equation}
where $\mathbf{x_1} \neq \mathbf{x_2} \neq \ldots \neq \mathbf{x_N}$ denote the locations of the Taub-NUT centers, and each $\omega_k$ in \eqref{eq:multicenteredTaubNUT} is defined analogously to the single KK monopole case.

\subsubsection*{$N$-coincident Kaluza-Klein monopoles}

Up to now we have discussed the global structure associated to having one or more \emph{distinct} KK monopoles. A natural question that arises then, is what happens if (some of) these sources coincide in spacetime. Following our discussion above, the solution we seek for can be described by the metric \eqref{eq:HawkingTaubNUT} with the functions
\begin{equation}\label{eq:NKKmonopolesfunctions}
    f(r)= 1+ \frac{NR}{r}\, ,\qquad \omega= N R \cos \theta d\phi\, .
\end{equation}
Notice that the far-away region looks again exactly like the one associated to the Taub-NUT configuration, with a local topology of the form $\mathbb{R}^3\times \mathbf{S}^1$, and where the circle is non-trivially fibered over the $\mathbf{S}^2 \hookrightarrow \mathbb{R}^3$ at infinity due to the $d\psi d\phi$ cross term in \eqref{eq:farawaygeometry}. On the other hand, when focusing on the near-horizon geometry, the corresponding line element reduces to
\begin{equation}\label{eq:nearhorizongeometryKKmonopoles}
    ds^2\, \sim\,  \frac{NR}{r}\left[ dr^2 + r^2 d\Omega_2^2 + r^2 \left( d\psi/N +\cos \theta d\phi \right)^2\right]\, ,\qquad \text{for}\ \ r \gg R\, .
\end{equation}
Therefore, for $N>1$ ---and upon redefining $\psi \to \psi N$--- we find that the metric becomes equivalent to that of flat $\mathbb{R}^4$. Topologically, however, since the new angular direction $\psi$ has periodicity equal to $4\pi/N$, one obtains instead the orbifold $\mathbb{R}^4/\mathbb{Z}_N$ \cite{Asano:2000mx}. Still, the solution behaves as an asymptotically locally euclidean (ALE) space \cite{Eguchi:1978gw}, and in particular it solves the gravitational equations of motion.

\subsection{5d $\mathcal{N}=1$ black holes with KK monopole charge}\label{ss:5dspinningBHs&Taub-NUT}

With this, we are now ready to revisit the 4d black holes examined in detail in Section \ref{s:other4dBHs}. As noted in the main text, these configurations can equivalently be understood, when seen from the perspective of M-theory compactified on a Calabi--Yau threefold, as a 5d BPS spinning black hole located at the center of the Taub-NUT geometry. Hence, within the underlying 5d $\mathcal{N}=1$ supergravity theory, these solutions are described by the following line element \cite{Gauntlett:2002nw}
\begin{equation}\label{eq:5dBHmetric}
    ds^2= - f(r)^{-2} \left( dt+ \frac{J_{\rm L}\, G_5}{4\pi\, p^0 R^2}\, \mathsf{a} \right)^2 + f(r) ds_{\rm TN}^2\, ,
\end{equation}
with
\begin{equation}
    f(r)= 1+ \frac{|Z_{\rm 5d}|\, G_5^{2/3}}{(4\pi)^{2/3}\, R\, r}\, , \qquad \mathsf{a} =  \left( 1+ \frac{p^0 R}{r}\right)\left( d\psi + p^0 \cos{\theta} d\phi\right) -d\psi\, ,
\end{equation}
where $\theta, \phi, \varphi$ are defined as in \eqref{eq:leftinv1forms}. In addition, the Taub-NUT metric reads as
\begin{equation}\label{eq:TaubNUT5dBH}
    ds_{\rm TN}^2=  \left( 1+ \frac{p^0 R}{r}\right) \left( dr^2 + r^2 d\Omega_2^2\right) + R^2 \left( 1+ \frac{p^0 R}{r}\right)^{-1} \left( d\psi + p^0 \cos{\theta} d\phi\right)^2\, .
\end{equation}
Note that this precisely corresponds to the $N$-coincident Kaluza-Klein monopole characterized by the functions \eqref{eq:NKKmonopolesfunctions}, with the identification $N=p^0$. The 5d black hole moreover exhibits a non-trivial graviphoton background \cite{Gauntlett:2002nw}\footnote{To be precise, the 5d graviphoton gauge field arises as the linear combination $\mathsf{V}=L_a A^a = \frac12 \mathcal{K}_{abc} L^b L^c A^a$. The latter appears in the supersymmetry variation of the spin-$\frac32$ gravitino, and its field strength is moreover defined as $\mathsf{T}= L_a F^a$ \cite{Larsen:2006xm}.}
\begin{equation}\label{eq:graviphotonfieldstrength}
    \mathsf{T} = \frac{\sqrt{3}}{2}\, d \wedge \left( f(r)^{-1} \left( dt+ \frac{J_{\rm L}\, G_5}{4\pi\, p^0\, R^2}\, \mathsf{a} \right)\right)\, ,
\end{equation}
with a generically non-vanishing $U(1)_L$ angular momentum $J_{\rm L}$. The former is sourced by the central charge $Z_{\rm 5d} = q^{\rm 5d}_a L^a$, which can be shown to depend solely on the black hole charges due to the attractor mechanism in 5d (minimal) supergravity, see e.g., \cite{Larsen:2006xm} and references therein. Importantly, the solution described by eqs. \eqref{eq:5dBHmetric}-\eqref{eq:TaubNUT5dBH} manifestly preserves a $U(1)_L \times SU(2)_R$ subgroup of isometries in five dimensions, thus constituting a BPS configuration \cite{Gauntlett:1998fz} of the kind we are interested in here. Geometrically, it interpolates between an asymptotically flat spacetime of topology $\mathbb{R}^{1,3}\times \mathbf{S}^1$ --- with the circle at infinity having total length equal to $4\pi R$, and a supersymmetric spinning black hole located at the center of an $\mathbb{R}^{1,4}/\mathbb{Z}_{p^0}$ orbifold. The latter arises close to the horizon locus, i.e., for $r \gtrsim 0$.

Furthermore, from \eqref{eq:5dBHmetric} it is easy to determine the various relevant thermodynamic quantities of the black object. Following \cite{Gaiotto:2005gf}, one may perform the change of coordinates $\varrho^2=r R$ and subsequently take the limit $R \to \infty$, which yields the approximate expression
\begin{equation}\label{eq:5dBHnearhorizongeometry}
\begin{aligned}
    ds^2\, \sim\,& - \left( 1+ \frac{|Z_{\rm 5d}|\, G_5^{2/3}}{(4\pi)^{2/3}\, \varrho^2} \right)^{-2} \left( dt+ \frac{J_{\rm L} G_5}{4\pi\, \varrho^2} \left( d\psi + p^0 \cos{\theta} d\phi\right)\right)^2\\
    &+\, 4 p^0 \left( 1+ \frac{|Z_{\rm 5d}|\, G_5^{2/3}}{(4\pi)^{2/3}\,\varrho^2} \right) \left( d\varrho^2 +\varrho^2 d\tilde{\Omega}_3^2\right)\, ,
    \end{aligned}
\end{equation}
with $d\tilde{\Omega}_3^2 = \frac14 ( d\Omega_2^2 + \left( d\psi/p^0 +\cos \theta d\phi\right)^2)$ denoting the metric on $\mathbf{S}^3/\mathbb{Z}_{p^0}$. Hence, for the particular case of black holes having $J_{\rm L}=0$, we deduce that the horizon radius behaves as $r_{h,\, 5d}^2 = (2/\pi)^{2/3} p^0 |Z_{\rm 5d}| G_5^{2/3}$, thus providing an entropy of the form
\begin{equation}\label{eq:5dBHentropy}
    \mathcal{S}_{\rm BH} = \frac{A_{\rm hor}}{4 G_5} = \frac{2 \pi^2 r_{h,\, 5d}^3}{4 p^0 G_5} = \pi \sqrt{p^0 |Z_{\rm 5d}|^3}\, ,
\end{equation}
where the additional $1/p^0$ factor comes from the reduced volume of $\mathbf{S}^3/\mathbb{Z}_{p^0}$ with respect to that of the round 3-sphere. Notice that, for large central charges, the BPS solution \eqref{eq:5dBHmetric} should be more accurately regarded as a four-dimensional black hole, since its radius is larger than the asymptotic value of the $\mathbf{S}^1$ component. Conversely, if $r_{h,\, 5d}$ is small compared to the circle at infinity, it must be seen as a 5d black hole at the center of a Taub-NUT geometry. Therefore, assuming we are in the second scenario, one may ask about the relative size of the black hole horizon and the compact $\mathbf{S}^1$, when evaluated at $r=0$. The latter can be easily determined from eqs. \eqref{eq:5dBHmetric} and \eqref{eq:TaubNUT5dBH} to be $\rho_5^2 = (2/\pi)^{2/3}\, |Z_{\rm 5d}| G_5^{2/3}/p^0$, which implies that
\begin{equation}\label{eq:alpharesult}
    \frac{\rho_5}{r_{h,\, 5d}} = \frac{1}{p^0} \, .
\end{equation}
Note that we have chosen to assign a different label to the circle radius than the one used in Sections \ref{s:BHs&EFTtransitions} and \ref{s:other4dBHs}, where it was determined through the D0-brane mass and denoted as $r_5$. The rationale behind this choice will become apparent in what follows.

\subsubsection{The 4d Perspective}\label{sss:4dperspective}

To connect with the four-dimensional analysis based on the attractor mechanism (cf. Section \ref{sss:2derivativeD2D6}), we must dimensionally reduce the 5d solution described above and employ the familiar Type IIA/M-theory dictionary. First, let us compute what is commonly referred to as the M-theory circle radius, $r_{5}$. This quantity, in turn, determines the physical mass of D0-brane states in Type IIA string theory. Hence, at any given point in spacetime, the latter adopts the following form
\begin{equation}
    r_{5} = e^\phi \ell_s \,,
\end{equation}
where $\phi$ is the 10d dilaton. Notice that, when evaluated in the D2-D6 black hole background, the M-theory radius presents a non-trivial spatial profile, which asymptotically yields $R_{5} = g_s \ell_s$, where $g_s$ denotes the string coupling. If we embed the metric (\ref{eq:5dBHmetric}) in 11d supergravity, we realize that in terms of the eleven-dimensional Planck length $\ell_{11} = g_s^{1/3} \ell_s$, $R_5$ is expressed as $g_s^{2/3} \ell_{11}$ \cite{Witten:1995ex}. By dimensionally reducing the non-rotating black hole metric along the direction $\bar{\psi} = \psi/2$, we obtain the four-dimensional Einstein frame line element
\begin{equation}\label{eq:metricEinTaubnut}
    ds^2_{4d, \, E} = - h(r)^{-1} dt^2 + h(r) (dr^2 + r^2 d\Omega_2^2) \,,
\end{equation}
with 
\begin{equation}
    h(r) = f(r)^{3/2} g(r)^{1/2} \,, \qquad  f(r)= 1+ \frac{|Z_{\rm 5d}|\, G_5^{2/3}}{(4\pi)^{2/3}\, R\, r} \,, \qquad g(r) = 1 + \frac{p_0 R}{r} \,.
\end{equation}
Similarly, the Kaluza--Klein scalar can be readily determined to be
\begin{equation}
    \varphi (r) = 2 R (f/g)^{1/2}\,.
\end{equation}
The latter, if measured in units of $\ell_{11}$, is related to the dilaton field by ${\varphi^{3/2} = e^{\phi} \ell_{11}^{3/2}}$ \cite{Ortin:2015hya}. Consequently, at the horizon, it takes the explicit value
\begin{equation}
    \left(\frac{\varphi}{\ell_{11}}\right)^{3/2} \bigg\rvert_{\text{hor}} =\, g_{\bar{\psi}\bar{\psi}}^{3/4} \bigg\rvert_{r = 0} =\,  \frac{ |Z_{5d}|^{3/4} G_5^{1/2}}{p_0^{3/4} \left(\ell_{11}\right)^{3/2}}\frac{4^{3/4}}{ (4\pi)^{1/2}}\, .
\end{equation}
On the other hand, at asymptotic infinity, we obtain instead
\begin{equation}
    \left(\frac{\varphi}{\ell_{11}}\right)^{3/2} \to\, \left(\frac{2R}{\ell_{11}}\right)^{3/2} = g_s\,,
\end{equation}
which fixes, in turn, $2R = R_5$. Furthermore, from (\ref{eq:metricEinTaubnut}) one may be easily compute the four-dimensional black hole entropy 
\begin{equation}
    \mathcal{S}_{\rm BH} = \frac{4 \pi r_{h, \,4d}^2}{4 G_4} = \pi \sqrt{p_0 |Z_{5d}|^3} \,,
\end{equation}
where we used that $G_5 = G_4 (2\pi R_5)$. Notice that this agrees with \eqref{eq:5dBHentropy}, as it should. Therefore, we deduce that the near-horizon behavior of $h r^2$ must take the form
\begin{equation} \label{eq:apradii}
    h r^2 \sim  \frac{G_5}{4\pi R}\sqrt{p_0 |Z_{5d}|^3} \equiv \frac{|\mathscr{Z}|^2 \kappa_4^2}{8\pi} \,,
\end{equation}
which prompts us to identify the 5d and 4d central charges as follows (recall that $8 \pi G_4 = \kappa_4^2$)\footnote{This can also be deduced directly from the map relating the 4d and 5d gauge charges, cf. footnote \ref{fnote:consistencycond5d}.}
\begin{equation}
    p_0^{1/4} |Z_{5d}|^{3/4} = |\mathscr{Z}|\, .
\end{equation}
All in all, we find that the 4d black hole radius can be conveniently written as
\begin{equation}
     r_{h,\,4d} = \frac{\kappa_4}{\sqrt{8\pi}}|\mathscr{Z}| \, ,
\end{equation}
whereas the M-theory circle radius is given instead by
\begin{equation}
    r_5 = \left(\frac{\varphi}{\ell_{11}}\right)^{3/2} \ell_s = \frac{\ell_s G_5^{1/2}}{\left(\ell_{11}\right)^{3/2}} \frac{|\mathscr{Z}|}{p^0}\frac{2^{3/2}}{ (4\pi)^{1/2}} = \frac{\ell_s (2R)^{1/2}}{\left(\ell_{11}\right)^{3/2}} \frac{|\mathscr{Z}|}{|Y^0|} \frac{\kappa_4}{\sqrt{8\pi}} \,.
\end{equation} 
However, using the fact that $R_5 = 2 R$ as well as $\ell_s^{2/3} R_5^{1/3} = \ell_{11}$, we end up with the following expression
\begin{equation}
    r_5 = \frac{|\mathscr{Z}|}{|Y^0|}\frac{\kappa_4}{\sqrt{8\pi}} \,,
\end{equation}
such that we recover precisely the ratio of scales obtained via the attractor mechanism, namely
\begin{equation}\label{eq:alpharesult4d}
    \frac{r_5}{r_{h,\,4d}} = \frac{2}{p^0} =  |\alpha| \,.
\end{equation}

\subsubsection{Connecting the 4d and 5d pictures}

Before we conclude this appendix, we would like to emphasize several interesting features that emerge from the discussion above. First, one might wonder why we have been careful to distinguish between the five-dimensional ratio $\rho_5/r_{h,\, 5d}$ (cf. \eqref{eq:alpharesult}) and the analogous one (i.e., $r_5/r_{h,\, 4d}$) computed in the 4d supergravity theory, see eq. \eqref{eq:alpharesult4d}. The key point here is that, since the 4d and 5d entropies match, the corresponding black hole radii must necessarily differ. More precisely, they satisfy
\begin{equation}
    \frac{4 \pi r_{h,\,4d}^2}{4 (G_5/4\pi R)} =  \frac{2 \pi^2 r_{h,\, 5d}^3}{4 p^0 G_5}\,,
\end{equation}
which implies
\begin{equation} \label{eq:comb1}
8 p^0 R\, r^2_{h,\,4d} = r^3_{h,\,5d} \,.
\end{equation}
Therefore, it cannot be that the ratio between the 5d and 4d radii with the M-theory circle $r_5$ give both $\alpha$ at the same time. Yet, in eqs. \eqref{eq:alpharesult} and \eqref{eq:alpharesult4d} we found two closely related expressions differing only in a factor of 2. To properly understand this, we need to clarify what precisely is the quantity denoted by $\rho_5$ that was introduced after \eqref{eq:5dBHentropy}. By looking at the metric \eqref{eq:TaubNUT5dBH}, we see that it behaves as
\begin{equation}
    \rho_5 (r) = \varphi (r) \ell_{11} = e^{2/3\phi}\ell_{11} \,,
\end{equation}
so it simply corresponds to the vacuum expectation value of the Kaluza-Klein scalar. Notice that, asymptotically, $\rho_5 = R_5 = r_5$. However, in general we have that $\rho_5 \ne r_5$, and in fact the actual relation between them is 
\begin{equation} \label{eq:comb2}
    \rho_5 = (r_5/\ell_s)^{2/3}\ell_{11} \, .
\end{equation}
Finally, by combining equations (\ref{eq:comb1}) and (\ref{eq:comb2}), we recover our previous result
\begin{equation}
    \frac{\rho_5^{2}}{r^2_{h,\,5d}}\bigg\rvert_{\text{hor}} = \frac{\alpha^{4/3}}{(4 p_0)^{2/3}} \frac{\ell_{11}^2}{R_5^{2/3}\ell_s^{4/3}} = \frac{1}{p_0^2} \, ,
\end{equation}
where $\rho_5$ has the explicit form (at the horizon)
\begin{equation}
    \rho_5^2 \to \frac{4 |Z_{5d}| G_5^{2/3}}{p_0 (4\pi)^{2/3}} \,.
\end{equation}
The conclusion is that $\alpha$ can be interpreted as either \emph{i)} the ratio between the 4d black hole radius and that of the M-theory circle $r_5$ (which is defined here as the inverse D0-brane mass, cf. eq. \eqref{eq:alpha=r_5/rh}), or rather as \emph{ii)} twice the quotient between the 5d black hole radius and the size of the Taub-NUT 1-cycle $\rho_5$, the latter being measured by the 4d KK scalar. It is important to note that, in the familiar flat 10d background where the duality between M-theory and Type IIA string theory is typically invoked, these two quantities ---namely $r_5$ and $\rho_5$--- are, in fact, identical. However, due to the different topologies of the near-horizon geometry when viewed from the 4d or 5d perspectives (i.e., AdS$_2 \times \mathbf{S}^2$ and AdS$_2 \times \mathbf{S}^3/\mathbb{Z}_{p^0}$, respectively), these quantities crucially deviate from each other in a way that is consistent with both the Type IIA/M-theory duality and the single-valuedness of the black hole entropy.

\bibliography{ref.bib}

\providecommand{\href}[2]{#2}\begingroup\raggedright\begin{thebibliography}{100}

\bibitem{Bekenstein:1972tm}
J.~D. Bekenstein, {\it {Black holes and the second law}},  {\em Lett. Nuovo
  Cim.} {\bf 4} (1972) 737--740.

\bibitem{Hawking:1975vcx}
S.~W. Hawking, {\it {Particle Creation by Black Holes}},  {\em Commun. Math.
  Phys.} {\bf 43} (1975) 199--220. [Erratum: Commun.Math.Phys. 46, 206 (1976)].

\bibitem{vandeHeisteeg:2022btw}
D.~van~de Heisteeg, C.~Vafa, M.~Wiesner, and D.~H. Wu, {\it {Moduli-dependent
  species scale}},  {\em Beijing J. Pure Appl. Math.} {\bf 1} (2024), no.~1
  1--41, [\href{http://arxiv.org/abs/2212.06841}{{\tt arXiv:2212.06841}}].

\bibitem{Cribiori:2022nke}
N.~Cribiori, D.~L\"ust, and G.~Staudt, {\it {Black hole entropy and
  moduli-dependent species scale}},  {\em Phys. Lett. B} {\bf 844} (2023)
  138113, [\href{http://arxiv.org/abs/2212.10286}{{\tt arXiv:2212.10286}}].

\bibitem{vandeHeisteeg:2023ubh}
D.~van~de Heisteeg, C.~Vafa, and M.~Wiesner, {\it {Bounds on Species Scale and
  the Distance Conjecture}},  {\em Fortsch. Phys.} {\bf 71} (2023), no.~10-11
  2300143, [\href{http://arxiv.org/abs/2303.13580}{{\tt arXiv:2303.13580}}].

\bibitem{vandeHeisteeg:2023dlw}
D.~van~de Heisteeg, C.~Vafa, M.~Wiesner, and D.~H. Wu, {\it {Species scale in
  diverse dimensions}},  {\em JHEP} {\bf 05} (2024) 112,
  [\href{http://arxiv.org/abs/2310.07213}{{\tt arXiv:2310.07213}}].

\bibitem{Castellano:2023aum}
A.~Castellano, A.~Herr\'aez, and L.~E. Ib\'a\~nez, {\it {On the species scale,
  modular invariance and the gravitational EFT expansion}},  {\em JHEP} {\bf
  12} (2024) 019, [\href{http://arxiv.org/abs/2310.07708}{{\tt
  arXiv:2310.07708}}].

\bibitem{Dvali:2007hz}
G.~Dvali, {\it {Black Holes and Large N Species Solution to the Hierarchy
  Problem}},  {\em Fortsch. Phys.} {\bf 58} (2010) 528--536,
  [\href{http://arxiv.org/abs/0706.2050}{{\tt arXiv:0706.2050}}].

\bibitem{Dvali:2009ks}
G.~Dvali and D.~Lust, {\it {Evaporation of Microscopic Black Holes in String
  Theory and the Bound on Species}},  {\em Fortsch. Phys.} {\bf 58} (2010)
  505--527, [\href{http://arxiv.org/abs/0912.3167}{{\tt arXiv:0912.3167}}].

\bibitem{Dvali:2010vm}
G.~Dvali and C.~Gomez, {\it {Species and Strings}},
  \href{http://arxiv.org/abs/1004.3744}{{\tt arXiv:1004.3744}}.

\bibitem{Dvali:2012uq}
G.~Dvali, C.~Gomez, and D.~Lust, {\it {Black Hole Quantum Mechanics in the
  Presence of Species}},  {\em Fortsch. Phys.} {\bf 61} (2013) 768--778,
  [\href{http://arxiv.org/abs/1206.2365}{{\tt arXiv:1206.2365}}].

\bibitem{Castellano:2024bna}
A.~Castellano, {\em {The Quantum Gravity Scale and the Swampland}}.
\newblock PhD thesis, U. Autonoma, Madrid (main), 2024.
\newblock \href{http://arxiv.org/abs/2409.10003}{{\tt arXiv:2409.10003}}.

\bibitem{Aoufia:2024awo}
C.~Aoufia, I.~Basile, and G.~Leone, {\it {Species scale, worldsheet CFTs and
  emergent geometry}},  {\em JHEP} {\bf 12} (2024) 111,
  [\href{http://arxiv.org/abs/2405.03683}{{\tt arXiv:2405.03683}}].

\bibitem{Calderon-Infante:2025ldq}
J.~Calder\'on-Infante, A.~Castellano, and A.~Herr\'aez, {\it {The Double EFT
  Expansion in Quantum Gravity}},  \href{http://arxiv.org/abs/2501.14880}{{\tt
  arXiv:2501.14880}}.

\bibitem{Castellano:2022bvr}
A.~Castellano, A.~Herr\'aez, and L.~E. Ib\'a\~nez, {\it {The emergence proposal
  in quantum gravity and the species scale}},  {\em JHEP} {\bf 06} (2023) 047,
  [\href{http://arxiv.org/abs/2212.03908}{{\tt arXiv:2212.03908}}].

\bibitem{Gregory:1993vy}
R.~Gregory and R.~Laflamme, {\it {Black strings and p-branes are unstable}},
  {\em Phys. Rev. Lett.} {\bf 70} (1993) 2837--2840,
  [\href{http://arxiv.org/abs/hep-th/9301052}{{\tt hep-th/9301052}}].

\bibitem{Gregory:1994bj}
R.~Gregory and R.~Laflamme, {\it {The Instability of charged black strings and
  p-branes}},  {\em Nucl. Phys. B} {\bf 428} (1994) 399--434,
  [\href{http://arxiv.org/abs/hep-th/9404071}{{\tt hep-th/9404071}}].

\bibitem{Horowitz:1996nw}
G.~T. Horowitz and J.~Polchinski, {\it {A Correspondence principle for black
  holes and strings}},  {\em Phys. Rev. D} {\bf 55} (1997) 6189--6197,
  [\href{http://arxiv.org/abs/hep-th/9612146}{{\tt hep-th/9612146}}].

\bibitem{Horowitz:1997jc}
G.~T. Horowitz and J.~Polchinski, {\it {Selfgravitating fundamental strings}},
  {\em Phys. Rev. D} {\bf 57} (1998) 2557--2563,
  [\href{http://arxiv.org/abs/hep-th/9707170}{{\tt hep-th/9707170}}].

\bibitem{Brustein:2021ifl}
R.~Brustein and Y.~Zigdon, {\it {Effective field theory for closed strings near
  the Hagedorn temperature}},  {\em JHEP} {\bf 04} (2021) 107,
  [\href{http://arxiv.org/abs/2101.07836}{{\tt arXiv:2101.07836}}].

\bibitem{Chen:2021emg}
Y.~Chen and J.~Maldacena, {\it {String scale black holes at large D}},  {\em
  JHEP} {\bf 01} (2022) 095, [\href{http://arxiv.org/abs/2106.02169}{{\tt
  arXiv:2106.02169}}].

\bibitem{Chen:2021dsw}
Y.~Chen, J.~Maldacena, and E.~Witten, {\it {On the black hole/string
  transition}},  {\em JHEP} {\bf 01} (2023) 103,
  [\href{http://arxiv.org/abs/2109.08563}{{\tt arXiv:2109.08563}}].

\bibitem{Urbach:2022xzw}
E.~Y. Urbach, {\it {String stars in anti de Sitter space}},  {\em JHEP} {\bf
  04} (2022) 072, [\href{http://arxiv.org/abs/2202.06966}{{\tt
  arXiv:2202.06966}}].

\bibitem{Balthazar:2022szl}
B.~Balthazar, J.~Chu, and D.~Kutasov, {\it {Winding Tachyons and Stringy Black
  Holes}},  \href{http://arxiv.org/abs/2204.00012}{{\tt arXiv:2204.00012}}.

\bibitem{Balthazar:2022hno}
B.~Balthazar, J.~Chu, and D.~Kutasov, {\it {On small black holes in string
  theory}},  {\em JHEP} {\bf 03} (2024) 116,
  [\href{http://arxiv.org/abs/2210.12033}{{\tt arXiv:2210.12033}}].

\bibitem{Ceplak:2023afb}
N.~\v{C}eplak, R.~Emparan, A.~Puhm, and M.~Toma\v{s}evi\'c, {\it {The
  correspondence between rotating black holes and fundamental strings}},  {\em
  JHEP} {\bf 11} (2023) 226, [\href{http://arxiv.org/abs/2307.03573}{{\tt
  arXiv:2307.03573}}].

\bibitem{Herraez:2024kux}
A.~Herr\'aez, D.~L\"ust, J.~Masias, and M.~Scalisi, {\it {On the Origin of
  Species Thermodynamics and the Black Hole - Tower Correspondence}},
  \href{http://arxiv.org/abs/2406.17851}{{\tt arXiv:2406.17851}}.

\bibitem{Albertini:2024hwi}
E.~Albertini, D.~Platt, and T.~Wiseman, {\it {Towards a uniqueness theorem for
  static black holes in Kaluza-Klein theory with small circle size}},
  \href{http://arxiv.org/abs/2410.20967}{{\tt arXiv:2410.20967}}.

\bibitem{Chu:2024ggi}
J.~Chu, {\it {From Black Strings to Fundamental Strings: Non-uniformity and
  Phase Transitions}},  \href{http://arxiv.org/abs/2410.23597}{{\tt
  arXiv:2410.23597}}.

\bibitem{Emparan:2024mbp}
R.~Emparan, M.~Sanchez-Garitaonandia, and M.~Toma\v{s}evi\'c, {\it {String
  Theory in a Pinch: Resolving the Gregory-Laflamme Singularity}},
  \href{http://arxiv.org/abs/2411.14998}{{\tt arXiv:2411.14998}}.

\bibitem{Ceplak:2024dxm}
N.~\v{C}eplak, R.~Emparan, A.~Puhm, and M.~Toma\v{s}evi\'c, {\it {Size and
  Shape of Rotating Strings and the Correspondence to Black Holes}},
  \href{http://arxiv.org/abs/2411.18690}{{\tt arXiv:2411.18690}}.

\bibitem{Bedroya:2024igb}
A.~Bedroya and D.~Wu, {\it {String stars in $d\geq 7$}},
  \href{http://arxiv.org/abs/2412.19888}{{\tt arXiv:2412.19888}}.

\bibitem{Chu:2025fko}
J.~Chu, {\it {Phases of String Stars in the Presence of a Spatial Circle}},
  \href{http://arxiv.org/abs/2501.03312}{{\tt arXiv:2501.03312}}.

\bibitem{Bogomolny:1975de}
E.~B. Bogomolny, {\it {Stability of Classical Solutions}},  {\em Sov. J. Nucl.
  Phys.} {\bf 24} (1976) 449.

\bibitem{Prasad:1975kr}
M.~K. Prasad and C.~M. Sommerfield, {\it {An Exact Classical Solution for the
  't Hooft Monopole and the Julia-Zee Dyon}},  {\em Phys. Rev. Lett.} {\bf 35}
  (1975) 760--762.

\bibitem{LopesCardoso:1998tkj}
G.~Lopes~Cardoso, B.~de~Wit, and T.~Mohaupt, {\it {Corrections to macroscopic
  supersymmetric black hole entropy}},  {\em Phys. Lett. B} {\bf 451} (1999)
  309--316, [\href{http://arxiv.org/abs/hep-th/9812082}{{\tt hep-th/9812082}}].

\bibitem{LopesCardoso:1999cv}
G.~Lopes~Cardoso, B.~de~Wit, and T.~Mohaupt, {\it {Deviations from the area law
  for supersymmetric black holes}},  {\em Fortsch. Phys.} {\bf 48} (2000)
  49--64, [\href{http://arxiv.org/abs/hep-th/9904005}{{\tt hep-th/9904005}}].

\bibitem{LopesCardoso:1999fsj}
G.~Lopes~Cardoso, B.~de~Wit, and T.~Mohaupt, {\it {Macroscopic entropy formulae
  and nonholomorphic corrections for supersymmetric black holes}},  {\em Nucl.
  Phys. B} {\bf 567} (2000) 87--110,
  [\href{http://arxiv.org/abs/hep-th/9906094}{{\tt hep-th/9906094}}].

\bibitem{LopesCardoso:1999xn}
G.~Lopes~Cardoso, B.~de~Wit, and T.~Mohaupt, {\it {Area law corrections from
  state counting and supergravity}},  {\em Class. Quant. Grav.} {\bf 17} (2000)
  1007--1015, [\href{http://arxiv.org/abs/hep-th/9910179}{{\tt
  hep-th/9910179}}].

\bibitem{Mohaupt:2000mj}
T.~Mohaupt, {\it {Black hole entropy, special geometry and strings}},  {\em
  Fortsch. Phys.} {\bf 49} (2001) 3--161,
  [\href{http://arxiv.org/abs/hep-th/0007195}{{\tt hep-th/0007195}}].

\bibitem{Ooguri:2004zv}
H.~Ooguri, A.~Strominger, and C.~Vafa, {\it {Black hole attractors and the
  topological string}},  {\em Phys. Rev. D} {\bf 70} (2004) 106007,
  [\href{http://arxiv.org/abs/hep-th/0405146}{{\tt hep-th/0405146}}].

\bibitem{Wald:1993nt}
R.~M. Wald, {\it {Black hole entropy is the Noether charge}},  {\em Phys. Rev.
  D} {\bf 48} (1993), no.~8 R3427--R3431,
  [\href{http://arxiv.org/abs/gr-qc/9307038}{{\tt gr-qc/9307038}}].

\bibitem{Iyer:1994ys}
V.~Iyer and R.~M. Wald, {\it {Some properties of Noether charge and a proposal
  for dynamical black hole entropy}},  {\em Phys. Rev. D} {\bf 50} (1994)
  846--864, [\href{http://arxiv.org/abs/gr-qc/9403028}{{\tt gr-qc/9403028}}].

\bibitem{Cribiori:2023ffn}
N.~Cribiori, D.~Lust, and C.~Montella, {\it {Species entropy and
  thermodynamics}},  {\em JHEP} {\bf 10} (2023) 059,
  [\href{http://arxiv.org/abs/2305.10489}{{\tt arXiv:2305.10489}}].

\bibitem{Calderon-Infante:2023uhz}
J.~Calder\'on-Infante, M.~Delgado, and A.~M. Uranga, {\it {Emergence of species
  scale black hole horizons}},  {\em JHEP} {\bf 01} (2024) 003,
  [\href{http://arxiv.org/abs/2310.04488}{{\tt arXiv:2310.04488}}].

\bibitem{Basile:2023blg}
I.~Basile, D.~L\"ust, and C.~Montella, {\it {Shedding black hole light on the
  emergent string conjecture}},  {\em JHEP} {\bf 07} (2024) 208,
  [\href{http://arxiv.org/abs/2311.12113}{{\tt arXiv:2311.12113}}].

\bibitem{Basile:2024dqq}
I.~Basile, N.~Cribiori, D.~Lust, and C.~Montella, {\it {Minimal black holes and
  species thermodynamics}},  {\em JHEP} {\bf 06} (2024) 127,
  [\href{http://arxiv.org/abs/2401.06851}{{\tt arXiv:2401.06851}}].

\bibitem{Bedroya:2024ubj}
A.~Bedroya, R.~K. Mishra, and M.~Wiesner, {\it {Density of States, Black Holes
  and the Emergent String Conjecture}},
  \href{http://arxiv.org/abs/2405.00083}{{\tt arXiv:2405.00083}}.

\bibitem{Calderon-Infante:2025pls}
J.~Calder\'on-Infante, M.~Delgado, Y.~Li, D.~Lust, and A.~M. Uranga, {\it
  {Classical Black Hole Probes of UV Scales}},
  \href{http://arxiv.org/abs/2502.03514}{{\tt arXiv:2502.03514}}.

\bibitem{Gopakumar:1998ii}
R.~Gopakumar and C.~Vafa, {\it {M theory and topological strings. 1.}},
  \href{http://arxiv.org/abs/hep-th/9809187}{{\tt hep-th/9809187}}.

\bibitem{Gopakumar:1998jq}
R.~Gopakumar and C.~Vafa, {\it {M theory and topological strings. 2.}},
  \href{http://arxiv.org/abs/hep-th/9812127}{{\tt hep-th/9812127}}.

\bibitem{Ferrara:1995ih}
S.~Ferrara, R.~Kallosh, and A.~Strominger, {\it {N=2 extremal black holes}},
  {\em Phys. Rev. D} {\bf 52} (1995) R5412--R5416,
  [\href{http://arxiv.org/abs/hep-th/9508072}{{\tt hep-th/9508072}}].

\bibitem{Strominger:1996kf}
A.~Strominger, {\it {Macroscopic entropy of N=2 extremal black holes}},  {\em
  Phys. Lett. B} {\bf 383} (1996) 39--43,
  [\href{http://arxiv.org/abs/hep-th/9602111}{{\tt hep-th/9602111}}].

\bibitem{Ferrara:1996dd}
S.~Ferrara and R.~Kallosh, {\it {Supersymmetry and attractors}},  {\em Phys.
  Rev. D} {\bf 54} (1996) 1514--1524,
  [\href{http://arxiv.org/abs/hep-th/9602136}{{\tt hep-th/9602136}}].

\bibitem{Ferrara:1996um}
S.~Ferrara and R.~Kallosh, {\it {Universality of supersymmetric attractors}},
  {\em Phys. Rev. D} {\bf 54} (1996) 1525--1534,
  [\href{http://arxiv.org/abs/hep-th/9603090}{{\tt hep-th/9603090}}].

\bibitem{Bodner:1990zm}
M.~Bodner, A.~C. Cadavid, and S.~Ferrara, {\it {(2,2) vacuum configurations for
  type IIA superstrings: N=2 supergravity Lagrangians and algebraic geometry}},
   {\em Class. Quant. Grav.} {\bf 8} (1991) 789--808.

\bibitem{deWit:1980lyi}
B.~de~Wit, J.~W. van Holten, and A.~Van~Proeyen, {\it {Structure of N=2
  Supergravity}},  {\em Nucl. Phys. B} {\bf 184} (1981) 77. [Erratum:
  Nucl.Phys.B 222, 516 (1983)].

\bibitem{deWit:1984wbb}
B.~de~Wit and A.~Van~Proeyen, {\it {Potentials and Symmetries of General Gauged
  N=2 Supergravity: Yang-Mills Models}},  {\em Nucl. Phys. B} {\bf 245} (1984)
  89--117.

\bibitem{deWit:1984rvr}
B.~de~Wit, P.~G. Lauwers, and A.~Van~Proeyen, {\it {Lagrangians of N=2
  Supergravity - Matter Systems}},  {\em Nucl. Phys. B} {\bf 255} (1985)
  569--608.

\bibitem{Cremmer:1984hj}
E.~Cremmer, C.~Kounnas, A.~Van~Proeyen, J.~P. Derendinger, S.~Ferrara,
  B.~de~Wit, and L.~Girardello, {\it {Vector Multiplets Coupled to N=2
  Supergravity: SuperHiggs Effect, Flat Potentials and Geometric Structure}},
  {\em Nucl. Phys. B} {\bf 250} (1985) 385--426.

\bibitem{Strominger:1990pd}
A.~Strominger, {\it {Special Geometry}},  {\em Commun. Math. Phys.} {\bf 133}
  (1990) 163--180.

\bibitem{Ceresole:1995ca}
A.~Ceresole, R.~D'Auria, and S.~Ferrara, {\it {The Symplectic structure of N=2
  supergravity and its central extension}},  {\em Nucl. Phys. B Proc. Suppl.}
  {\bf 46} (1996) 67--74, [\href{http://arxiv.org/abs/hep-th/9509160}{{\tt
  hep-th/9509160}}].

\bibitem{Craps:1997gp}
B.~Craps, F.~Roose, W.~Troost, and A.~Van~Proeyen, {\it {What is special Kahler
  geometry?}},  {\em Nucl. Phys. B} {\bf 503} (1997) 565--613,
  [\href{http://arxiv.org/abs/hep-th/9703082}{{\tt hep-th/9703082}}].

\bibitem{Craps:1997nv}
B.~Craps, F.~Roose, W.~Troost, and A.~Van~Proeyen, {\it {Special Kahler
  geometry: Does there exist a prepotential?}},  {\em NATO Sci. Ser. C} {\bf
  520} (1999) 449--454, [\href{http://arxiv.org/abs/hep-th/9712092}{{\tt
  hep-th/9712092}}].

\bibitem{Bershadsky:1993ta}
M.~Bershadsky, S.~Cecotti, H.~Ooguri, and C.~Vafa, {\it {Holomorphic anomalies
  in topological field theories}},  {\em Nucl. Phys. B} {\bf 405} (1993)
  279--304, [\href{http://arxiv.org/abs/hep-th/9302103}{{\tt hep-th/9302103}}].

\bibitem{Bershadsky:1993cx}
M.~Bershadsky, S.~Cecotti, H.~Ooguri, and C.~Vafa, {\it {Kodaira-Spencer theory
  of gravity and exact results for quantum string amplitudes}},  {\em Commun.
  Math. Phys.} {\bf 165} (1994) 311--428,
  [\href{http://arxiv.org/abs/hep-th/9309140}{{\tt hep-th/9309140}}].

\bibitem{Antoniadis:1993ze}
I.~Antoniadis, E.~Gava, K.~S. Narain, and T.~R. Taylor, {\it {Topological
  amplitudes in string theory}},  {\em Nucl. Phys. B} {\bf 413} (1994)
  162--184, [\href{http://arxiv.org/abs/hep-th/9307158}{{\tt hep-th/9307158}}].

\bibitem{Antoniadis:1995zn}
I.~Antoniadis, E.~Gava, K.~S. Narain, and T.~R. Taylor, {\it {N=2 type II
  heterotic duality and higher derivative F terms}},  {\em Nucl. Phys. B} {\bf
  455} (1995), no.~1-2 109--130,
  [\href{http://arxiv.org/abs/hep-th/9507115}{{\tt hep-th/9507115}}].

\bibitem{deWit:1979dzm}
B.~de~Wit, J.~W. van Holten, and A.~Van~Proeyen, {\it {Transformation Rules of
  N=2 Supergravity Multiplets}},  {\em Nucl. Phys. B} {\bf 167} (1980) 186.

\bibitem{Bergshoeff:1980is}
E.~Bergshoeff, M.~de~Roo, and B.~de~Wit, {\it {Extended Conformal
  Supergravity}},  {\em Nucl. Phys. B} {\bf 182} (1981) 173--204.

\bibitem{deRoo:1980mm}
M.~de~Roo, J.~W. van Holten, B.~de~Wit, and A.~Van~Proeyen, {\it {Chiral
  Superfields in $N=2$ Supergravity}},  {\em Nucl. Phys. B} {\bf 173} (1980)
  175--188.

\bibitem{Candelas:1990pi}
P.~Candelas and X.~de~la Ossa, {\it {Moduli Space of {Calabi-Yau} Manifolds}},
  {\em Nucl. Phys. B} {\bf 355} (1991) 455--481.

\bibitem{Bastianelli:2008cu}
F.~Bastianelli, J.~M. Davila, and C.~Schubert, {\it {Gravitational corrections
  to the Euler-Heisenberg Lagrangian}},  {\em JHEP} {\bf 03} (2009) 086,
  [\href{http://arxiv.org/abs/0812.4849}{{\tt arXiv:0812.4849}}].

\bibitem{Dunne:2004nc}
G.~V. Dunne, {\em {Heisenberg-Euler effective Lagrangians: Basics and
  extensions}}, pp.~445--522.
\newblock 6, 2004.
\newblock \href{http://arxiv.org/abs/hep-th/0406216}{{\tt hep-th/0406216}}.

\bibitem{Chadha:1977my}
S.~Chadha and P.~Olesen, {\it {On Borel Singularities in Quantum Field
  Theory}},  {\em Phys. Lett. B} {\bf 72} (1977) 87--90.

\bibitem{Hattab:2024ewk}
J.~Hattab and E.~Palti, {\it {Non-perturbative topological string theory on
  compact Calabi-Yau manifolds from M-theory}},  {\em JHEP} {\bf 04} (2025)
  017, [\href{http://arxiv.org/abs/2408.09255}{{\tt arXiv:2408.09255}}].

\bibitem{Hattab:2024ssg}
J.~Hattab and E.~Palti, {\it {Notes on integrating out M2 branes}},  {\em Eur.
  Phys. J. C} {\bf 85} (2025), no.~1 107,
  [\href{http://arxiv.org/abs/2410.15809}{{\tt arXiv:2410.15809}}].

\bibitem{Schwinger:1951nm}
J.~S. Schwinger, {\it {On gauge invariance and vacuum polarization}},  {\em
  Phys. Rev.} {\bf 82} (1951) 664--679.

\bibitem{Dedushenko:2014nya}
M.~Dedushenko and E.~Witten, {\it {Some Details On The Gopakumar-Vafa and
  Ooguri-Vafa Formulas}},  {\em Adv. Theor. Math. Phys.} {\bf 20} (2016)
  1--133, [\href{http://arxiv.org/abs/1411.7108}{{\tt arXiv:1411.7108}}].

\bibitem{Moore:2004fg}
G.~W. Moore, {\it {Strings and Arithmetic}},  in {\em {Les Houches School of
  Physics: Frontiers in Number Theory, Physics and Geometry}}, pp.~303--359,
  2007.
\newblock \href{http://arxiv.org/abs/hep-th/0401049}{{\tt hep-th/0401049}}.

\bibitem{Behrndt:1996jn}
K.~Behrndt, G.~Lopes~Cardoso, B.~de~Wit, R.~Kallosh, D.~Lust, and T.~Mohaupt,
  {\it {Classical and quantum N=2 supersymmetric black holes}},  {\em Nucl.
  Phys. B} {\bf 488} (1997) 236--260,
  [\href{http://arxiv.org/abs/hep-th/9610105}{{\tt hep-th/9610105}}].

\bibitem{Behrndt:1998eq}
K.~Behrndt, G.~Lopes~Cardoso, B.~de~Wit, D.~Lust, T.~Mohaupt, and W.~A. Sabra,
  {\it {Higher order black hole solutions in N=2 supergravity and Calabi-Yau
  string backgrounds}},  {\em Phys. Lett. B} {\bf 429} (1998) 289--296,
  [\href{http://arxiv.org/abs/hep-th/9801081}{{\tt hep-th/9801081}}].

\bibitem{LopesCardoso:2000qm}
G.~Lopes~Cardoso, B.~de~Wit, J.~Kappeli, and T.~Mohaupt, {\it {Stationary BPS
  solutions in N=2 supergravity with R**2 interactions}},  {\em JHEP} {\bf 12}
  (2000) 019, [\href{http://arxiv.org/abs/hep-th/0009234}{{\tt
  hep-th/0009234}}].

\bibitem{Witten:1988xj}
E.~Witten, {\it {Topological Sigma Models}},  {\em Commun. Math. Phys.} {\bf
  118} (1988) 411.

\bibitem{Witten:1991zz}
E.~Witten, {\it {Mirror manifolds and topological field theory}},  {\em AMS/IP
  Stud. Adv. Math.} {\bf 9} (1998) 121--160,
  [\href{http://arxiv.org/abs/hep-th/9112056}{{\tt hep-th/9112056}}].

\bibitem{Labastida:1991qq}
J.~M.~F. Labastida and P.~M. Llatas, {\it {Topological matter in
  two-dimensions}},  {\em Nucl. Phys. B} {\bf 379} (1992) 220--258,
  [\href{http://arxiv.org/abs/hep-th/9112051}{{\tt hep-th/9112051}}].

\bibitem{Labastida:1994ss}
J.~M.~F. Labastida and M.~Marino, {\it {Type B topological matter,
  Kodaira-Spencer theory, and mirror symmetry}},  {\em Phys. Lett. B} {\bf 333}
  (1994) 386--395, [\href{http://arxiv.org/abs/hep-th/9405151}{{\tt
  hep-th/9405151}}].

\bibitem{Ferrara:1997yr}
S.~Ferrara, {\it {Bertotti-Robinson geometry and supersymmetry}},  in {\em
  {12th Italian Conference on General Relativity and Gravitational Physics}},
  pp.~135--147, 1, 1997.
\newblock \href{http://arxiv.org/abs/hep-th/9701163}{{\tt hep-th/9701163}}.

\bibitem{Ferrara:1977ij}
S.~Ferrara, M.~Kaku, P.~K. Townsend, and P.~van Nieuwenhuizen, {\it {Gauging
  the Graded Conformal Group with Unitary Internal Symmetries}},  {\em Nucl.
  Phys. B} {\bf 129} (1977) 125--134.

\bibitem{VanProeyen:1983wk}
A.~Van~Proeyen, {\it {Superconformal tensor calculus in N=1 and N=2
  superrgavity}},  in {\em {19th Winter School and Workshop on Theoretical
  Physics: Supersymmetry and Supergravity}}, 4, 1983.

\bibitem{deWit:1984hw}
B.~de~Wit, {\it {Introduction to Supergravity}},  in {\em {Spring School on
  Supergravity and Supersymmetry}}, 6, 1984.

\bibitem{Robles-Llana:2006vct}
D.~Robles-Llana, F.~Saueressig, and S.~Vandoren, {\it {String loop corrected
  hypermultiplet moduli spaces}},  {\em JHEP} {\bf 03} (2006) 081,
  [\href{http://arxiv.org/abs/hep-th/0602164}{{\tt hep-th/0602164}}].

\bibitem{deWit:2010za}
B.~de~Wit, S.~Katmadas, and M.~van Zalk, {\it {New supersymmetric
  higher-derivative couplings: Full N=2 superspace does not count!}},  {\em
  JHEP} {\bf 01} (2011) 007, [\href{http://arxiv.org/abs/1010.2150}{{\tt
  arXiv:1010.2150}}].

\bibitem{Murthy:2013xpa}
S.~Murthy and V.~Reys, {\it {Quantum black hole entropy and the holomorphic
  prepotential of N=2 supergravity}},  {\em JHEP} {\bf 10} (2013) 099,
  [\href{http://arxiv.org/abs/1306.3796}{{\tt arXiv:1306.3796}}].

\bibitem{deWit:1999fp}
B.~de~Wit, B.~Kleijn, and S.~Vandoren, {\it {Superconformal hypermultiplets}},
  {\em Nucl. Phys. B} {\bf 568} (2000) 475--502,
  [\href{http://arxiv.org/abs/hep-th/9909228}{{\tt hep-th/9909228}}].

\bibitem{Strominger:1996sh}
A.~Strominger and C.~Vafa, {\it {Microscopic origin of the Bekenstein-Hawking
  entropy}},  {\em Phys. Lett. B} {\bf 379} (1996) 99--104,
  [\href{http://arxiv.org/abs/hep-th/9601029}{{\tt hep-th/9601029}}].

\bibitem{Zaffaroni:2019dhb}
A.~Zaffaroni, {\it {AdS black holes, holography and localization}},  {\em
  Living Rev. Rel.} {\bf 23} (2020), no.~1 2,
  [\href{http://arxiv.org/abs/1902.07176}{{\tt arXiv:1902.07176}}].

\bibitem{deWit:1992wf}
B.~de~Wit, F.~Vanderseypen, and A.~Van~Proeyen, {\it {Symmetry structure of
  special geometries}},  {\em Nucl. Phys. B} {\bf 400} (1993) 463--524,
  [\href{http://arxiv.org/abs/hep-th/9210068}{{\tt hep-th/9210068}}].

\bibitem{Harvey:1995fq}
J.~A. Harvey and G.~W. Moore, {\it {Algebras, BPS states, and strings}},  {\em
  Nucl. Phys. B} {\bf 463} (1996) 315--368,
  [\href{http://arxiv.org/abs/hep-th/9510182}{{\tt hep-th/9510182}}].

\bibitem{Katz:1999xq}
S.~H. Katz, A.~Klemm, and C.~Vafa, {\it {M theory, topological strings and
  spinning black holes}},  {\em Adv. Theor. Math. Phys.} {\bf 3} (1999)
  1445--1537, [\href{http://arxiv.org/abs/hep-th/9910181}{{\tt
  hep-th/9910181}}].

\bibitem{Grimm:2017okk}
T.~W. Grimm, K.~Mayer, and M.~Weissenbacher, {\it {Higher derivatives in Type
  II and M-theory on Calabi-Yau threefolds}},  {\em JHEP} {\bf 02} (2018) 127,
  [\href{http://arxiv.org/abs/1702.08404}{{\tt arXiv:1702.08404}}].

\bibitem{Marino:1998pg}
M.~Marino and G.~W. Moore, {\it {Counting higher genus curves in a Calabi-Yau
  manifold}},  {\em Nucl. Phys. B} {\bf 543} (1999) 592--614,
  [\href{http://arxiv.org/abs/hep-th/9808131}{{\tt hep-th/9808131}}].

\bibitem{Faber:1998gsw}
C.~Faber and R.~Pandharipande, {\it {Hodge integrals and Gromov-Witten
  theory}},  {\em Inventiones mathematicae} {\bf 139} (2000), no.~1 173--199,
  [\href{http://arxiv.org/abs/math/9810173}{{\tt math/9810173}}].

\bibitem{Gauntlett:2002nw}
J.~P. Gauntlett, J.~B. Gutowski, C.~M. Hull, S.~Pakis, and H.~S. Reall, {\it
  {All supersymmetric solutions of minimal supergravity in five- dimensions}},
  {\em Class. Quant. Grav.} {\bf 20} (2003) 4587--4634,
  [\href{http://arxiv.org/abs/hep-th/0209114}{{\tt hep-th/0209114}}].

\bibitem{Shmakova:1996nz}
M.~Shmakova, {\it {Calabi-Yau black holes}},  {\em Phys. Rev. D} {\bf 56}
  (1997) 540--544, [\href{http://arxiv.org/abs/hep-th/9612076}{{\tt
  hep-th/9612076}}].

\bibitem{Marchesano:2023thx}
F.~Marchesano, L.~Melotti, and L.~Paoloni, {\it {On the moduli space curvature
  at infinity}},  {\em JHEP} {\bf 02} (2024) 103,
  [\href{http://arxiv.org/abs/2311.07979}{{\tt arXiv:2311.07979}}].

\bibitem{Marchesano:2024tod}
F.~Marchesano, L.~Melotti, and M.~Wiesner, {\it {Asymptotic curvature
  divergences and non-gravitational theories}},
  \href{http://arxiv.org/abs/2409.02991}{{\tt arXiv:2409.02991}}.

\bibitem{Castellano:2024gwi}
A.~Castellano, F.~Marchesano, L.~Melotti, and L.~Paoloni, {\it {The Moduli
  Space Curvature and the Weak Gravity Conjecture}},
  \href{http://arxiv.org/abs/2410.10966}{{\tt arXiv:2410.10966}}.

\bibitem{CMP}
A.~Castellano, F.~Marchesano, and L.~Paoloni, ``{To appear}.''

\bibitem{Ferrara:1997tw}
S.~Ferrara, G.~W. Gibbons, and R.~Kallosh, {\it {Black holes and critical
  points in moduli space}},  {\em Nucl. Phys. B} {\bf 500} (1997) 75--93,
  [\href{http://arxiv.org/abs/hep-th/9702103}{{\tt hep-th/9702103}}].

\bibitem{Shenker:1990uf}
S.~H. Shenker, {\it {The Strength of nonperturbative effects in string
  theory}},  in {\em {Cargese Study Institute: Random Surfaces, Quantum Gravity
  and Strings}}, pp.~809--819, 8, 1990.

\bibitem{Pasquetti:2010bps}
S.~Pasquetti and R.~Schiappa, {\it {Borel and Stokes Nonperturbative Phenomena
  in Topological String Theory and c=1 Matrix Models}},  {\em Annales Henri
  Poincare} {\bf 11} (2010) 351--431,
  [\href{http://arxiv.org/abs/0907.4082}{{\tt arXiv:0907.4082}}].

\bibitem{4dBHs}
A.~Castellano and M.~Zatti, ``{To appear}.''

\bibitem{Witten:1995zh}
E.~Witten, {\it {Some comments on string dynamics}},  in {\em {STRINGS 95:
  Future Perspectives in String Theory}}, pp.~501--523, 7, 1995.
\newblock \href{http://arxiv.org/abs/hep-th/9507121}{{\tt hep-th/9507121}}.

\bibitem{Cadavid:1995bk}
A.~C. Cadavid, A.~Ceresole, R.~D'Auria, and S.~Ferrara, {\it
  {Eleven-dimensional supergravity compactified on Calabi-Yau threefolds}},
  {\em Phys. Lett. B} {\bf 357} (1995) 76--80,
  [\href{http://arxiv.org/abs/hep-th/9506144}{{\tt hep-th/9506144}}].

\bibitem{Maldacena:1996ky}
J.~M. Maldacena, {\em {Black holes in string theory}}.
\newblock PhD thesis, Princeton U., 1996.
\newblock \href{http://arxiv.org/abs/hep-th/9607235}{{\tt hep-th/9607235}}.

\bibitem{Maldacena:1997de}
J.~M. Maldacena, A.~Strominger, and E.~Witten, {\it {Black hole entropy in M
  theory}},  {\em JHEP} {\bf 12} (1997) 002,
  [\href{http://arxiv.org/abs/hep-th/9711053}{{\tt hep-th/9711053}}].

\bibitem{Cappelli:1986hf}
A.~Cappelli, C.~Itzykson, and J.~B. Zuber, {\it {Modular invariant partition
  functions in two dimensions}},  {\em Nucl. Phys. B} {\bf 280} (1987)
  445--465.

\bibitem{DiFrancesco:1997nk}
P.~Di~Francesco, P.~Mathieu, and D.~Senechal, {\em {Conformal Field Theory}}.
\newblock Graduate Texts in Contemporary Physics. Springer-Verlag, New York,
  1997.

\bibitem{Miyaoka1987}
Y.~Miyaoka, {\it The chern classes and kodaira dimension of a minimal variety},
   in {\em Algebraic Geometry (Sendai, 1985)}, vol.~10 of {\em Advanced Studies
  in Pure Mathematics}, pp.~449--476.
\newblock 1987.

\bibitem{kanazawa2013}
A.~Kanazawa and P.~M.~H. Wilson, {\it {Trilinear forms and Chern classes of
  Calabi-Yau threefolds}},  {\em Osaka Journal of Mathematics 51 no. 1, (2014)
  203–213} (2013) [\href{http://arxiv.org/abs/1201.3266}{{\tt
  arXiv:1201.3266}}].

\bibitem{Vafa:1997gr}
C.~Vafa, {\it {Black holes and Calabi-Yau threefolds}},  {\em Adv. Theor. Math.
  Phys.} {\bf 2} (1998) 207--218,
  [\href{http://arxiv.org/abs/hep-th/9711067}{{\tt hep-th/9711067}}].

\bibitem{Harvey:1998bx}
J.~A. Harvey, R.~Minasian, and G.~W. Moore, {\it {NonAbelian tensor multiplet
  anomalies}},  {\em JHEP} {\bf 09} (1998) 004,
  [\href{http://arxiv.org/abs/hep-th/9808060}{{\tt hep-th/9808060}}].

\bibitem{deAntonioMartin:2012bi}
A.~de~Antonio~Martin, T.~Ortin, and C.~S. Shahbazi, {\it {The FGK formalism for
  black p-branes in d dimensions}},  {\em JHEP} {\bf 05} (2012) 045,
  [\href{http://arxiv.org/abs/1203.0260}{{\tt arXiv:1203.0260}}].

\bibitem{Meessen:2012su}
P.~Meessen, T.~Ortin, J.~Perz, and C.~S. Shahbazi, {\it {Black holes and black
  strings of N=2, d=5 supergravity in the H-FGK formalism}},  {\em JHEP} {\bf
  09} (2012) 001, [\href{http://arxiv.org/abs/1204.0507}{{\tt
  arXiv:1204.0507}}].

\bibitem{Gomez-Fayren:2023wxk}
C.~Gomez-Fayren, P.~Meessen, T.~Ortin, and M.~Zatti, {\it {Wald entropy in
  Kaluza-Klein black holes}},  {\em JHEP} {\bf 08} (2023) 039,
  [\href{http://arxiv.org/abs/2305.01742}{{\tt arXiv:2305.01742}}].

\bibitem{Sen:2005wa}
A.~Sen, {\it {Black hole entropy function and the attractor mechanism in higher
  derivative gravity}},  {\em JHEP} {\bf 09} (2005) 038,
  [\href{http://arxiv.org/abs/hep-th/0506177}{{\tt hep-th/0506177}}].

\bibitem{Kraus:2005vz}
P.~Kraus and F.~Larsen, {\it {Microscopic black hole entropy in theories with
  higher derivatives}},  {\em JHEP} {\bf 09} (2005) 034,
  [\href{http://arxiv.org/abs/hep-th/0506176}{{\tt hep-th/0506176}}].

\bibitem{Castro:2007sd}
A.~Castro, J.~L. Davis, P.~Kraus, and F.~Larsen, {\it {5D attractors with
  higher derivatives}},  {\em JHEP} {\bf 04} (2007) 091,
  [\href{http://arxiv.org/abs/hep-th/0702072}{{\tt hep-th/0702072}}].

\bibitem{Antoniadis:1997eg}
I.~Antoniadis, S.~Ferrara, R.~Minasian, and K.~S. Narain, {\it {R**4 couplings
  in M and type II theories on Calabi-Yau spaces}},  {\em Nucl. Phys. B} {\bf
  507} (1997) 571--588, [\href{http://arxiv.org/abs/hep-th/9707013}{{\tt
  hep-th/9707013}}].

\bibitem{Marino:2024tbx}
M.~Marino, {\it {Les Houches lectures on non-perturbative topological
  strings}},  \href{http://arxiv.org/abs/2411.16211}{{\tt arXiv:2411.16211}}.

\bibitem{Hattab:2024chf}
J.~Hattab and E.~Palti, {\it {Emergent potentials and non-perturbative open
  topological strings}},  {\em JHEP} {\bf 10} (2024) 195,
  [\href{http://arxiv.org/abs/2408.12302}{{\tt arXiv:2408.12302}}].

\bibitem{Hattab:2024yol}
J.~Hattab and E.~Palti, {\it {On Calabi-Yau Manifolds at Strong Topological
  String Coupling}},  {\em Fortsch. Phys.} {\bf 72} (2024), no.~12 2400199,
  [\href{http://arxiv.org/abs/2409.01721}{{\tt arXiv:2409.01721}}].

\bibitem{Lin:2024jug}
P.~Lin and G.~Shiu, {\it {Schwinger Effect of Extremal Reissner-Nordstr\"om
  Black Holes}},  \href{http://arxiv.org/abs/2409.02197}{{\tt
  arXiv:2409.02197}}.

\bibitem{Gendler:2020dfp}
N.~Gendler and I.~Valenzuela, {\it {Merging the weak gravity and distance
  conjectures using BPS extremal black holes}},  {\em JHEP} {\bf 01} (2021)
  176, [\href{http://arxiv.org/abs/2004.10768}{{\tt arXiv:2004.10768}}].

\bibitem{Heidenreich:2020upe}
B.~Heidenreich, {\it {Black Holes, Moduli, and Long-Range Forces}},  {\em JHEP}
  {\bf 11} (2020) 029, [\href{http://arxiv.org/abs/2006.09378}{{\tt
  arXiv:2006.09378}}].

\bibitem{Heidenreich:2024dmr}
B.~Heidenreich and M.~Lotito, {\it {Proving the Weak Gravity Conjecture in
  Perturbative String Theory, Part I: The Bosonic String}},
  \href{http://arxiv.org/abs/2401.14449}{{\tt arXiv:2401.14449}}.

\bibitem{Gaiotto:2005gf}
D.~Gaiotto, A.~Strominger, and X.~Yin, {\it {New connections between 4-D and
  5-D black holes}},  {\em JHEP} {\bf 02} (2006) 024,
  [\href{http://arxiv.org/abs/hep-th/0503217}{{\tt hep-th/0503217}}].

\bibitem{Kallosh:1996vy}
R.~Kallosh, A.~Rajaraman, and W.~K. Wong, {\it {Supersymmetric rotating black
  holes and attractors}},  {\em Phys. Rev. D} {\bf 55} (1997) R3246--R3249,
  [\href{http://arxiv.org/abs/hep-th/9611094}{{\tt hep-th/9611094}}].

\bibitem{Larsen:2006xm}
F.~Larsen, {\it {The Attractor Mechanism in Five Dimensions}},  {\em Lect.
  Notes Phys.} {\bf 755} (2008) 249--281,
  [\href{http://arxiv.org/abs/hep-th/0608191}{{\tt hep-th/0608191}}].

\bibitem{Schwartz:2014sze}
M.~D. Schwartz, {\em {Quantum Field Theory and the Standard Model}}.
\newblock Cambridge University Press, 3, 2014.

\bibitem{Kim:2003qp}
S.~P. Kim and D.~N. Page, {\it {Schwinger pair production in electric and
  magnetic fields}},  {\em Phys. Rev. D} {\bf 73} (2006) 065020,
  [\href{http://arxiv.org/abs/hep-th/0301132}{{\tt hep-th/0301132}}].

\bibitem{Dunne:2002qg}
G.~V. Dunne and C.~Schubert, {\it {Two loop selfdual Euler-Heisenberg
  Lagrangians. 2. Imaginary part and Borel analysis}},  {\em JHEP} {\bf 06}
  (2002) 042, [\href{http://arxiv.org/abs/hep-th/0205005}{{\tt
  hep-th/0205005}}].

\bibitem{Dunne:2001pp}
G.~V. Dunne and C.~Schubert, {\it {Closed form two loop Euler-Heisenberg
  Lagrangian in a selfdual background}},  {\em Phys. Lett. B} {\bf 526} (2002)
  55--60, [\href{http://arxiv.org/abs/hep-th/0111134}{{\tt hep-th/0111134}}].

\bibitem{Dunne:2002qf}
G.~V. Dunne and C.~Schubert, {\it {Two loop selfdual Euler-Heisenberg
  Lagrangians. 1. Real part and helicity amplitudes}},  {\em JHEP} {\bf 08}
  (2002) 053, [\href{http://arxiv.org/abs/hep-th/0205004}{{\tt
  hep-th/0205004}}].

\bibitem{apostol2012modular}
T.~Apostol, {\em Modular Functions and Dirichlet Series in Number Theory}.
\newblock Graduate Texts in Mathematics. Springer New York, 2012.

\bibitem{Ahlfors1966}
L.~V. Ahlfors, {\em Complex Analysis}.
\newblock McGraw-Hill Book Company, 2~ed.

\bibitem{Castellano:2025yur}
A.~Castellano, D.~L{\"u}st, C.~Montella, and M.~Zatti, {\it {Quantum Calabi-Yau
  Black Holes and Non-Perturbative D0-brane Effects}},
  \href{http://arxiv.org/abs/2505.15920}{{\tt arXiv:2505.15920}}.

\bibitem{Lee:2019wij}
S.-J. Lee, W.~Lerche, and T.~Weigand, {\it {Emergent strings from infinite
  distance limits}},  {\em JHEP} {\bf 02} (2022) 190,
  [\href{http://arxiv.org/abs/1910.01135}{{\tt arXiv:1910.01135}}].

\bibitem{Gabriel_2000}
C.~Gabriel and P.~Spindel, {\it Quantum charged fields in (1+1) rindler space},
   {\em Annals of Physics} {\bf 284} (Sept., 2000) 263–335.

\bibitem{Friedmann:2002gx}
T.~Friedmann and H.~L. Verlinde, {\it {Schwinger pair creation of Kaluza-Klein
  particles: Pair creation without tunneling}},  {\em Phys. Rev. D} {\bf 71}
  (2005) 064018, [\href{http://arxiv.org/abs/hep-th/0212163}{{\tt
  hep-th/0212163}}].

\bibitem{Russo:2009ga}
J.~G. Russo, {\it {On Schwinger Pair Creation in Gravity and in Closed
  Superstring Theory}},  {\em JHEP} {\bf 03} (2009) 080,
  [\href{http://arxiv.org/abs/0901.1664}{{\tt arXiv:0901.1664}}].

\bibitem{Cano:2018hut}
P.~A. Cano, P.~F. Ram\'\i{}rez, and A.~Ruip\'erez, {\it {The small black hole
  illusion}},  {\em JHEP} {\bf 03} (2020) 115,
  [\href{http://arxiv.org/abs/1808.10449}{{\tt arXiv:1808.10449}}].

\bibitem{Dyson:1952tj}
F.~J. Dyson, {\it {Divergence of perturbation theory in quantum
  electrodynamics}},  {\em Phys. Rev.} {\bf 85} (1952) 631--632.

\bibitem{Dorigoni:2014hea}
D.~Dorigoni, {\it {An Introduction to Resurgence, Trans-Series and Alien
  Calculus}},  {\em Annals Phys.} {\bf 409} (2019) 167914,
  [\href{http://arxiv.org/abs/1411.3585}{{\tt arXiv:1411.3585}}].

\bibitem{white2010asymptotic}
R.~White, {\em Asymptotic Analysis of Differential Equations}.
\newblock Asymptotic Analysis of Differential Equations. Imperial College
  Press, 2010.

\bibitem{Bender:1971gu}
C.~M. Bender and T.~T. Wu, {\it {Large order behavior of Perturbation theory}},
   {\em Phys. Rev. Lett.} {\bf 27} (1971) 461.

\bibitem{Bender:1973rz}
C.~M. Bender and T.~T. Wu, {\it {Anharmonic oscillator. 2: A Study of
  perturbation theory in large order}},  {\em Phys. Rev. D} {\bf 7} (1973)
  1620--1636.

\bibitem{Collins:1977dw}
J.~C. Collins and D.~E. Soper, {\it {Large Order Expansion in Perturbation
  Theory}},  {\em Annals Phys.} {\bf 112} (1978) 209--234.

\bibitem{Zinn-Justin:1980oco}
J.~Zinn-Justin, {\it {Perturbation Series at Large Orders in Quantum Mechanics
  and Field Theories: Application to the Problem of Resummation}},  {\em Phys.
  Rept.} {\bf 70} (1981) 109.

\bibitem{LeGuillou:1990nq}
J.~C. Le~Guillou and J.~Zinn-Justin, eds., {\em {Large order behavior of
  perturbation theory}}.
\newblock 1990.

\bibitem{ecalle1981fonctions}
J.~Ecalle, {\em Les fonctions resurgentes. 1. Les alg{\`e}bres de fonctions
  r{\'e}surgentes}.
\newblock No.~parte 3,v. 1 in Fonctions r{\'e}surgentes. Univ. de Paris-Sud,
  D{\'e}p. de Math{\'e}matique, 1981.

\bibitem{Gu:2023mgf}
J.~Gu, A.-K. Kashani-Poor, A.~Klemm, and M.~Marino, {\it {Non-perturbative
  topological string theory on compact Calabi-Yau 3-folds}},  {\em SciPost
  Phys.} {\bf 16} (2024), no.~3 079,
  [\href{http://arxiv.org/abs/2305.19916}{{\tt arXiv:2305.19916}}].

\bibitem{Gu:2021ize}
J.~Gu and M.~Marino, {\it {Peacock patterns and new integer invariants in
  topological string theory}},  {\em SciPost Phys.} {\bf 12} (2022), no.~2 058,
  [\href{http://arxiv.org/abs/2104.07437}{{\tt arXiv:2104.07437}}].

\bibitem{Marino:2015yie}
M.~Mari\~no, {\em {Instantons and Large N}: {An Introduction to
  Non-Perturbative Methods in Quantum Field Theory}}.
\newblock Cambridge University Press, 9, 2015.

\bibitem{Taub:1950ez}
A.~H. Taub, {\it {Empty space-times admitting a three parameter group of
  motions}},  {\em Annals Math.} {\bf 53} (1951) 472--490.

\bibitem{Newman:1963yy}
E.~Newman, L.~Tamburino, and T.~Unti, {\it {Empty space generalization of the
  Schwarzschild metric}},  {\em J. Math. Phys.} {\bf 4} (1963) 915.

\bibitem{Hawking:1976jb}
S.~W. Hawking, {\it {Gravitational Instantons}},  {\em Phys. Lett. A} {\bf 60}
  (1977) 81.

\bibitem{Sorkin:1983ns}
R.~D. Sorkin, {\it {Kaluza-Klein Monopole}},  {\em Phys. Rev. Lett.} {\bf 51}
  (1983) 87--90.

\bibitem{Gross:1983hb}
D.~J. Gross and M.~J. Perry, {\it {Magnetic Monopoles in Kaluza-Klein
  Theories}},  {\em Nucl. Phys. B} {\bf 226} (1983) 29--48.

\bibitem{Sen:1997js}
A.~Sen, {\it {Dynamics of multiple Kaluza-Klein monopoles in M and string
  theory}},  {\em Adv. Theor. Math. Phys.} {\bf 1} (1998) 115--126,
  [\href{http://arxiv.org/abs/hep-th/9707042}{{\tt hep-th/9707042}}].

\bibitem{Majumdar:1947eu}
S.~D. Majumdar, {\it {A class of exact solutions of Einstein's field
  equations}},  {\em Phys. Rev.} {\bf 72} (1947) 390--398.

\bibitem{Papaetrou:1947ib}
A.~Papaetrou, {\it {A Static solution of the equations of the gravitational
  field for an arbitrary charge distribution}},  {\em Proc. Roy. Irish Acad. A}
  {\bf 51} (1947) 191--204.

\bibitem{Hartle:1972ya}
J.~B. Hartle and S.~W. Hawking, {\it {Solutions of the Einstein-Maxwell
  equations with many black holes}},  {\em Commun. Math. Phys.} {\bf 26} (1972)
  87--101.

\bibitem{Asano:2000mx}
M.~Asano, {\it {Compactification and identification of branes in the
  Kaluza-Klein monopole backgrounds}},
  \href{http://arxiv.org/abs/hep-th/0003241}{{\tt hep-th/0003241}}.

\bibitem{Eguchi:1978gw}
T.~Eguchi and A.~J. Hanson, {\it {Selfdual Solutions to Euclidean Gravity}},
  {\em Annals Phys.} {\bf 120} (1979) 82.

\bibitem{Gauntlett:1998fz}
J.~P. Gauntlett, R.~C. Myers, and P.~K. Townsend, {\it {Black holes of D = 5
  supergravity}},  {\em Class. Quant. Grav.} {\bf 16} (1999) 1--21,
  [\href{http://arxiv.org/abs/hep-th/9810204}{{\tt hep-th/9810204}}].

\bibitem{Witten:1995ex}
E.~Witten, {\it {String theory dynamics in various dimensions}},  {\em Nucl.
  Phys. B} {\bf 443} (1995) 85--126,
  [\href{http://arxiv.org/abs/hep-th/9503124}{{\tt hep-th/9503124}}].

\bibitem{Ortin:2015hya}
T.~Ortin, {\em {Gravity and Strings}}.
\newblock Cambridge Monographs on Mathematical Physics. Cambridge University
  Press, 2nd ed.~ed., 7, 2015.

\end{thebibliography}\endgroup
\bibliographystyle{JHEP}

\end{document}